\documentclass{emulateapj}
\usepackage{graphicx}
\usepackage{natbib}

\shorttitle{Extended Halo of Centaurus~A}
\shortauthors{Crnojevi\'c et al.}

\begin{document}


\title{The extended halo of Centaurus~A: uncovering
  satellites, streams and substructures\altaffilmark{*}}


\author{D. Crnojevi\'c\altaffilmark{1}, D. J. Sand\altaffilmark{1}, K. Spekkens\altaffilmark{2}, N. Caldwell\altaffilmark{3}, P. Guhathakurta\altaffilmark{4}, B. McLeod\altaffilmark{3}, A. Seth\altaffilmark{5}, J. D. Simon\altaffilmark{6}, J. Strader\altaffilmark{7}, E. Toloba\altaffilmark{1,4}}





\begin{abstract}
We present the widest-field resolved stellar map to date of the closest
($D\sim3.8$~Mpc) massive elliptical galaxy NGC~5128 (Centaurus~A; Cen~A), extending out to
a projected galactocentric radius of $\sim150$~kpc. The dataset is part of 
our ongoing Panoramic Imaging Survey of Centaurus and Sculptor 
(PISCeS) utilizing the Magellan/Megacam imager. 
We resolve a population of old red giant branch stars down to
$\sim1.5$~mag below the tip
of the red giant branch, reaching surface brightness limits 
as low as $\mu_{V,0}\sim32$~mag~arcsec$^{-2}$. The resulting spatial stellar
density map highlights a plethora of previously unknown streams, shells, and 
satellites, including the first tidally disrupting dwarf around Cen~A
(CenA-MM-Dw3), which underline its active accretion history.
We report 13 previously unknown dwarf satellite candidates,
of which 9 are confirmed to be at the
distance of Cen~A (the remaining 4 are not resolved into stars),
with magnitudes in the range $M_V=-7.2$ to $-13.0$, 
central surface brightness values of $\mu_{V,0}=25.4-26.9$~mag~arcsec$^{-2}$, 
and half-light radii of $r_h=0.22-2.92$~kpc. These values are in line with 
Local Group dwarfs but also lie at the faint/diffuse end of their
distribution; interestingly, CenA-MM-Dw3
has similar properties to the recently discovered ultra-diffuse
galaxies in Virgo and Coma.
Most of the new dwarfs are fainter than the previously known Cen~A 
satellites. The newly discovered dwarfs and halo
substructures are discussed in light of their stellar
populations, and they are compared to those discovered by the
PAndAS survey of M31.

\end{abstract}


\altaffiltext{*}{This paper includes data gathered with the 6.5 meter Magellan Telescopes located at Las Campanas Observatory, Chile.}
\altaffiltext{1}{Department of Physics, Texas Tech University, Box 41051, Lubbock, TX 79409-1051, USA; \email{denija.crnojevic@ttu.edu}}
\altaffiltext{2}{Department of Physics, Royal Military College of Canada, Box 17000, Station Forces, Kingston, Ontario K7L 7B4, Canada}
\altaffiltext{3}{Harvard-Smithsonian Center for Astrophysics, Cambridge, MA 02138, USA}
\altaffiltext{4}{UCO/Lick Observatory, University of California, Santa Cruz, 1156 High Street, Santa Cruz, CA 95064, USA}
\altaffiltext{5}{Department of Physics and Astronomy, University of Utah, Salt Lake City, UT 84112, USA}
\altaffiltext{6}{Observatories of the Carnegie Institution for Science, 813 Santa Barbara Street, Pasadena, CA 91101, USA}
\altaffiltext{7}{Department of Physics and Astronomy, Michigan State University, East Lansing, MI 48824, USA}


\section{Introduction}

The currently favored $\Lambda$Cold Dark Matter ($\Lambda$CDM) cosmological 
scenario is successful at reproducing the observed properties of
large-scale structures, but it encounters challenges at
galactic scales, particularly at the low-luminosity end \citep[e.g.,][]{weinberg13}. 
The $\Lambda$CDM cosmology predicts a bottom-up hierarchical assembly of
structures, with the smallest halos critically contributing to the build-up of the most massive
ones. The number and physical properties of stellar streams associated with accreted
halos are major predictions of $\Lambda$CDM \citep[e.g.,][]{bullock05, johnston08, cooper10},
and since dynamical timescales at large galactocentric distances are comparable to the age of the 
Universe, accretion events can be effectively caught in the act in the outskirts of galaxies. 

Disrupting satellites and streams are clearly detected in the Milky Way (MW) and M31 
\citep[e.g.,][]{ibata94, belokurov06a, grillmair06, grillmair09, bonaca12, grillmair13, 
bernard14, ibata14, koposov14, belokurov15}, but there are significant differences 
between them despite their
shared environment and similar mass: the halo of M31 seems to present a substantially larger number of
substructures than the MW (with the caveat that we might be missing some of the
substructure of the MW due to our position within it), indicating an 
enhanced level of interaction and/or a more active
and prolonged accretion history \citep{richardson08, deason13, pillepich14, dorman15}. 
The stochasticity of the hierarchical assembly process implies that
a broad diversity of halo to halo substructure is expected at any
given halo mass \citep[e.g.,][]{purcell07} due to different accretion histories
\citep[e.g.,][]{johnston08, busha10}, and the relative contributions of in-situ star formation
and disrupted satellites remain poorly understood \citep[e.g.,][]{pillepich14, tissera14, lu15}.
While we can study galaxies within the Local Group (LG) in the most detail due to their proximity, 
they may not be representative of generic MW-mass halos.

Streams/substructures are known to be a common feature in the halos of
giant galaxies beyond the LG \citep{mouhcine10, atkinson13, martinez10, duc15}, 
thus lending credence to $\Lambda$CDM models. However, their mere detection only provides a qualitative 
comparison set with simulations:
more quantitative observational contraints from a wide range of galaxies are urgently needed
to verify predictions for the assembly of galaxy halos at a deeper level.
Halos and their satellites and substructures in galaxies beyond the LG are 
extremely challenging to survey because of their intrinsic faintness. 
Integrated light studies suffer from degeneracies in age, metallicity and extinction and, 
despite being able to detect low surface brightness features and galaxies
\citep[e.g][]{atkinson13, merritt14, martinez10, duc15, koda15, mihos15, mueller15, munoz15}, 
they cannot easily constrain distances. The direct imaging of individual stars
is thus the ultimate way of revealing the information locked in outer halos. From their
position in the color-magnitude diagram (CMD), it is possible to derive the stars' physical
properties, such as metallicity and/or age, with generally less severe
degeneracies with respect to integrated studies, as well as accurate distances. 
{\it Hubble Space Telescope} (HST) imaging can make strategic 
contributions by drilling deep in selected, narrow fields \citep[e.g.,][]{dalcanton09, radburn11,
rejkuba05, rejkuba14, monachesi15}, but this strategy lacks the field of view to perform panoramic 
assessments of galaxy halos. 
Only wide-field imagers on ground-based telescopes can resolve stars over large areas 
for nearby ($D < 10$~Mpc) galaxies, but few such studies have been undertaken to date and 
they cover at most a few tens of kpc into the target halos 
\citep{mouhcine10, bailin11, barker11, greggio14}. A few pioneering works are however underway to 
survey larger portions of the spiral M81 \citep{chiboucas09, chiboucas13, okamoto15} 
and the dwarf spiral/irregular NGC~3109 \citep{sand15}.

At the same time, the discovery of an entirely new class of faint MW and M31
satellites ($M_V\gtrsim-8.0$) one decade ago has opened unexpected perspectives on galaxy 
formation at these low luminosities/masses
\citep[e.g.,][]{willman05, belokurov07, mcconnachie09, martin13}, and 
continues to keep astronomers engaged with the most recent Dark Energy Survey (DES) and 
Panoramic Survey Telescope and Rapid Response System (Pan-STARRS) discoveries
\citep[e.g.,][]{bechtol15, drlica15, kim15a, kim15b, koposov15,
  laevens15a, laevens15b, martin15}.
Despite our increasing knowledge of these lowest-mass systems,
the infamous ``missing satellite'' and ``too big to fail'' problems 
\citep[e.g.,][]{moore99, boylan11} are still calling for a definitive solution.
The former problem, i.e., the overprediction of the number of dark matter subhalos
around a MW halo in simulations compared to the observed numbers, 
can be effectively tackled by extending the census of low-mass satellites 
in environments beyond the LG.

We have addressed these challenges by starting an ambitious wide-field observational
campaign (Panoramic Imaging Survey of Centaurus and Sculptor, or PISCeS) 
to search for faint satellites and signs of hierarchical structure formation 
in the halos of two nearby galaxies of different morphologies, the
spiral NGC~253 ($D \sim 3.5$~Mpc; 
\citealt{radburn11}) and the elliptical NGC~5128, or Centaurus~A
(Cen~A, $D \sim 3.8$~Mpc; 
\citealt{harrisg09}). The former (similar in mass to the MW; \citealt{kara05}) is the 
dominant galaxy in the loose, elongated Sculptor group, while the latter 
(slightly more massive than the MW; \citealt{woodley07}) resides in a rich, more 
dynamically evolved group and shows signs of a recent gas-rich accretion event,
thus we are probing opposite types of environment.
Our project aims at resolving the extended halos of these galaxies
out to projected radii ($\sim$150 kpc) similar to those reached by the Pan-Andromeda Archaeological 
Survey \citep[PAndAS, see][]{mcconnachie09, ibata14} and the Spectroscopic and Photometric 
Landscape of Andromeda's Stellar Halo (SPLASH; \citealt{guata05,
guata06, gilbert12}) around M31.
PISCeS will complement and extend similar deep halo work in other
nearby galaxy systems, sampling a range of morphologies and
environments, with the goal of obtaining quantitative constraints on
theoretical predictions.
The PISCeS discoveries of two faint satellites around each of NGC~253 and Cen~A are
already reported in \citet{crnojevic14b}, \citet{sand14} and
\citet{toloba15}.

This paper presents the first general overview of the PISCeS survey
for Cen~A, within the coverage obtained to date ($\sim$70\% of the total
planned area).
In \S \ref{sec:sur} we describe our survey and the data reduction process, 
while in \S \ref{sec:glob} we present the global properties of the
stellar populations and spatial stellar density map of the halo of Cen~A.
In \S \ref{sec:cmds_subs} we focus on the new streams and substructures in 
the halo of Cen~A and on their distances, as well as on the properties 
(surface brightness profile, structural parameters, luminosity) of its 
most robust newly discovered satellites, i.e., those confirmed to be
at the distance of Cen~A.
In \S \ref{sec:concl} we discuss the current results and the future 
contributions stemming from the PISCeS survey, which will include an
investigation of the halo of Cen~A profile, a detailed analysis of its
satellites' stellar populations, and the derivation of the satellite 
luminosity function of Cen~A in light of our detection limits.


\section{The PISCeS survey} \label{sec:sur}

The outer halo of Cen~A has been targeted as part of the PISCeS survey with the 
Megacam imager \citep{mcleod15} on the Magellan Clay 6.5-m telescope, Las Campanas
Observatory. Megacam has a $\sim24'\times24'$ field-of-view and a binned
pixel scale of 0\farcs16. Over the course of 25 nights
during the first semesters of 2010, 2011, 2013, 2014 and 2015, we have observed 80
fields around Cen~A, out of which 5 had to be completely discarded due to poor
observing conditions. The survey to date covers $\sim$11~deg$^2$
and reaches out to a projected radius of $\sim$150~kpc in the North and East
direction and $\sim$100~kpc in the South and West direction, with an additional 
extension along the North-West minor axis of Cen~A out to $\sim$200~kpc.
The final year of observations is planned to be 2016, when we will reach the complete
survey area of $\sim$15~deg$^2$ by targeting the regions South and
South-West of Cen~A out to $\sim$150~kpc.  

A typical pointing taken in good seeing conditions is observed for 
$6\times300$~sec in both $g$- and $r$-band, except for a few cases 
where fewer individual exposures per filter were taken. 
The median seeing throughout the survey to date has been $\sim$0\farcs65,
the best/worst seeing being $\sim$0\farcs5/1\farcs3 in both bands.
In poorer seeing conditions ($\gtrsim$0\farcs9) or in the presence of 
clouds, we increase the exposure time 
by up to a factor of two. The exposure times and seeing conditions for each
pointing will be fully reported in a future contribution, together
with the final photometric catalogues. 
The individual exposures have been dithered in order
to fill in the gaps between the 36 Megacam CCDs.
The reduction process is performed 
by the Smithsonian Astrophysical Observatory Telescope Data Center, 
using a dedicated code developed by M. Conroy, J. Roll and B. McLeod,
and details can be found in \citet{mcleod15}. Briefly, 
after standard pre-processing steps (overscan, bias, dark
correction, flat-fielding, cosmic ray rejection, defringing, 
illumination correction, astrometric solution),
the individual exposures are regridded and combined into stacks using
SWARP \citep{bertin10}.

We perform point spread function (PSF)-fitting photometry on each of
the stacked final images (rather than on the individual exposures), 
using the suite of programs DAOPHOT and 
ALLFRAME \citep{stetson87, stetson94}. To construct a reliable PSF
model, we pick $\sim300-400$ isolated, non-saturated bright stars
across the Megacam field-of-view and allow the model solution to vary
as a function of position. The typical variation of the PSF across the 
field-of-view is on the order of $\sim5\%-10\%$. We set our source detection 
limit to 3~$\sigma$ above the background level, and derive
photometry for each band with ALLSTAR. The catalogues for each band
are matched with DAOMATCH/DAOMASTER 
\citep{stetson93}, and then a final round of photometry is performed
simultaneously in both bands through ALLFRAME, in order to increase
photometric depth.
Given the wide range of observing conditions, the quality cuts applied
to the photometric catalogues are tailored to each pointing, and 
fluctuate around sharpness values of $\left|sharp\right|<3$ and around 
$\chi<1.5$. These cuts eliminate most contaminant galaxies.

Once the clean photometric catalogues are obtained for each pointing, we
proceed to calibrate the instrumental magnitudes to the Sloan Digital 
Sky Survey (SDSS) system. Overlaps between adjacent pointings are 
part of our observing strategy, and are used to calibrate those
pointings aquired under non-photometric conditions. Out of the 
25 nights of observations, 8 were photometric: during each of these nights
we observed two to three equatorial SDSS fields 
at different airmasses. We linearly derive zeropoints and color terms and
their relative airmass dependence, and apply the
calibration to all the pointings taken on that photometric night. For
the remainder of the pointings, we propagate the
calibration based on the overlapping regions,
which are $\sim$2~arcmin wide. We derive pointing-to-pointing 
zeropoints and color terms from these overlaps, using on average 
$\sim100-300$ reference bright stars with calibrated magnitudes in the range 
$19<r<22$ and with photometric errors smaller than 0.05~mag.
We ultimately compare pointings at opposite sides of the surveyed area in order to
check our calibration: the overall median calibration 
uncertainty is estimated to be $\sim0.03$~mag for each band across the entire
survey. 
We finally trim the overlapping edges for adjacent pointings, keeping
the photometry from the deepest one within each considered pair.

\begin{figure*}
 \centering
\includegraphics[width=8cm]{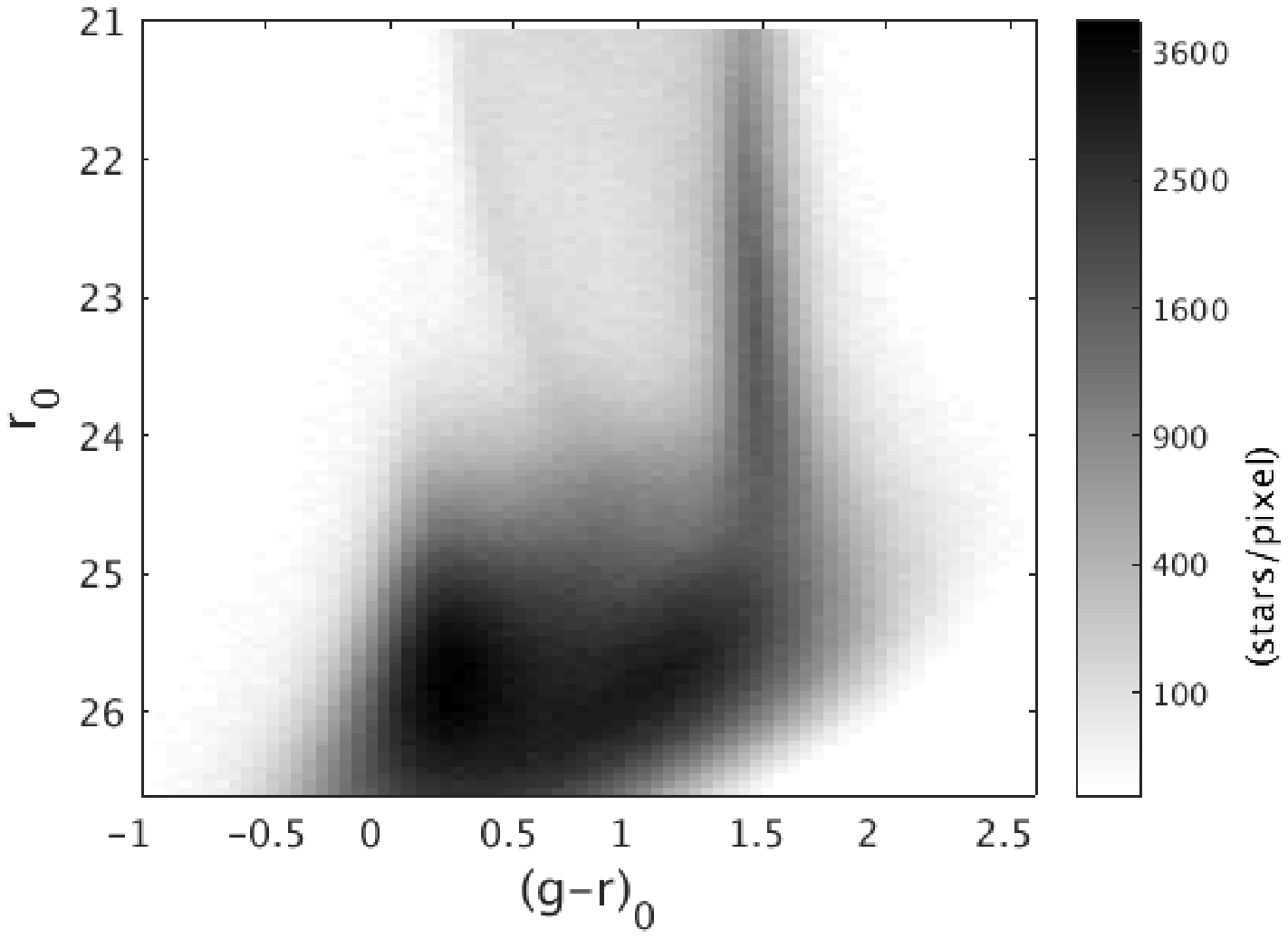}
\includegraphics[width=8cm]{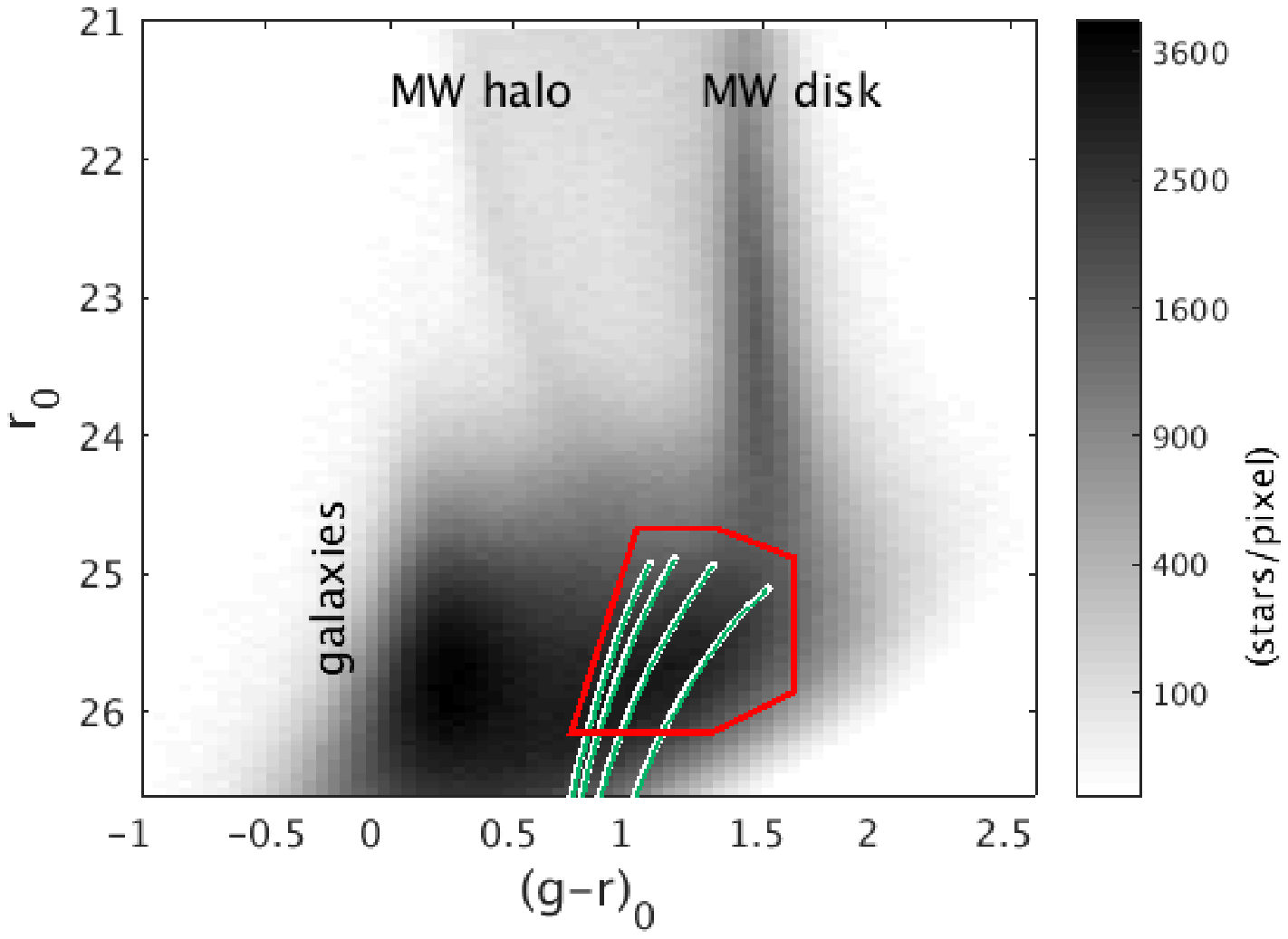}
\caption{\emph{Left panel}. De-reddened Hess diagram of the entire 
Cen~A stellar catalogue to date.
The CMD is binned into $0.05\times0.05$~mag pixels in both magnitude and color
and displayed on a square root scale. 
\emph{Right panel}. We overlay Dartmouth isochrones with 
a fixed age of 12~Gyr and varying metallicity from [Fe/H]$=-2.5$ to $-1.0$ 
(from left to right) in 0.5~dex steps, 
shifted to the distance of Cen~A to indicate the position of old RGB stars. The red
box indicates our selection for RGB stars, and the accompanying RGB stellar map 
of Cen~A is presented in Fig.~\ref{spat_glob}. The contaminant sequences are clearly
identified with respect to the RGB population of Cen~A: foreground stars are mostly found above the TRGB at  
$(g-r)_0$$\sim$0.3 (Galactic halo) and $\sim$1.5 (Galactic disk), and
their numbers drop significantly at fainter magnitudes, while 
unresolved background galaxies are concentrated blueward of the RGB around
$(g-r)_0$$\sim0.2$.}
\label{cmd_glob}
\end{figure*}

The final calibrated catalogues are dereddened on a source-by-source
basis by interpolating the \citet{schlegel98} extinction maps and 
the \citet{schlafly11} correction coefficients: $A_g=3.303\times E(B-V)$ and
$A_r=2.285\times E(B-V)$.
The median reddening over the surveyed area is $E(B-V)$$\sim$0.12,
with spatial variations of up to $\sim15\%$.
For the rest of the paper we will be assuming dereddened $g_0$ and $r_0$
magnitudes.  In Fig.~\ref{cmd_glob}, we present the global 
CMD of Cen~A, which we discuss in further detail in Sect.~\ref{sec:glob}.

\subsection{Artificial star experiments}

Given the varying observing conditions, the photometric depth is not 
uniform across the survey area (see also next section). It is thus critical to assess the 
photometric uncertainties and incompleteness to meaningfully interpret our wide-field dataset.
For each pointing, we inject a number of fake stars between 5 and 10
times the number of recovered bona-fide sources (i.e., after applying the quality
cuts to the photometric catalogues), distributed into tens of experiments in order not to increase
stellar crowding artificially. The minimum number of fake stars
produced is $\sim$$10^6$ per pointing, and they are homogeneously distributed across
the field-of-view of the pointing. The fake stars have a similar color-magnitude
distribution to that of the observed sources, except for a deeper
extension at faint magnitudes (down to $\sim$2~mag fainter than the
faintest real recovered stars), so as to take into account those faint
objects that are upscattered in the observed CMD due to noise. 
Their photometry is derived exactly in the same way as for the real data, and the same
quality cuts and calibration are applied. 

\begin{figure}
 \centering
\includegraphics[width=7cm]{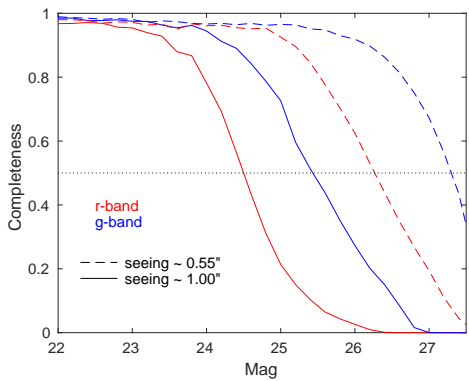}
\caption{Completeness curves in both $g$- (blue lines) and $r$-band 
(red lines) for two pointings within the PISCeS survey taken under 
different seeing conditions (as indicated). The dotted black line denotes 
the 50 per cent completeness level.}
\label{incompl}
\end{figure}

The overall photometric incompleteness is a product of
observing conditions (seeing and transparency), local stellar crowding, and spatial coverage
(e.g., holes in the stellar distribution are often present due to
saturated foreground stars). 
The color-averaged $50\%$ completeness limit per pointing 
varies from $r_0$$\sim$24.5--26.2 and $g_0$$\sim$25.4--27.3.
In Fig.~\ref{incompl} we show the completeness curves in
both bands for a pointing taken under good seeing conditions ($\sim$0\farcs55),
as well as one with poorer seeing ($\sim$1\farcs00).


\section{Global resolved stellar halo properties} \label{sec:glob}

\subsection{Global color-magnitude diagram} 
 
In Fig.~\ref{cmd_glob} we show the dereddened global color-magnitude diagram
(CMD) across the PISCeS survey, containing $\sim5\times 10^6$
stellar sources. We bin the CMD into $0.05\times0.05$~mag pixels in
color-magnitude space, and display it in the form of a Hess density
diagram (square root scale).
In the right panel, we overplot solar-scaled Dartmouth
isochrones in the SDSS filter system \citep{dotter08}. The models are chosen
to have a fixed old age (12~Gyr) and varying [Fe/H] 
(from $-2.5$ to $-1.0$ in 0.5~dex steps), to illustrate the 
position of old red giant branch (RGB) stars at the distance of Cen~A.

The contaminant sources are highlighted by this CMD: 1) Galactic
foreground stars are distributed in two almost vertical sequences at
magnitudes brighter than the tip of the RGB (TRGB) in Cen~A, at colors
$(g-r)_0$$\sim$0.3 (Galactic halo) and $\sim$1.5 (Galactic disk); and 
2) unresolved background galaxies are concentrated in a magnitude 
range similar to that of Cen~A RGB stars and blueward of it, i.e., 
$(g-r)_0$$\sim0.2$. The presence of intermediate-age
stellar populations ($\sim1-8$~Gyr) is indicated by asymptotic giant
branch (AGB) stars above the TRGB (old AGB stars at these magnitudes are 
far less numerous); while this region of the CMD is heavily 
contaminated by Galactic disk stars, they can still be detected as local 
statistical overdensities with respect to the contaminants. Similarly, 
unresolved galaxies at blue colors overlap the locus of young massive stars, 
if they are present. The possible young and intermediate-age populations will not 
be further quantified in this contribution. 

In order to investigate the halo of Cen~A and its substructures on such
large scales, we have to adopt an RGB selection box that minimizes the
fraction of contaminants (see above) and at the same time minimizes
the pointing-to-pointing variations due to inevitably varying
observing conditions, i.e., photometric incompleteness. We therefore 
select the red box drawn in Fig.~\ref{cmd_glob}, which has the following
coordinates: $(g-r)_0=(0.68, 0.94, 1.27, 1.58, 1.58, 1.25)$ and
$r_0=(26.10, 24.63, 24.63, 24.84, 25.80, 26.10)$.
This box encompasses RGB stars with $-2.5\lesssim$[Fe/H]$\lesssim-0.7$ 
and reaches $\sim1.2$~mag below the metal-poor TRGB ($r_0=26.1$). We allow 
the RGB box to extend $\sim0.3$~mag above the TRGB as well (corresponding to a distance
$\sim300$~kpc closer than the nominal Cen~A distance), in order to take 
into account possible distance gradients in the halo of Cen~A, as well
as photometric uncertainties.
 
\begin{figure*}
 \centering
\includegraphics[width=12cm]{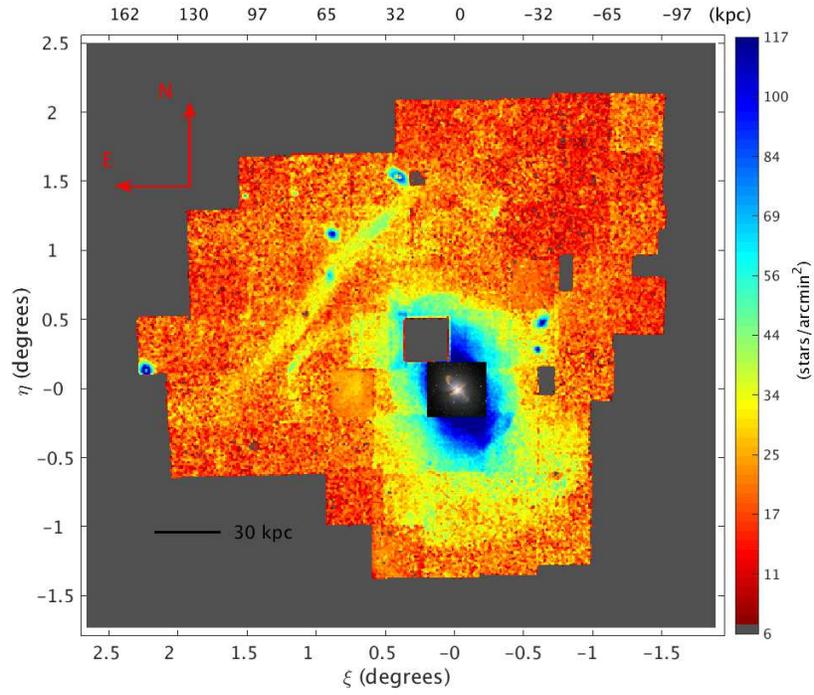}
\includegraphics[width=12cm]{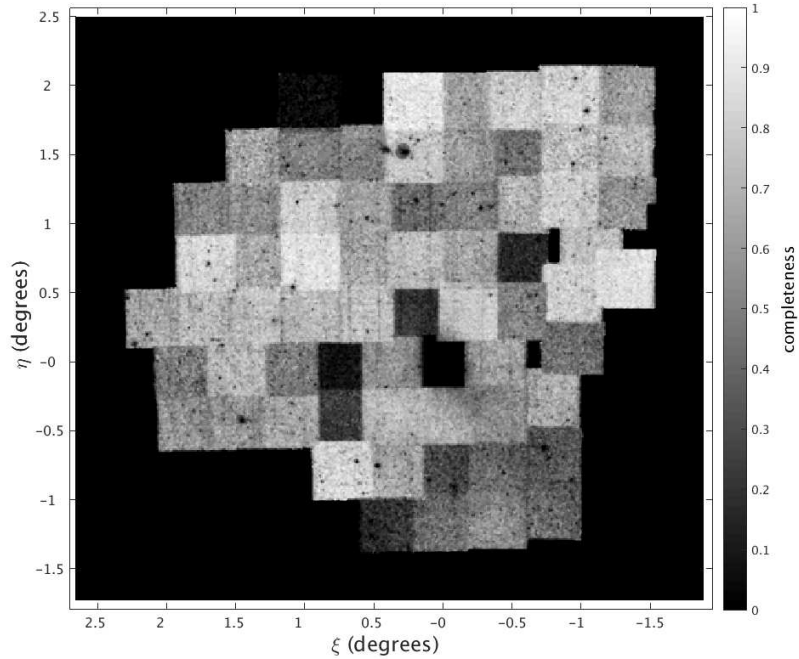}
\caption{\emph{Upper panel}. Density map of Cen~A RGB stars, in standard coordinates centered on Cen~A. 
Stars from the red selection box illustrated in Fig.~\ref{cmd_glob} are placed 
into $3\times3$~arcmin$^2$ bins, and the counts are corrected for incompleteness 
(see Sect.~\ref{sec:maps} for details). The density 
scale is shown on the right, while the physical scale is reported on 
the upper axis. The central pointing is highly incomplete due to extreme 
stellar crowding, and we replace it with a composite color image of 
Cen~A (credit: ESO/WFI (Optical); MPIfR/ESO/APEX/A.Weiss et al. 
(Submillimetre); NASA/CXC/CfA/R.Kraft et al. (X-ray); 
image at http://www.eso.org/public/images/eso0903a/).
\emph{Lower panel}. Completeness map for the region shown in the upper panel,
for stars in the RGB selection box (the central pointing has been omitted).
The colorbar covers completeness values from $0\%$ (0) to $100\%$ (1).}
\label{spat_glob}
\end{figure*}

\begin{figure*}
 \centering
\includegraphics[width=18cm]{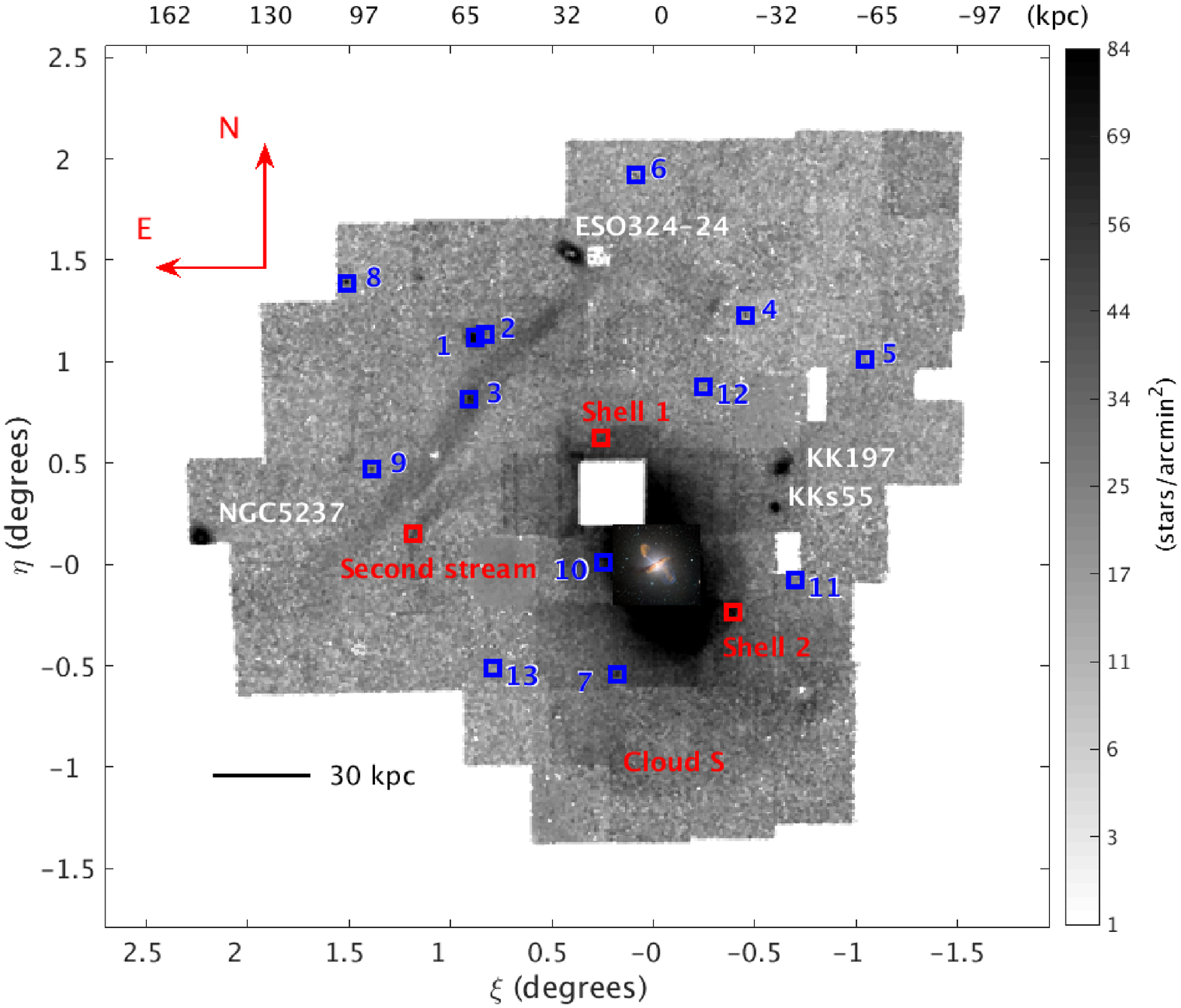}
\caption{Same as the upper panel of Fig.~\ref{spat_glob}: we label our 13 newly 
discovered dwarf candidates in blue (from CenA-MM-Dw1 to CenA-MM-Dw13, where the
label on the plot corresponds to the number of the dwarf), and the 4 previously 
known dwarfs in white. The red labels indicate the most prominent
stream and shell features in the halo of Cen~A.}
\label{spat_zoom}
\end{figure*}

\subsection{Red giant branch spatial density map} \label{sec:maps}

We show the RGB spatial density map of Cen~A in Fig.~\ref{spat_glob},
aimed at highlighting the substructure in its extended halo.
RGB stars within the selection box drawn in Fig.~\ref{cmd_glob}
are binned into $3\times3$~arcmin$^2$ bins
(this value was chosen after visual inspection of sample
maps produced with a range of bin sizes, with the aim of 
balancing spatial resolution and homogeneity).
In this map we exclude seven additional pointings 
obtained in 2010, which extend along the minor axis of Cen~A in the 
NW direction, and two separate pointings 
in the South targeting two known Cen~A dwarfs (KK196 and KK203); 
we will present these data in a future contribution. 
We also leave out two additional pointings, centered at 
$(\xi=1.0^{\circ},\eta=1.9^{\circ})$ and $(0.2^{\circ},0.3^{\circ})$,
because of their low completeness; the latter, in particular,
is located in the very central, high density region, which would make 
its contrast with the adjacent pointings even more apparent.
The completeness map for this region is also shown in the same figure,
and we derive it by counting the number of recovered fake stars with respect 
to the number of injected fake stars within the RGB selection box
in $3\times3$~arcmin$^2$ spatial bins, for each pointing.
The completeness map illustrates the wide range of observing conditions under which 
these pointings were taken. 
The RGB map is corrected by weighting the derived RGB star counts in each spatial bin 
by the relative completeness value, whenever the latter is higher than $20\%$.
For lower completeness values, we increase the stellar density per bin to match
the median density of surrounding bins within a $40<r<50$~bin
annulus. 

Our study is the first to contiguously map the extended halo 
of any elliptical galaxy in resolved stars out to such large
galactocentric distances and down to such low surface brightness limits
($\mu_{V,0}$$\sim$32~mag~arcsec$^{-2}$). This surface brightness limit
is an approximate value (i.e., it varies from pointing to pointing) 
obtained from the field-subtracted surface brightness
profiles of the newly discovered dwarfs (see Sect.~\ref{sec:structlum}).
The central regions of Cen~A (the innermost $\sim$1$\times$1~deg$^2$ in
Fig.~\ref{spat_glob}) have previously been the target of 
several integrated light studies \citep[see, e.g.,][]{israel98, peng02},
which expose signs of a recent merger event and of additional shells
by tweaking the contrast between surface brightness features.
We clearly resolve some of these previously known features into stars, which appear 
saturated in our RGB map as the contrast was tuned for much fainter 
features in the outer halo of the galaxy.  
Even so, a number of newly uncovered shells with embedded
overdensities show minor accretion events caught in the act -- 
see positions $(\xi=0.3^{\circ},\eta=0.6^{\circ})$ and 
$(-0.4^{\circ},-0.25^{\circ})$, labeled Shell 1 and
Shell 2, respectively, in Fig.~\ref{spat_zoom}.

In addition, the unprecedented areal coverage clearly shows the 
presence of an old, relatively metal-poor population across the whole survey,
confirming prior results with smaller fields-of-view  
\citep[e.g.,][]{crnojevic13, rejkuba14}, along with a plethora of
previously unknown streams and substructures. The overdensities in the RGB spatial
map not only uncover extremely low surface brightness features, but
also allow us to discover very faint dwarfs around Cen~A
\citep[see also][]{crnojevic14b}.
Within the survey area there are only four previously known dwarfs,
i.e., ESO324-24 $(\xi=0.3^{\circ},\eta=1.6^{\circ})$, NGC5237 $(2.2^{\circ},0.1^{\circ})$, 
KKs55 $(-0.7^{\circ},0.3^{\circ})$, and KK197 $(-0.75^{\circ},0.5^{\circ})$ 
\citep[see, e.g.,][]{kara05, kara07}, and they are labeled in white in Fig.~\ref{spat_zoom}. 
In the same area, we found 13 dwarf candidates (blue labels), which allow us to significantly extend
the faint end of the satellite luminosity function of Cen~A (see Sect.~\ref{sec:cmds_sats}).
At the same time, red squares in Fig. ~\ref{spat_zoom} are placed on the
most prominent new streams and shells in the outer halo of Cen~A (see Sect.~\ref{sec:cmds_subs}).
We will discuss both the satellites and the substructures in the next section,
and we additionally show a zoomed-in region of
the large RGB map in Fig.~\ref{spat_box}, focused on
the northern portion of the halo of Cen~A.

Finally, we stress that the inevitable residual contamination from foreground
and background sources in our RGB selection box does not significantly
affect the presented density map. We have derived similar maps for the
contaminant sources and found them to be smooth overall, and will
present these maps in a forthcoming paper on the smooth halo of Cen~A
vs its accreted substructure.


\section{Newly discovered satellites and halo substructures} \label{sec:cmds_subs}

Within the current survey footprint, we identified 13 candidate
satellites of Cen~A and several halo substructures. We first perform a visual 
inspection on smoothed images, and subsequently complement the search by identifying
overdensities in the RGB density map.
The location of the new satellites/substructures with respect to Cen~A is shown 
in Fig.~\ref{spat_zoom} (blue and red labels). 
Of the 13 candidate satellites, we have already presented the pair of dwarfs CenA-MM-Dw1 and
CenA-MM-Dw2 in \citet{crnojevic14b}. Seven additional candidates 
(CenA-MM-Dw3 to CenA-MM-Dw9, named in order of discovery)
are robustly confirmed from the unambiguous presence of RGB stars at the
distance of Cen~A in their CMDs, while four (CenA-MM-Dw10 to CenA-MM-Dw13) are only detected as
surface brightness enhancements in the images and do not have 
a significant resolved counterpart (i.e., we cannot constrain their distances
with the method presented in the next subsection).
These latter diffuse objects might be Cen~A satellites fainter than our detection
limit, such that the number of RGB stars in our covered magnitude
range is not large enough for them to show up as overdensities, or else they might
be background galaxies. We report their positions in Fig.~\ref{spat_zoom}
and in Table~\ref{tab2}, but we do not consider them further here.

We compute distances for all the newly discovered features in halo of Cen~A, and
additionally derive surface brightness profiles and structural parameters for
the robust new candidate dwarfs. We briefly describe our methods in the following
subsection, and then discuss the derived properties of our discoveries 
in more detail.

\tabletypesize{\scriptsize}
\begin{deluxetable} {lcc}
\tablecolumns{3}
\tablecaption{Coordinates of unresolved dwarf candidates (Fig.~\ref{spat_zoom}).}

\tablehead{\colhead{Dwarf candidate}  & \colhead{RA (h:m:s)} &\colhead{Dec (d:m:s)}}\\
  
\startdata
Dw10 & 13:26:49.2 & $-43$:00:00\\
Dw11 & 13:21:40.1 & $-43$:04:57\\
Dw12 & 13:24:10.9 & $-42$:08:23\\
Dw13 & 13:29:51.7 & $-43$:31:10\\
\enddata
\label{tab2}
\end{deluxetable}

\subsection{Methods} \label{sec:methods}

\subsubsection{Distances} \label{sec:distance}

Distances throughout this paper are derived with the TRGB method 
\citep[e.g.,][]{dacosta90, lee93, Salaris02, rizzi07},
as already described in \citet{crnojevic14b}. Briefly, the luminosity of the brightest
old, metal-poor RGB stars is nearly constant in red bands, and can 
thus be used as a standard candle.
We employ a Sobel detection filter to highlight the position of the sharp transition
in the $r$-band luminosity function that indicates the TRGB. The luminosity function
is corrected for field contamination, and it only includes those stars
with colors $0.9<(g-r)_0<1.5$ (corresponding to isochrones with 
metallicities [Fe/H]$=-2.5$ to $-1.0$) in order to avoid 
contamination from blue unresolved galaxies.  
The derived $r_{0,\rm TRGB}$ values are then transformed into distance
moduli by assuming $M_r^{\rm TRGB}=-3.01\pm0.10$ \citep[see][]{sand14};
the resulting errors mainly depend on 
photometric uncertainties and on the TRGB value calibration uncertainty.
As we will measure distances to the 
satellites and to many surface brightness features in our Cen~A RGB map, we tabulate 
all of our results in Table~\ref{tab0} and Table~\ref{tab1}, referencing directly the 
features seen in Fig.~\ref{spat_box}.

\subsubsection{Structural parameters and luminosities}  \label{sec:structlum}

\begin{figure*}
 \centering
\includegraphics[width=12.5cm]{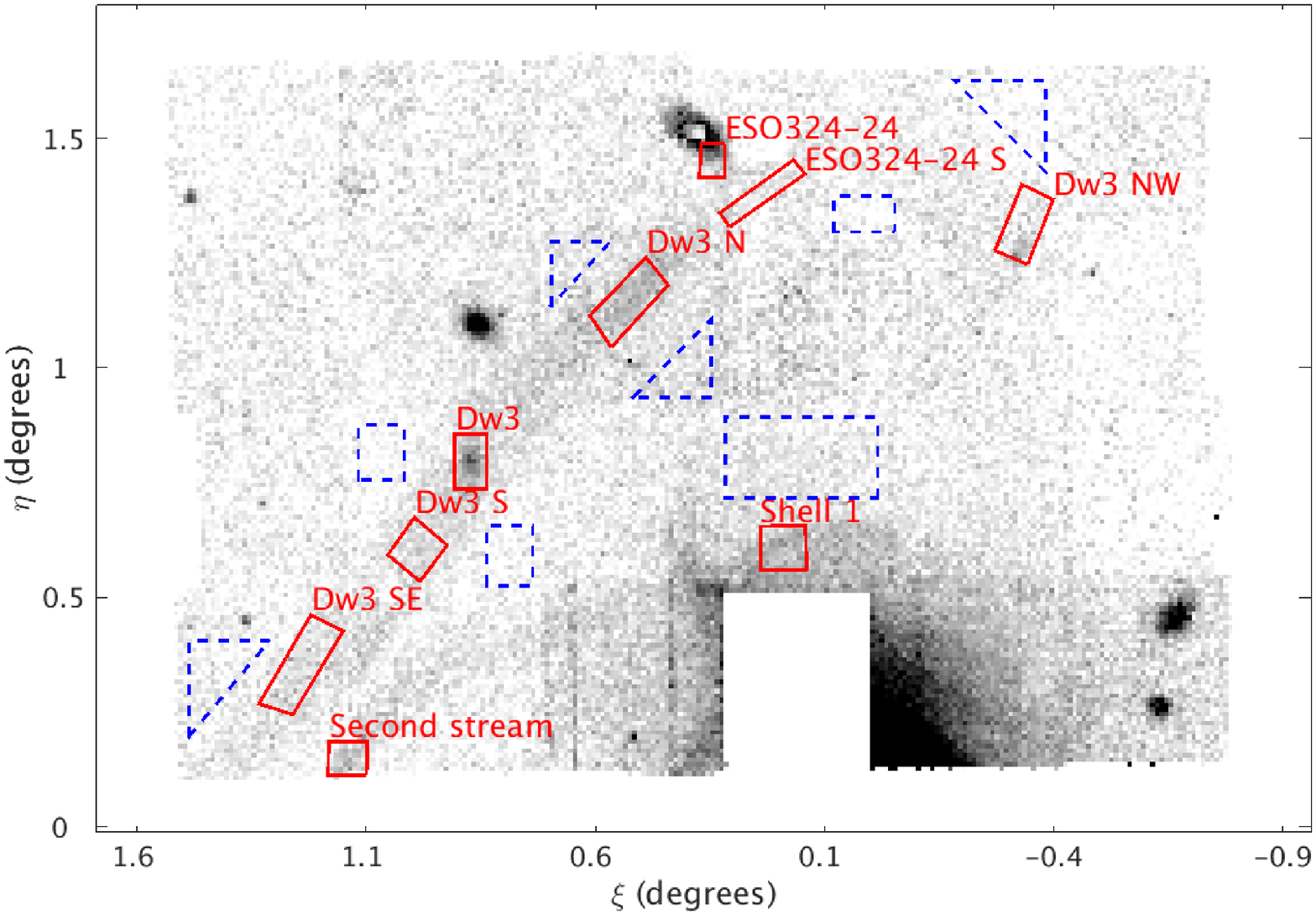}
\raisebox{0.4\height}{\includegraphics[width=5.cm]{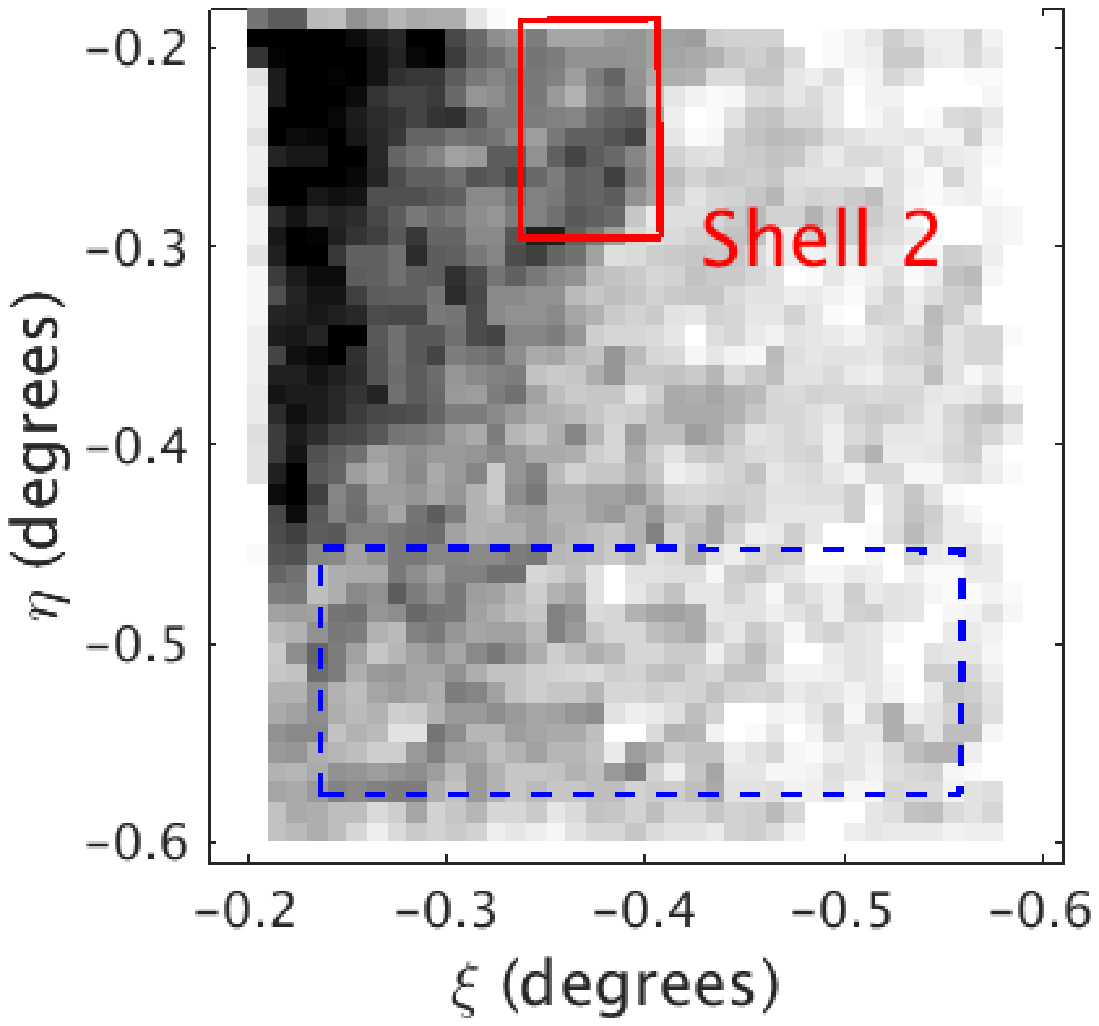}}
\caption{\emph{Left panel}. RGB density map for a zoomed-in region containing 
the most prominent substructures discovered in the northern 
portion of the survey (see Fig~\ref{spat_zoom} and Sect.~\ref{sec:cmds_subs}). 
Red boxes indicate the regions for which CMDs are drawn in 
Fig.~\ref{dw16_figs}--\ref{cmds_spati3};
blue dashed boxes indicate the field regions chosen on a pointing-to-pointing 
basis for the red boxes. \emph{Right panel}. Same as left panel,
for the pointing containing Shell 2 (located to the South-West of the 
center in Fig.~\ref{spat_zoom}).
}
\label{spat_box}
\end{figure*}

For the candidate dwarfs, we compute the ellipticity and position angle of their RGB stellar
distribution with the method of moments, as illustrated in detail in, e.g., 
\citet{crnojevic14a}. 
To derive their surface brightness profiles, we sum up the flux of RGB stars from our
selection box as a function of projected elliptical radius,
correct the results for incompleteness (averaged within the RGB box), divide by 
the annulus area, subtract the field level as derived from the regions shown 
in Fig.~\ref{spat_box} (see next subsection for details), 
and convert into surface brightness. In order to account for
unresolved light, we furthermore derive the surface brightness of the
innermost datapoint (i.e., within a 0.5~arcmin radius for CenA-MM-Dw3, 
and a 0.2~arcmin radius for the other dwarfs, see Figs.~\ref{dw16_figs}--\ref{dw18_figs})
via direct aperture photometry on the $r$-band image. The aperture 
photometry is performed after masking foreground/background objects, 
and the median pixel value (i.e., background level) for that image is subtracted.
We subsequently shift the first point of the profile derived from
individual RGB stars to match the central aperture photometry value. 
With this method we are able to efficiently trace the extremely low surface brightness 
in the outer regions of our dwarf targets via resolved RGB stars \citep[see,
e.g.,][]{crnojevic14a}. The resulting profiles are shown in 
the bottom panels of Figs.~\ref{dw16_figs}--\ref{dw18_figs}. 
We fit the exponential profiles via least squares minimization 
to the surface brightness profiles, and obtain values for the half-light radius 
($r_h$) and the central surface brightness ($\mu_{r,0}$, which is 
translated into $\mu_{V,0}$ by adopting the \citealt{jester05} conversion factors).
Finally, we compute absolute magnitudes by integrating the best-fit exponential profile.
All the derived quantities are found in Table~\ref{tab1}.

\begin{figure*}
 \centering
\includegraphics[width=12cm]{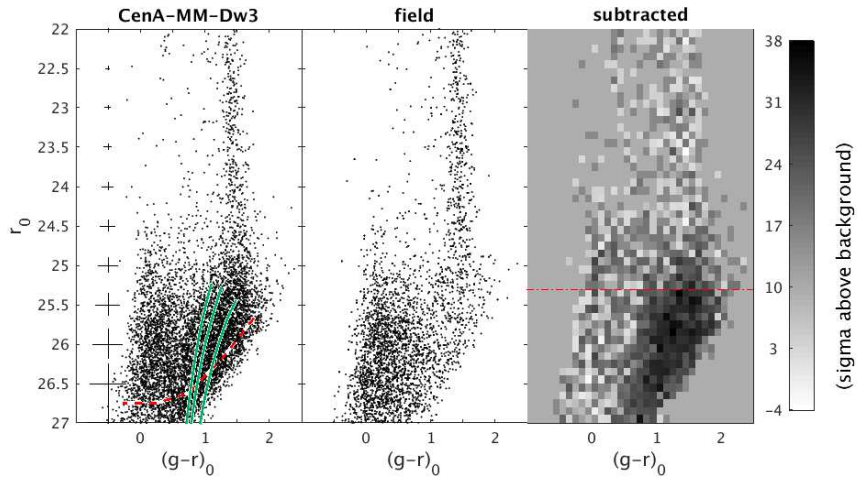}
\caption{Dereddened color-magnitude diagram
for CenA-MM-Dw3 (region labeled as Dw3 in Fig.~\ref{spat_box}). We overplot
isochrones shifted to the measured distance of the dwarf,
with a 12~Gyr age and metallicities
[Fe/H]$=-2.0$, $-1.5$ and $-1.0$ (from left to right). The red dashed line indicates 
the $50\%$ completeness level, and photometric errors as derived 
from artificial star tests are drawn on the left side of the CMD.
A background field CMD drawn from the same pointing (dashed blue
rectangles next to Dw3 in Fig.~\ref{spat_box}) and
rescaled to the box area of CenA-MM-Dw3 is shown for comparison 
in the \emph{middle panel}. The \emph{right panel} is a Hess 
diagram of the CMD of CenA-MM-Dw3 after background subtraction. The red 
dot-dashed line indicates the derived TRGB magnitude.
}
\label{dw16_cmd}
\end{figure*}

\subsection{CenA-MM-Dw3 and its extended tidal tails} 

The RGB maps of Cen~A show a clearly disrupting dwarf galaxy with tidal tails spanning 
over 1.5~degrees, which we dub CenA-MM-Dw3 (see Figs.~\ref{spat_glob} and \ref{spat_zoom}).
A zoom-in RGB density map of the North-East portion of the survey, containing 
CenA-MM-Dw3, is presented in Fig.~\ref{spat_box}. 
We select the remnant of CenA-MM-Dw3 and a few regions along its stream from 
the map in Fig.~\ref{spat_box} (red polygons) in order to have a closer look at their 
stellar populations, and we label them e.g., `Dw3 S', 
`Dw3 N', etc. The central coordinates for the selected regions are 
reported in Table~\ref{tab0}.
We extract CMDs for these regions and show them in Figs.~\ref{dw16_cmd}, 
\ref{cmds_spati}, \ref{cmds_spati2}.
Given that the substructures under study have been observed within
different Megacam pointings, at different projected distances from the
center of Cen~A, and under a range of conditions, we will plot the CMD of each one
together with a local ``field'' (i.e., background plus foreground, 
plus a likely Cen~A halo contribution) CMD extracted from the same pointing:
the closest blue dashed polygon to each of the red boxes, or an average of the 
two closest polygons, is adopted as its field (see Fig.~\ref{spat_box}),
appropriately scaled to the target substructure area. 
 
In Fig.~\ref{dw16_cmd} we plot the CMD of the
main galaxy remnant, which shows a prominent, red RGB. Besides these old stellar
populations, CenA-MM-Dw3 also has an overdensity of sources above the TRGB 
which may possibly indicate a population of intermediate-age (a few Gyr), luminous AGB stars.  
No blue main sequence stars are apparent, indicating no recent star formation.

\subsubsection{Distance to CenA-MM-Dw3} \label{sec:dw3distance}

The main body of CenA-MM-Dw3 lies 80~kpc in projection from Cen~A.
For this remnant we find a distance of
$r_{0,\rm TRGB}=25.31\pm0.16$, which translates
into a distance modulus of $(m-M)_0=28.32\pm0.19$.
This value implies that CenA-MM-Dw3 is located slightly behind Cen~A,
at a distance of $4.61\pm0.42$~Mpc.
At this distance, the highest surface brightness portion of
the tails of CenA-MM-Dw3 stretches for at least $\sim120$~kpc 
(i.e., $\sim1.5$~deg on the sky), or more if they are not co-planar.
As a cross-check, we additionally derive the distance modulus of Cen~A 
from our many selected field regions (see Fig.~\ref{spat_box}), i.e., for stars
belonging to the outer halo of Cen~A. We obtain a range of values bracketed
by $(m-M)_0=27.82\pm0.19$ and $(m-M)_0=28.22\pm0.19$ and with a mean
value of $(m-M)_0=28.03\pm0.15$: this is in agreement with the average
of literature values obtained via 
several methodologies \citep[$(m-M)_0=27.91\pm0.05$;][]{harrisg09}. 
Adopting the mean distance value to Cen~A from our dataset, 
the 3D distance of CenA-MM-Dw3 to Cen~A is $580\pm500$~kpc.

The distances of CenA-MM-Dw3 and Cen~A itself are consistent to within 2~$\sigma$.  
Nonetheless, we consider the possibility of the latter not being the main disturber 
of CenA-MM-Dw3.
The next closest relatively massive galaxy 
is NGC5408 ($M_V\sim-17.0$), an irregular galaxy located at $4.8\pm0.5$~Mpc and $\sim6$~deg 
away from CenA-MM-Dw3 \citep{kara07}. The 3D distance between 
CenA-MM-Dw3 and NGC5408 is $530\pm200$~kpc. While not significantly 
different from the distance between CenA-MM-Dw3 and Cen~A, the latter
is much more massive and is thus more plausibly the culprit for the 
heavy tidal disruption experienced by our target.
In addition, the radial velocity of NGC5408 is lower than that of Cen~A
by $\sim10\%$, such that a radial velocity measurement for CenA-MM-Dw3 
might help solve this question.
Only detailed simulations will be able to shed 
light onto these possibilities. If Cen~A is indeed the main
perturber of CenA-MM-Dw3, the morphology of this long stream can additionally constrain 
the dark matter halo mass of Cen~A \citep[e.g.,][]{amorisco15, pearson15}.

\subsubsection{Distances along the stream of CenA-MM-Dw3} 

We further characterize the properties of CenA-MM-Dw3
by deriving distances along its stream. We select two regions to the 
South-East (dubbed Dw3 S and SE in
Fig.~\ref{spat_box}), a region encompassing the higher density feature
along its northern elongation (Dw3 N), a region located in the same
direction and just South of the dwarf galaxy ESO324-24
(ESO324-24 S), and the very low density region to the far
North-West of CenA-MM-Dw3 (Dw3 NW).

\begin{figure*}
 \centering
\includegraphics[width=12cm]{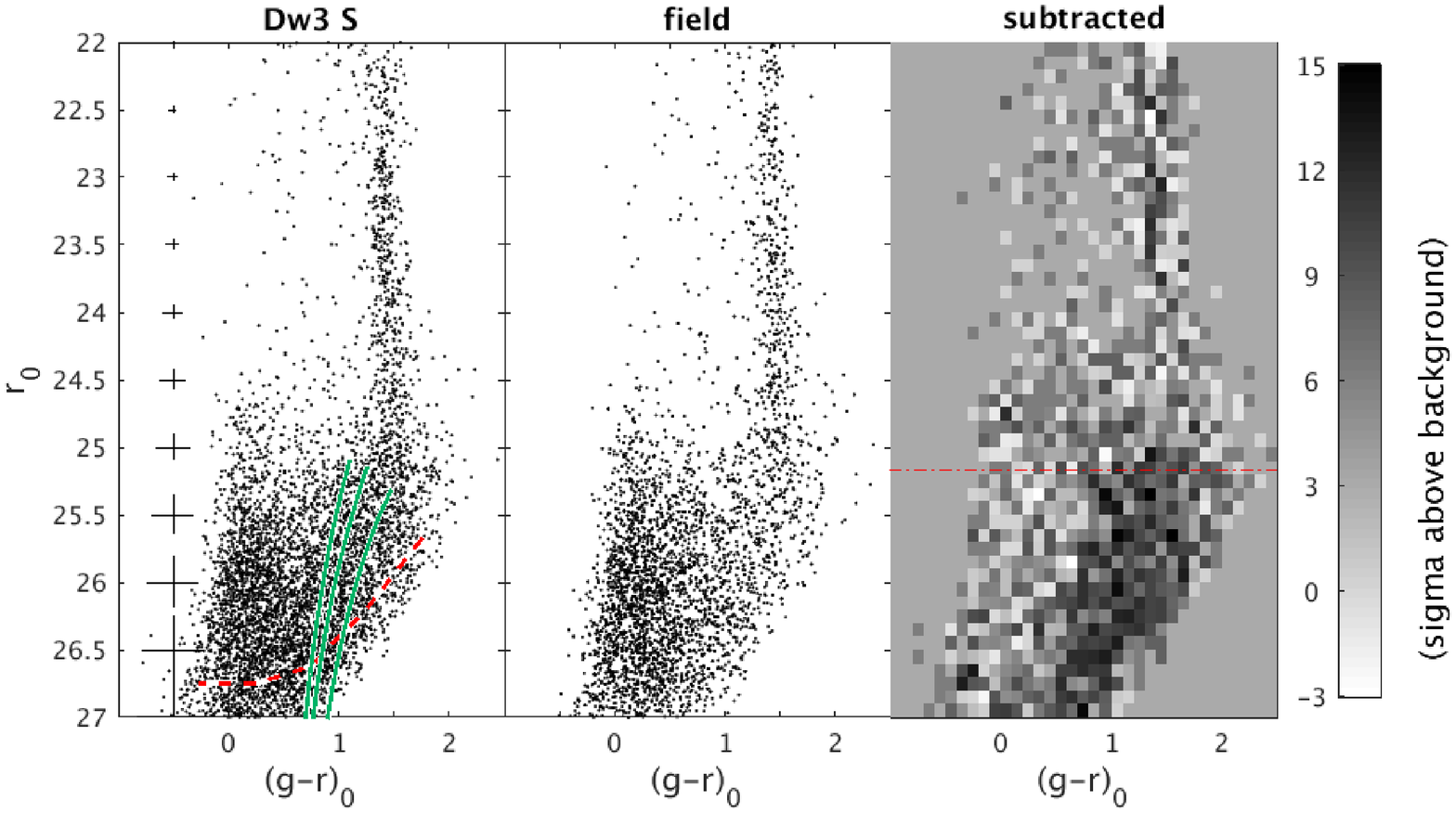}
\includegraphics[width=12cm]{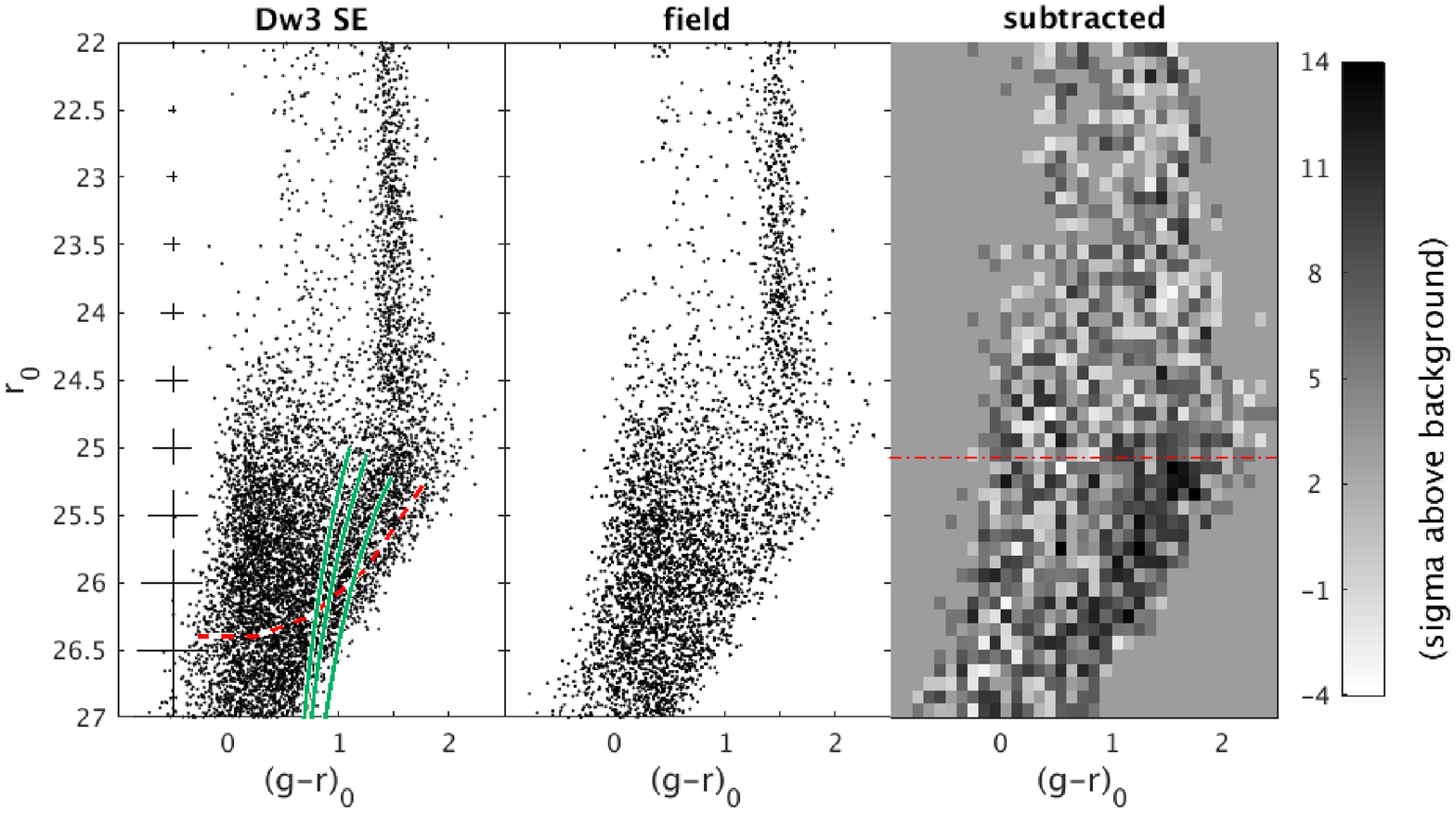}
\includegraphics[width=12cm]{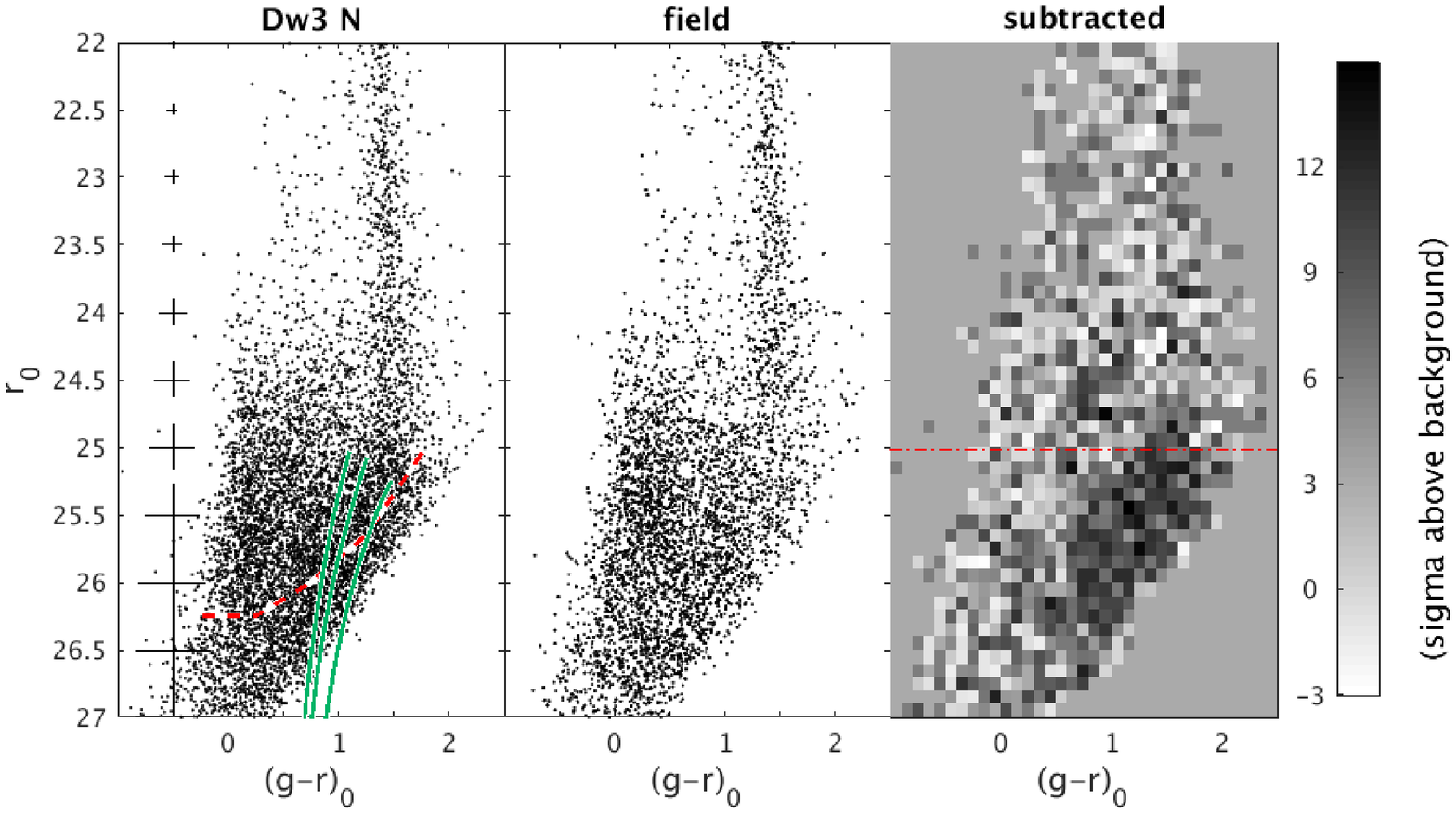}
\caption{Dereddened color-magnitude diagrams
for three regions along the stream of CenA-MM-Dw3 (rectangles labeled 
Dw3 S, Dw3 SE, Dw3 N in Fig.~\ref{spat_box}).
The panels are the same as in Fig.~\ref{dw16_cmd}, while the incompleteness
levels, photometric uncertainties and adopted field regions
change on a pointing-to-pointing basis. The TRGB is
recomputed for each of the selected regions and the 
isochrones are shifted to the appropriate distance. The
selected stream regions are consistent with the distance 
found for CenA-MM-Dw3. 
}
\label{cmds_spati}
\end{figure*}

The CMDs of Dw3 S, Dw3 SE and Dw3 NW are shown in 
Fig.~\ref{cmds_spati}. Despite a varying completeness level, the RGBs
in these regions are comparable to the one shown in
Fig.~\ref{dw16_cmd} for the remnant of CenA-MM-Dw3. 
We derive TRGB values and distance moduli as follows: $r_{0,\rm TRGB}=25.17\pm0.19$
and $(m-M)_0=28.18\pm0.21$ for Dw3 S,
$r_{0,\rm TRGB}=25.08\pm0.16$ and $(m-M)_0=28.09\pm0.19$ for
Dw3 SE, $r_{0,\rm TRGB}=25.11\pm0.17$ and $(m-M)_0=28.12\pm0.19$
for Dw3 N. The derived values are all consistent with each other,
and with the distance modulus of CenA-MM-Dw3 within the errorbars
($(m-M)_0=28.32\pm0.19$). 

\begin{figure*}
 \centering
\includegraphics[width=12cm]{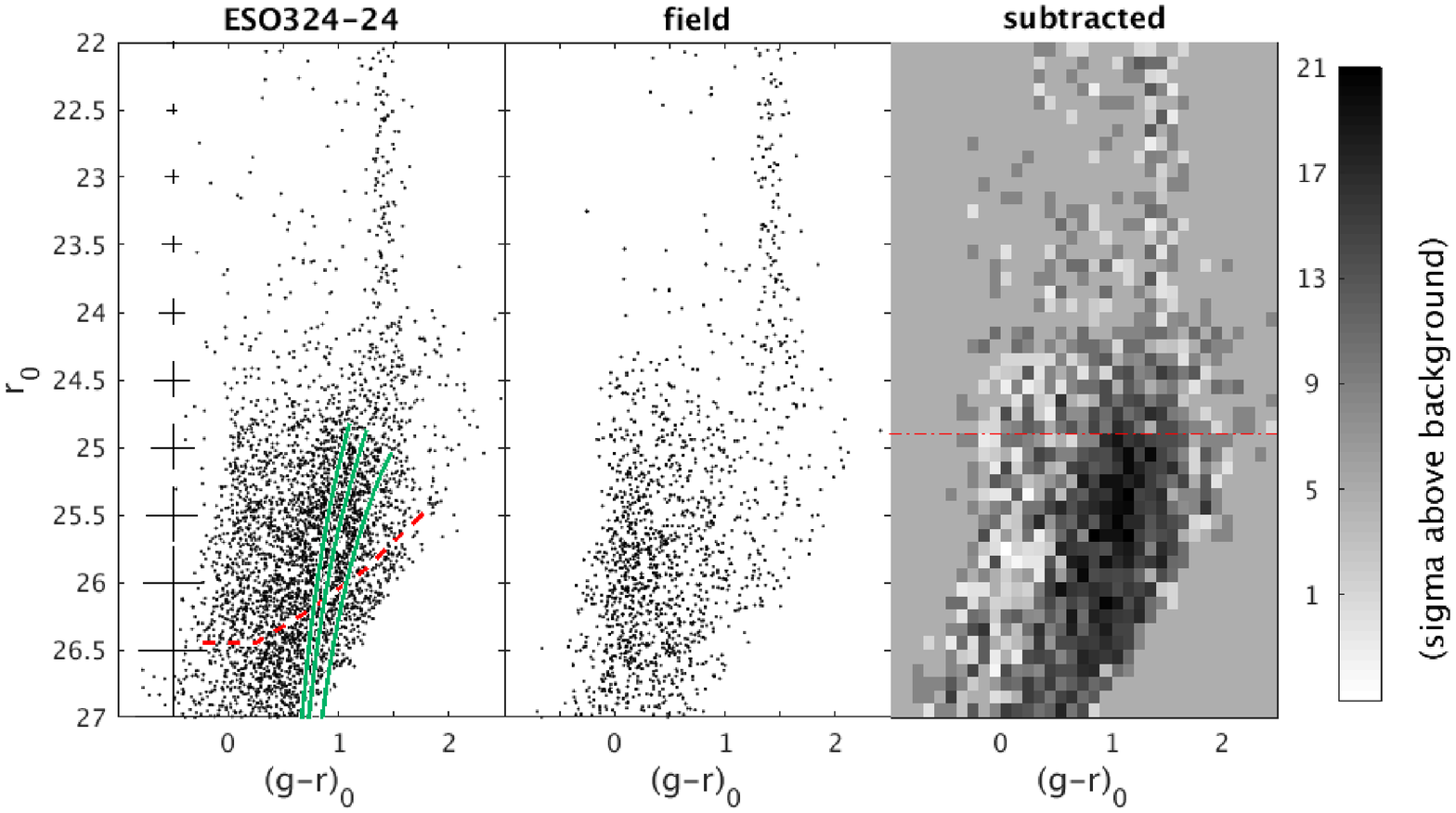}
\includegraphics[width=12cm]{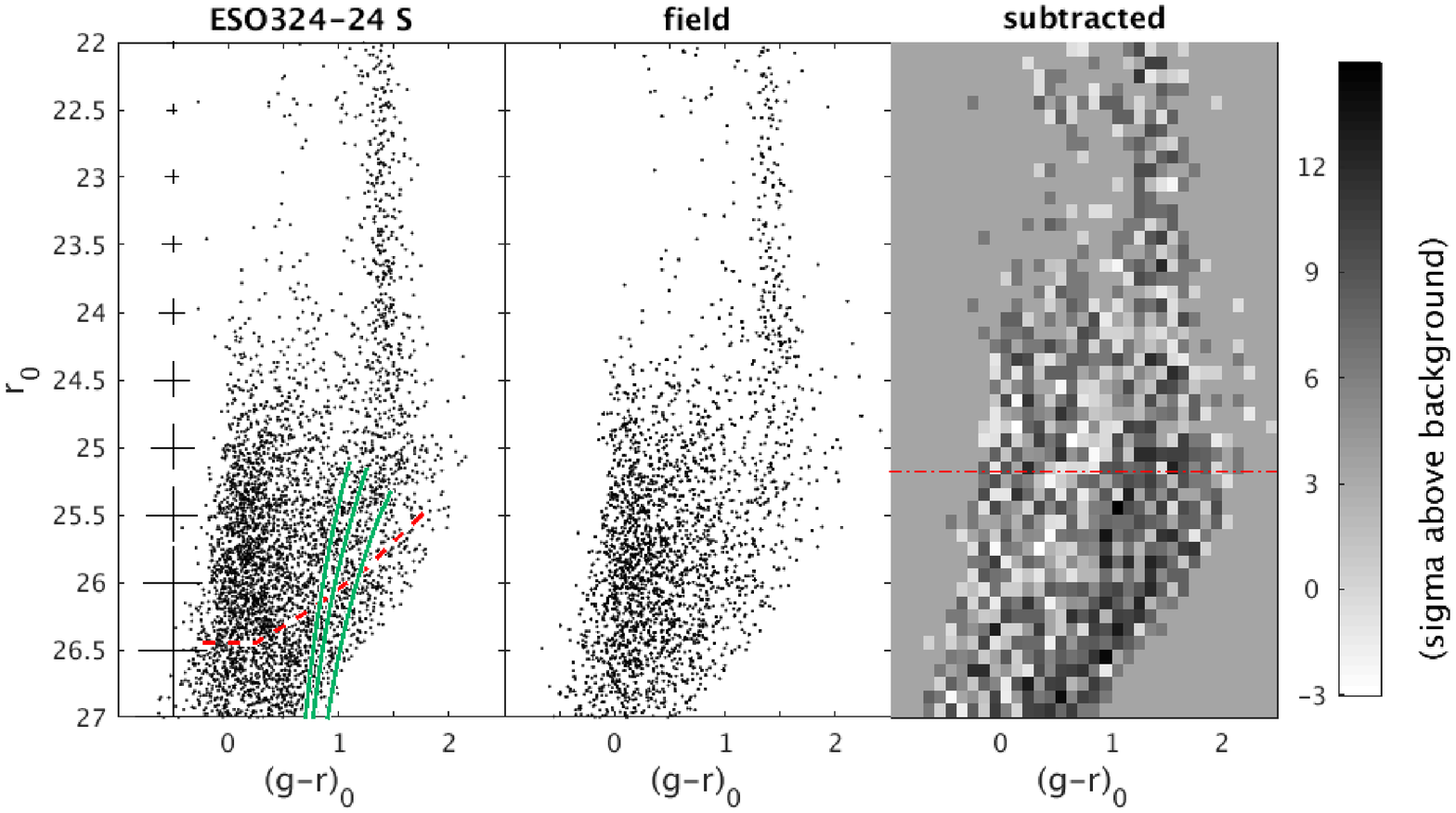}
\includegraphics[width=12cm]{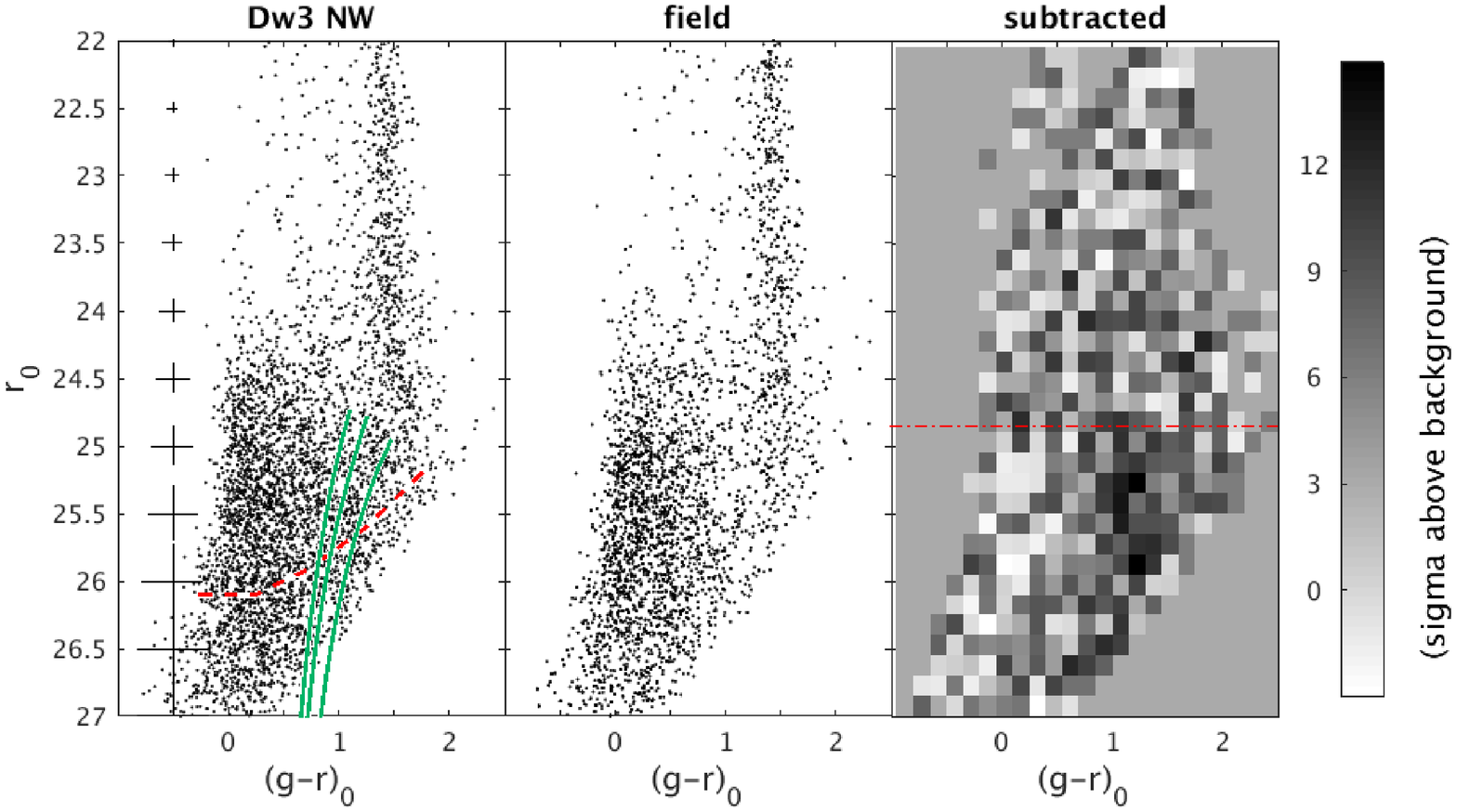}
\caption{CMDs for the previously known Cen~A satellite ESO324-24 and 
for two stream regions located to its
immediate South and to its far West (see Fig.~\ref{spat_box}).
Note that the adopted
color/magnitude bins for the subtracted Hess diagram of Dw3 NW are larger
than for the other regions because of low counts.
Based on the difference in the derived TRGB
values, we argue that the region immediately to the South
of ESO324-24 still belongs
to the stream of CenA-MM-Dw3, while the region to the North-West might
constitute a separate tidal feature in the halo of Cen~A.
}
\label{cmds_spati2}
\end{figure*}

Given the proximity of the northwestern portion of the stream to
the dwarf irregular ESO324-24 ($M_B\sim-14.9$; \citealt{cote09}), 
we investigate a possible spatial overlap in their stellar populations.
ESO324-24 presents a perturbed HI morphology: its asymmetric gas tail 
(pointing towards the NE), together with its gas kinematics,
suggest it is undergoing ram pressure stripping from 
the northern radio lobe of Cen~A \citep{johnson15}.
However, its resolved stellar content does not reveal clear signs
of distortion.
The CMDs we discuss are shown in Fig.~\ref{cmds_spati2}.
First, we compute the distance for a region in the outskirts
of ESO324-24: in this way we avoid the high
density central regions (stellar crowding is the culprit for
the hole seen at the center of this galaxy, see e.g., 
Fig.~\ref{spat_box}), as well as the most recent star
formation pockets closer to the central parts of this dwarf
\citep[e.g.,][]{cote09}.
We obtain $r_{0,\rm TRGB}=24.90\pm0.16$ and $(m-M)_0=27.91\pm0.19$, in
good agreement with the $(m-M)_0=27.84\pm0.05$ listed in 
\citet{jacobs09}.
The stream portion below ESO324-24 (ESO324-24 S) has 
$r_{0,\rm TRGB}=25.18\pm0.19$, thus $(m-M)_0=28.19\pm0.20$, which is
slightly closer to the rest of the stream rather than to
the values of ESO324-24.
This suggest that this low surface brightness elongation is not
associated with the latter, but with CenA-MM-Dw3. The lack of further
density enhancements on the opposite side of ESO324-24, or of obvious
signs of distortion in its main body, support this interpretation.
Finally, for the north-western overdensity (Dw3 NW) we obtain 
$r_{0,\rm TRGB}=24.81\pm0.17$, thus $(m-M)_0=27.82\pm0.19$:
this feature is closer (along the line of sight) than, and probably 
does not belong to, the stream of CenA-MM-Dw3.

\tabletypesize{\scriptsize}
\begin{deluxetable*} {lcccc}
\tablecolumns{5}
\tablecaption{Central coordinates of and distances to halo substructures (see Fig.~\ref{spat_box}).}

\tablehead{\colhead{Substructure}  & \colhead{RA (h:m:s)} & \colhead{Dec (d:m:s)} & \colhead{$(m-M)_0$ (mag)} &\colhead{D (Mpc)}}\\
  
\startdata
Cen~A\tablenotemark{a}& 13:25:27.6 & $-43$:01:09 &$27.91\pm0.05$ &$3.80\pm0.10$ \\
Cen~A (this work) &--&-- &$28.03\pm0.15$ &$4.04\pm0.29$ \\
CenA-MM-Dw3 &13:30:21.5&$-42$:11:33 &$28.32\pm0.19$ &$4.61\pm0.42$ \\
Dw3 S & 13:31:02.6 & $-42$:21:54 &$28.18\pm0.21$ &$4.33\pm0.44$ \\
Dw3 SE & 13:32:18.0 & $-42$:36:54&$28.09\pm0.19$ &$4.15\pm0.44$ \\
Dw3 N & 13:28:27.6 & $-41$:51:00&$28.12\pm0.19$ &$4.20\pm0.39$ \\
ESO324-24 & 13:27:32.4 & $-41$:33:00&$27.91\pm0.19$ &$3.82\pm0.36$ \\
ESO324-24 S & 13:27:03.6& $-41$:37:30&$28.19\pm0.20$ &$4.35\pm0.41$ \\
Dw3 NW & 13:23:57.6 & $-41$:42:00&$27.82\pm0.19$ &$3.66\pm0.34$ \\
Second stream & 13:31:54.0 & $-41$:51:00&$28.03\pm0.19$ &$4.04\pm0.37$ \\
Shell 1 & 13:26:45.6 & $-42$:22:30&$28.43\pm0.20$ &$4.75\pm0.46$ \\
Shell 2 & 13:23:21.6 & $-43$:14:06&$28.20\pm0.20$ &$4.36\pm0.40$ \\
\enddata
\label{tab0}
\tablenotetext{a}{Value from \citet{harrisg09}.}
\end{deluxetable*}

\begin{figure*}
 \centering
\includegraphics[width=17.cm]{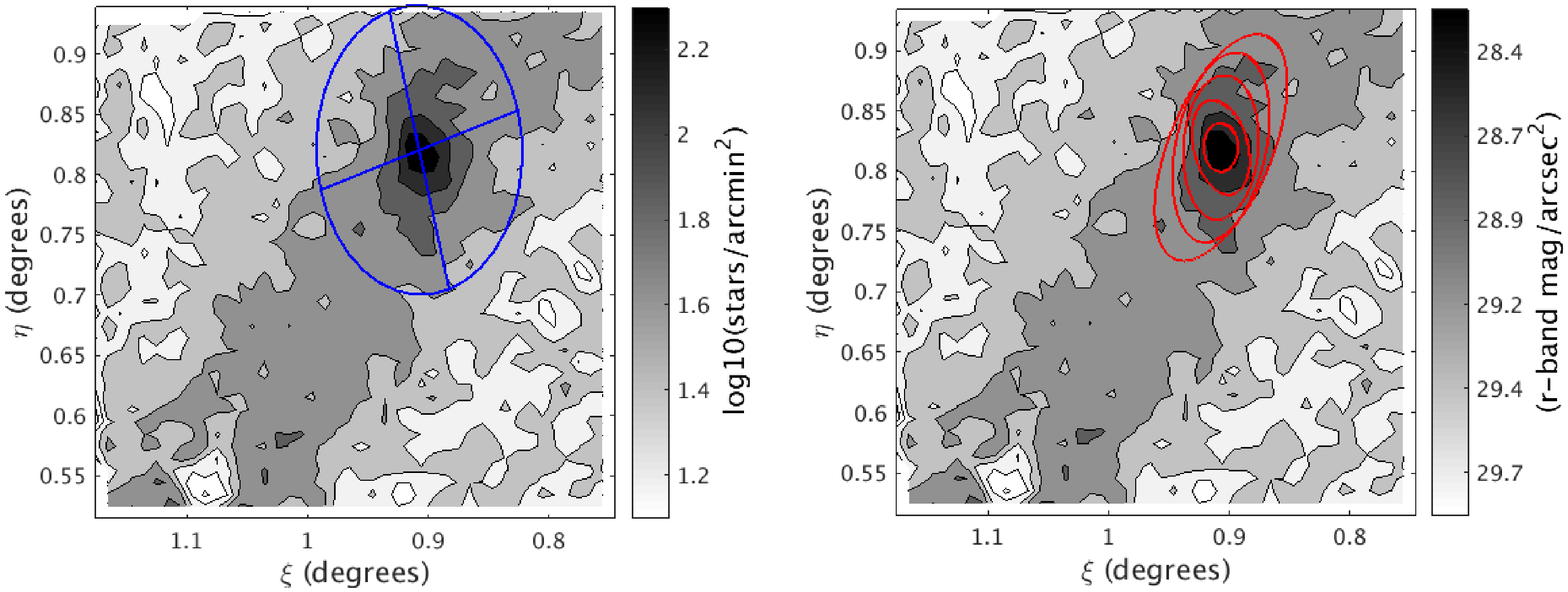}
\includegraphics[width=8.cm]{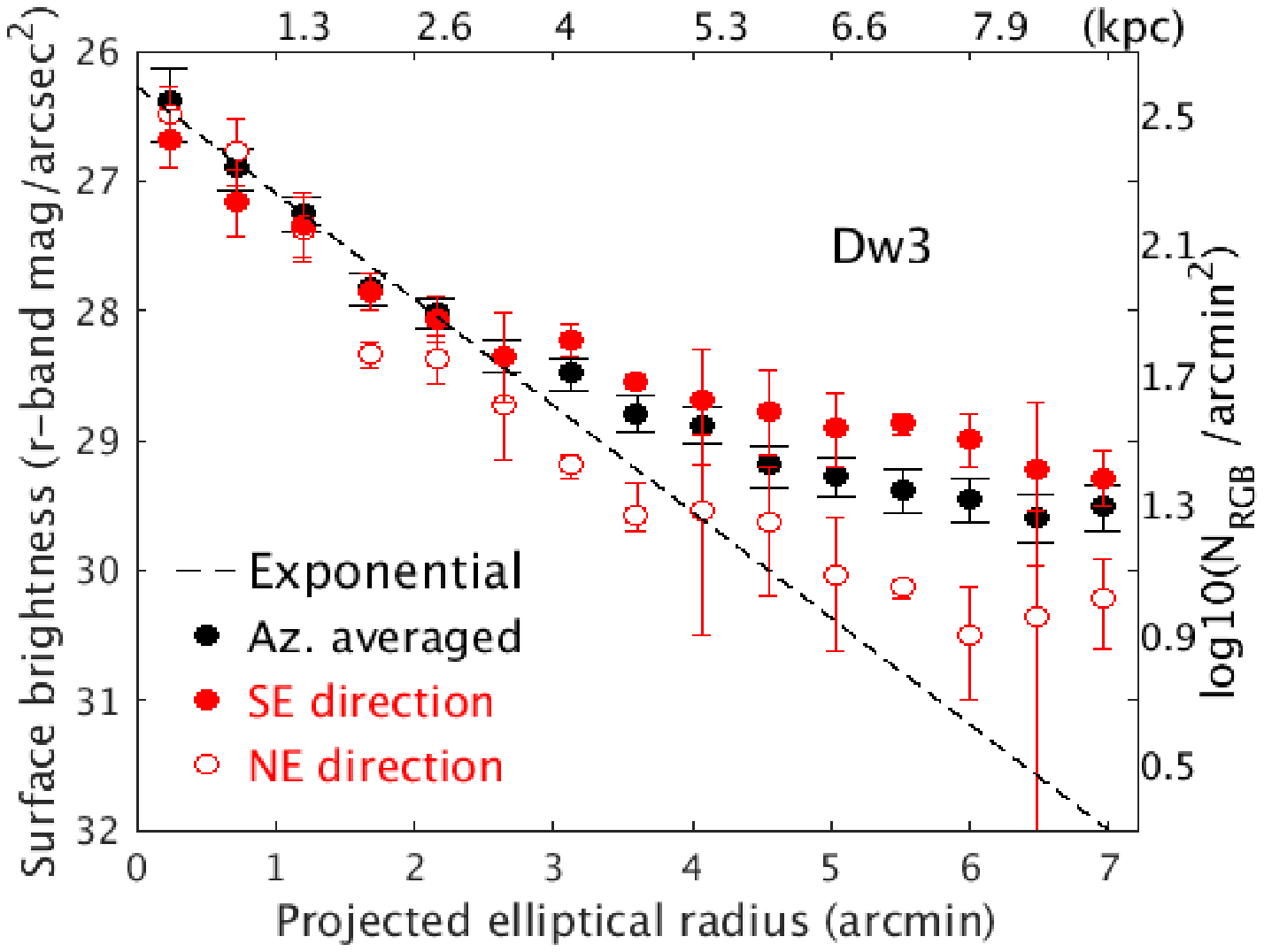}
\caption{
\emph{Upper panels}. Density map of CenA-MM-Dw3 RGB stars,
after completeness correction.
The color scales indicate RGB stellar density (\emph{left}) and 
$r$-band surface brightness (\emph{right}).
The left panel shows the four azimuthal wedges 
that divide the remnant of CenA-MM-Dw3 into ``tidal tail-free'' (NE and SW) 
and ``tidal tail'' (NW and SE) regions. For the right panel,
position angle and ellipticity are recomputed for radial annuli 
with semimajor axis lengths of 1.2, 2.4, 3.6, 4.8, and 6.0~arcmin 
and red ellipses are drawn accordingly.
\emph{Lower panel}. Surface brightness profiles in $r$-band for CenA-MM-Dw3,
where the average ellipticity has been adopted
(see Table~\ref{tab1}). The azimuthally-averaged profile is shown with filled 
black points, while the profiles derived in two different directions 
(SE and NE wedges, see upper left panel)
are shown with red symbols, filled for the
tidal-tail wedge and hollow for the wedge perpendicular to
that. The profiles have been corrected
for incompleteness and the field level has been subtracted. The best-fit
exponential profile is indicated with a black dashed line and only fits
datapoints within the innermost 3~arcmin to avoid the tidal tails.
}
\label{dw16_figs}
\end{figure*}

\subsubsection{Structural parameters and luminosity }  \label{sec:dw_lum}

We draw RGB isodensity contour maps (after incompleteness correction) 
of the pointing containing CenA-MM-Dw3 in the upper panels of 
Fig.~\ref{dw16_figs}. The adopted
RGB selection box is similar to the one drawn in Fig.~\ref{cmd_glob},
but slightly adjusted to follow the $50\%$ completeness limits for
this pointing (the color-averaged values are $r_0\sim26.2$ and $g_0\sim26.9$).

We compute the ellipticity and position angle as a function of radius
for the remnant of CenA-MM-Dw3. The derived values are adopted to draw ellipses
for increasingly large annuli centered on CenA-MM-Dw3 
(see Fig.~\ref{dw16_figs}). While the main body of
the remnant appears almost circular, the outer annuli show a
progressive elongation culminating in the tidal tails, and
the major axis of the stellar distribution twists to end up almost
perpendicularly to the direction of the main body. In Table~\ref{tab1}
we report the average ellipticity derived for the innermost 2~arcmin
of the galaxy.

To derive a surface brightness profile, we first subdivide the galaxy into 
four wedges with the same area, two of which are aligned with the tails. 
We then derive an azimuthally averaged surface brightness profile, as well as 
profiles in two directions, one along and one perpendicular to 
the tails (SE and NE wedges, respectively, in the upper left panel of 
Fig.~\ref{dw16_figs}).
The resulting profiles are shown in 
the bottom panel of Fig.~\ref{dw16_figs}. The SE and NE profiles
remain comparable to each other out to a radius of $\sim2.5$~arcmin,
beyond which the SE profile (i.e., the tidal tail direction) consistently
remains above the NE one.
The exponential profile is a good fit to the ``unperturbed'' 
surface brightness profile (even though it clearly underestimates the global profile)
and returns a very large half-light radius ($r_{h}=2.92\pm0.20$~kpc)
together with a remarkably low central surface brightness
($\mu_{r,0}=26.3\pm0.1$~mag~arcsec$^{-2}$, or
$\mu_{V,0}=26.7\pm0.1$~mag~arcsec$^{-2}$). The absolute magnitude of 
CenA-MM-Dw3 is obtained as $M_V=-13.0\pm0.4$, which is comparable
to that of the disrupting MW satellite Sagittarius ($M_V=-13.5\pm0.4$).
Given the advanced state of disruption of
this galaxy, this luminosity is likely much less than the undisrupted 
luminosity of the galaxy. 

As a rough estimate of CenA-MM-Dw3's original luminosity, 
we quantify the number of RGB stars along its tails, approximately in between 
the Dw3 SE and the ESO324-24 S boxes (see Fig.~\ref{spat_box}).
This number is $\sim8$ times higher than the number of RGB stars in the
Dw3 box: assuming that the tails entirely belong to 
CenA-MM-Dw3, this implies that its total unperturbed magnitude
might have been $\sim2$~mag brighter than our estimate for the remnant 
(i.e., $M_V\sim-15.0$).


\begin{figure*}
 \centering
\includegraphics[width=12cm]{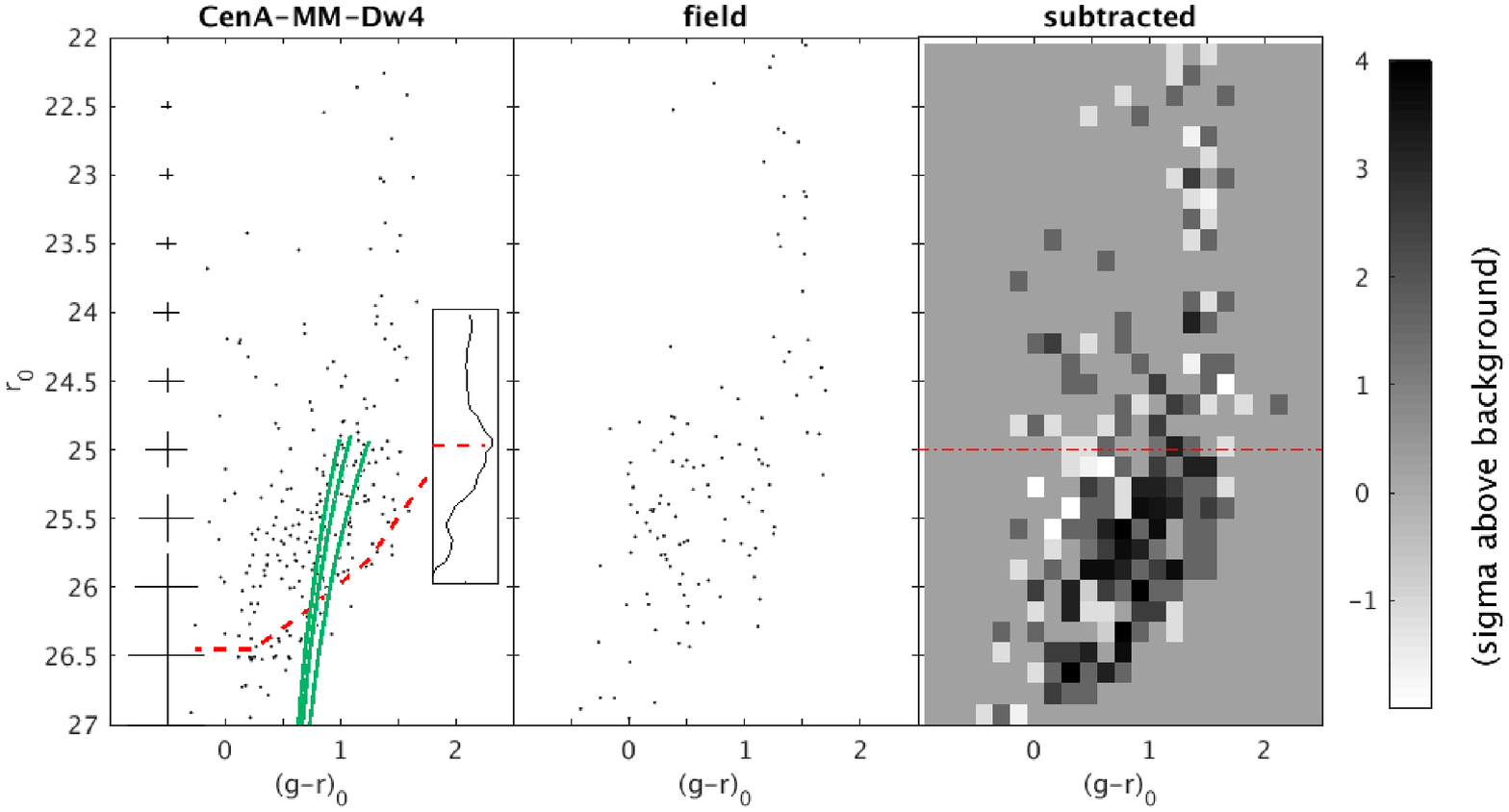}
\includegraphics[width=6.cm]{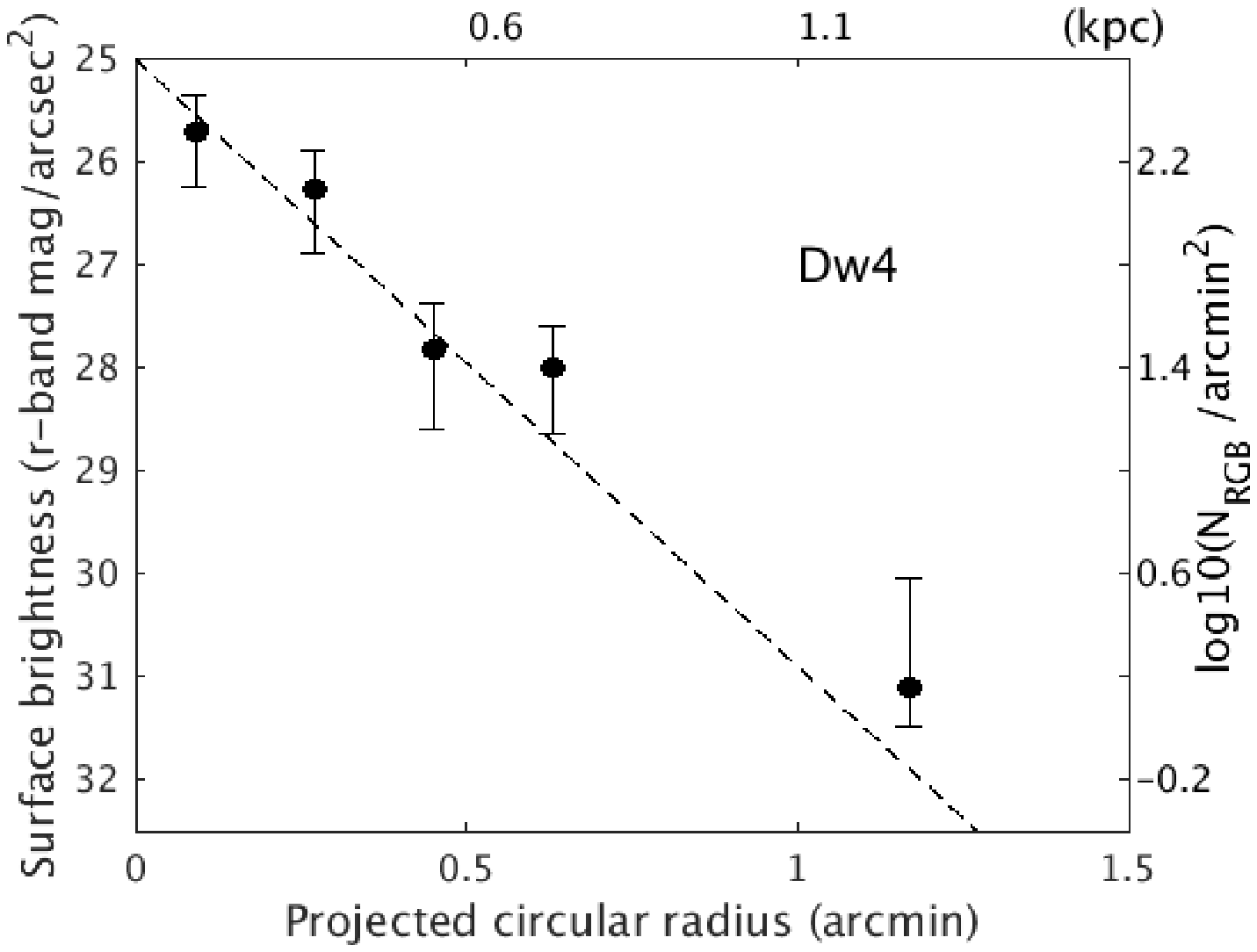}
\includegraphics[width=6.cm]{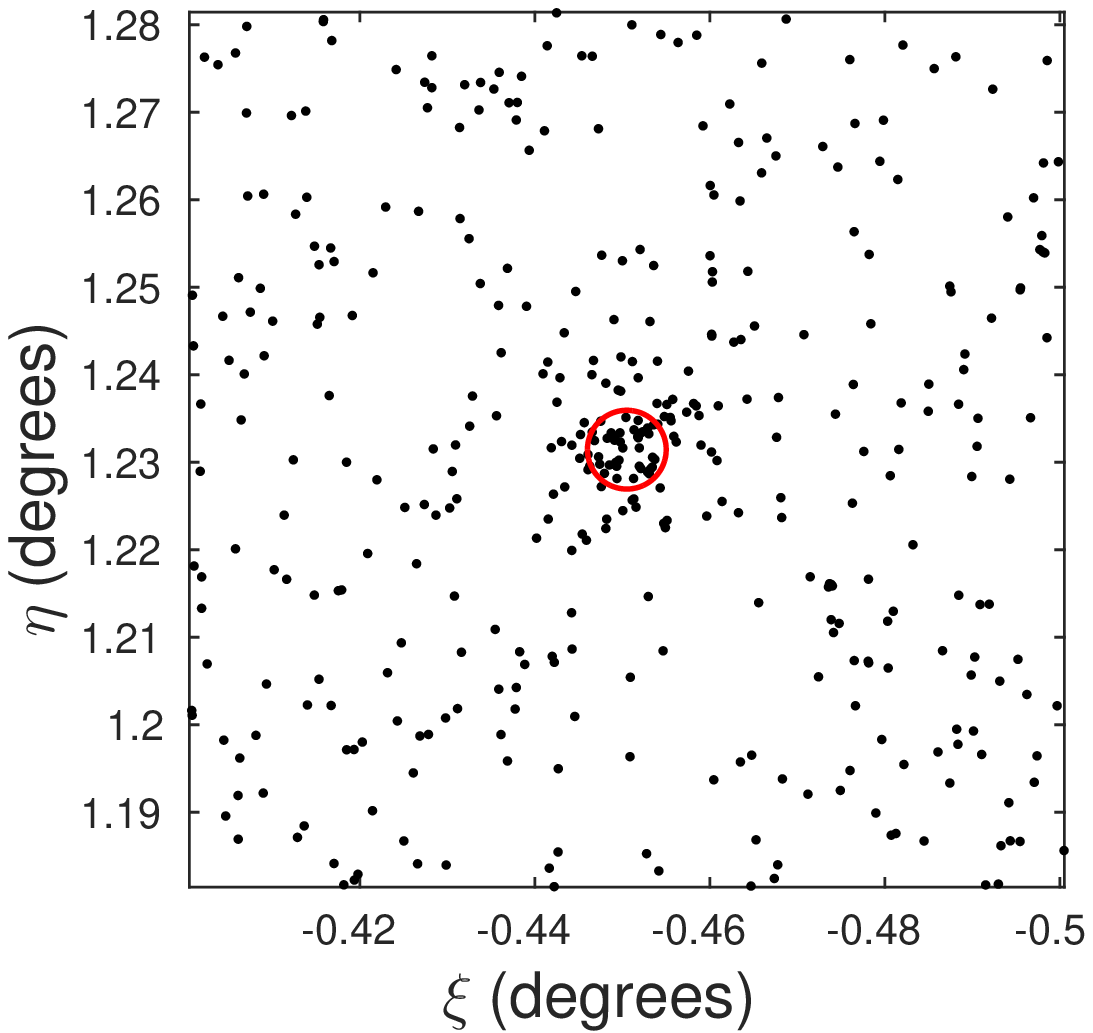}
\caption{\emph{Upper panels.} Dereddened color-magnitude diagram
for stars within a box of $0.6\times0.6$~arcmin$^2$ centered on 
CenA-MM-Dw4 (\emph{left panel}).
The isochrones are shifted to the measured 
distance of the dwarf, with a 12~Gyr age and metallicities
[Fe/H]$=-2.5$, $-2.0$ and $-1.5$. The red dashed line indicates 
the $50\%$ completeness level, and photometric errors as derived 
from artificial star tests are drawn on the left side of the CMD.
The inset plot shows the luminosity function after convolution
with a Sobel filter, and the derived TRGB magnitude (red dashed line).
A background field CMD drawn from the same pointing and
rescaled in area is shown for comparison 
in the \emph{middle panel}. The \emph{right panel} is a background-subtracted Hess 
diagram. The red dot-dashed line indicates the derived TRGB magnitude.
\emph{Lower left panel}. Surface brightness profile in $r$-band
as a function of radius. The profile has been corrected
for incompleteness and the field level has been subtracted. The best-fit
exponential profile is indicated with a black dashed line.
\emph{Lower right panel}. $3\times3$~arcmin$^2$ cutout of the 
RGB stars spatial distribution in the Magellan/Megacam pointing 
containing CenA-MM-Dw4, centered on the dwarf. The red circle is at 
the dwarf's half-light radius.
}
\label{dw3_figs}
\end{figure*}

\begin{figure*}
 \centering
\includegraphics[width=12cm]{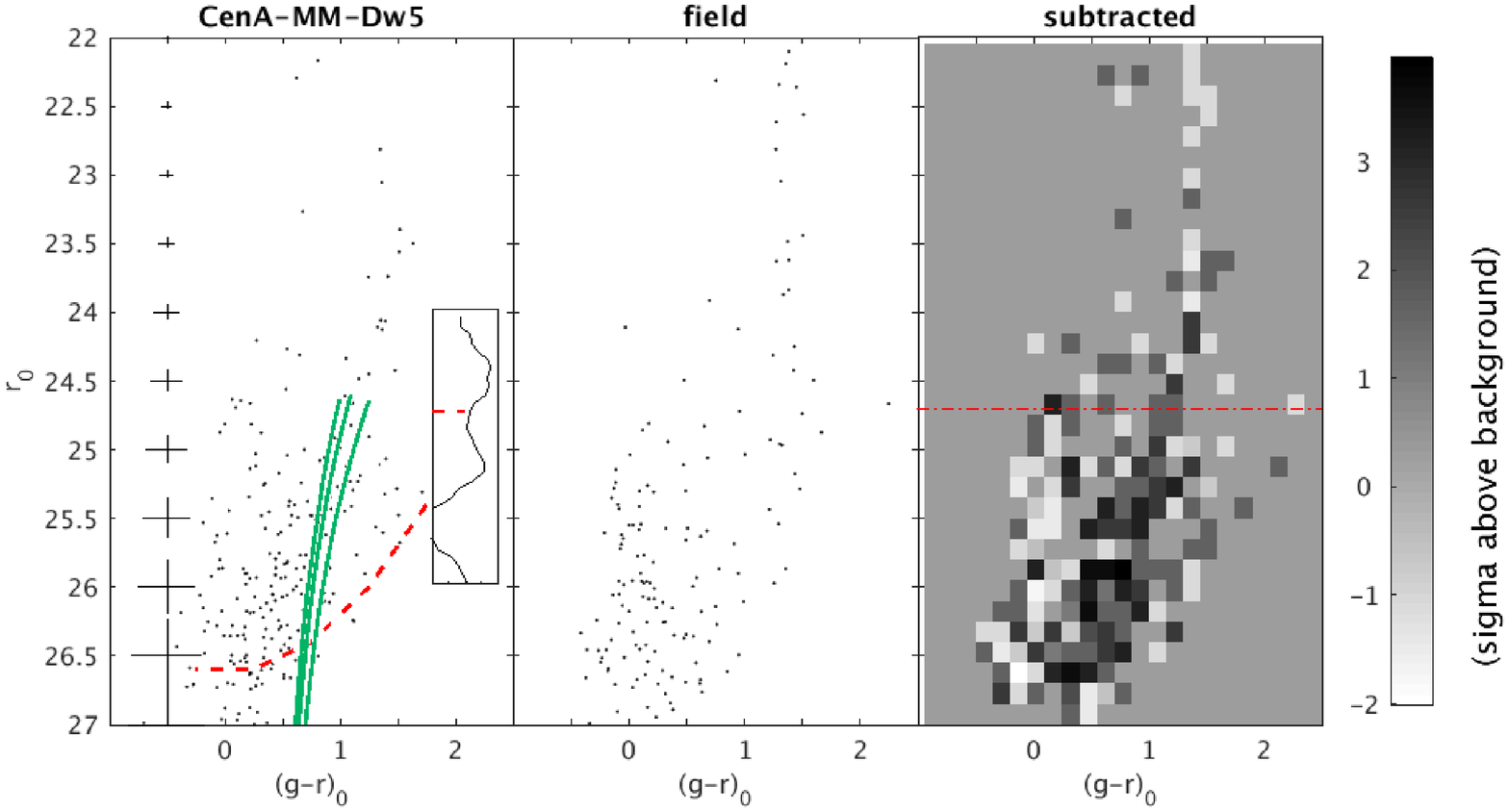}
\includegraphics[width=6.cm]{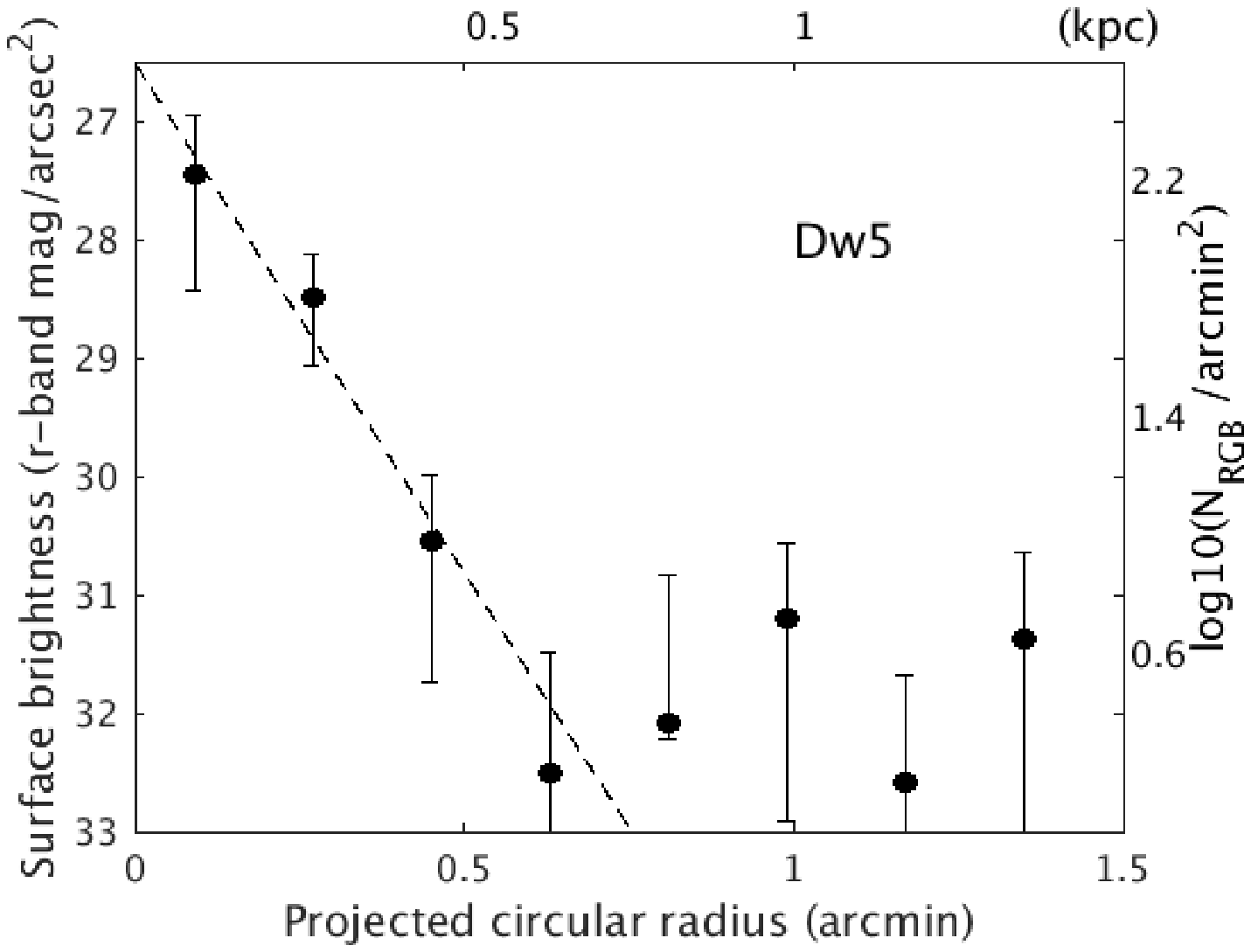}
\includegraphics[width=6.cm]{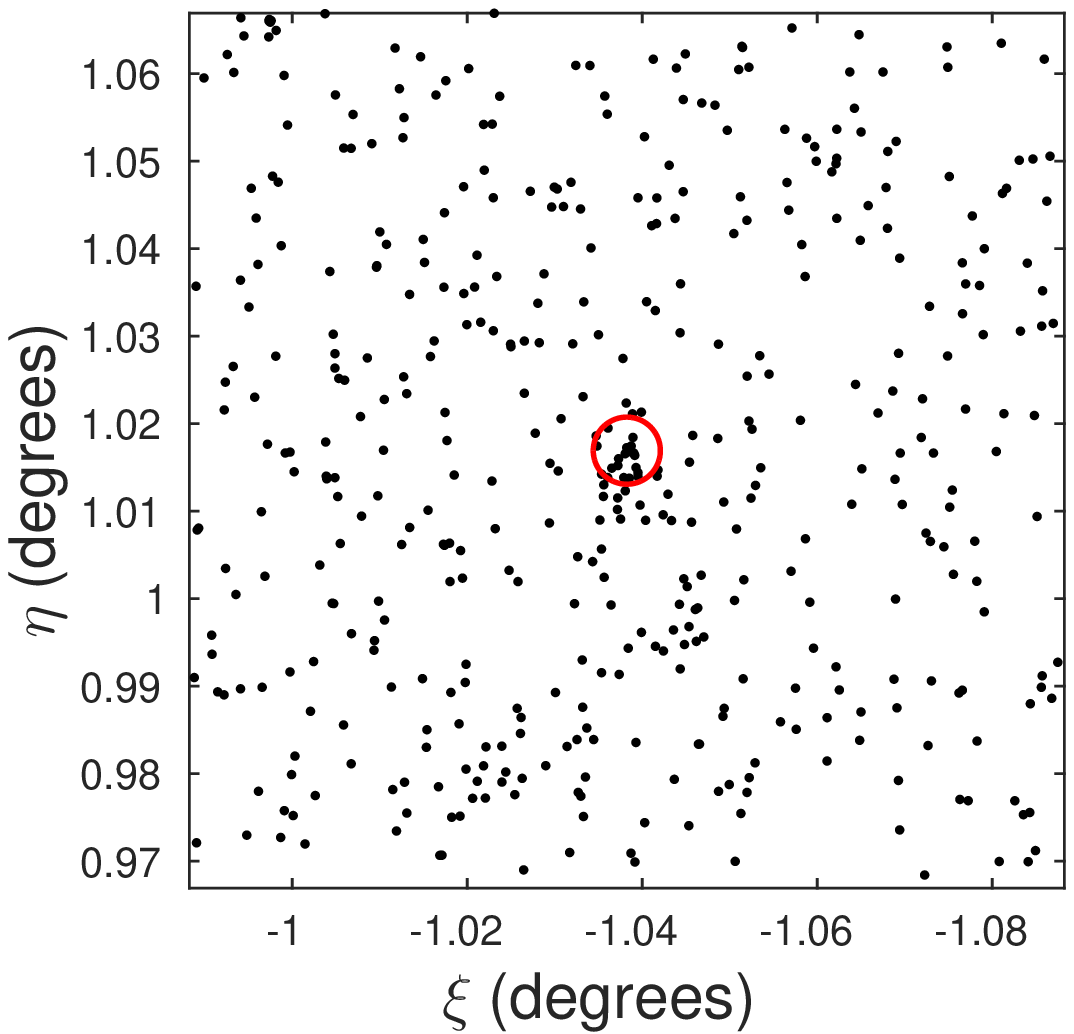}
\caption{Same as Fig~\ref{dw3_figs}, for CenA-MM-Dw5.
In this case the luminosity function presents two comparable
maxima, thus we take the TRGB to be the average of their values.}
\label{dw4_figs}
\end{figure*}

\begin{figure*}
 \centering
\includegraphics[width=12cm]{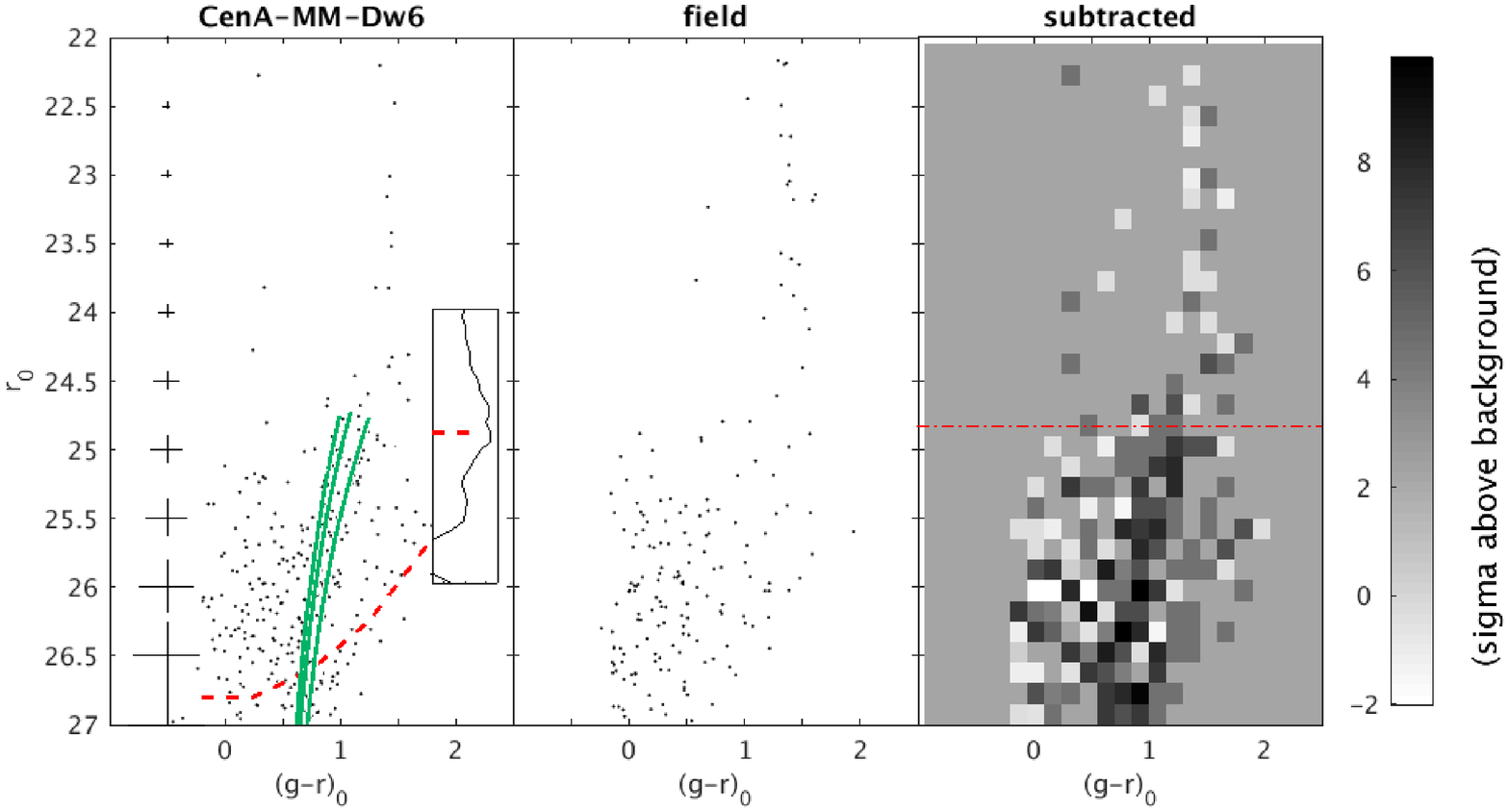}
\includegraphics[width=6.cm]{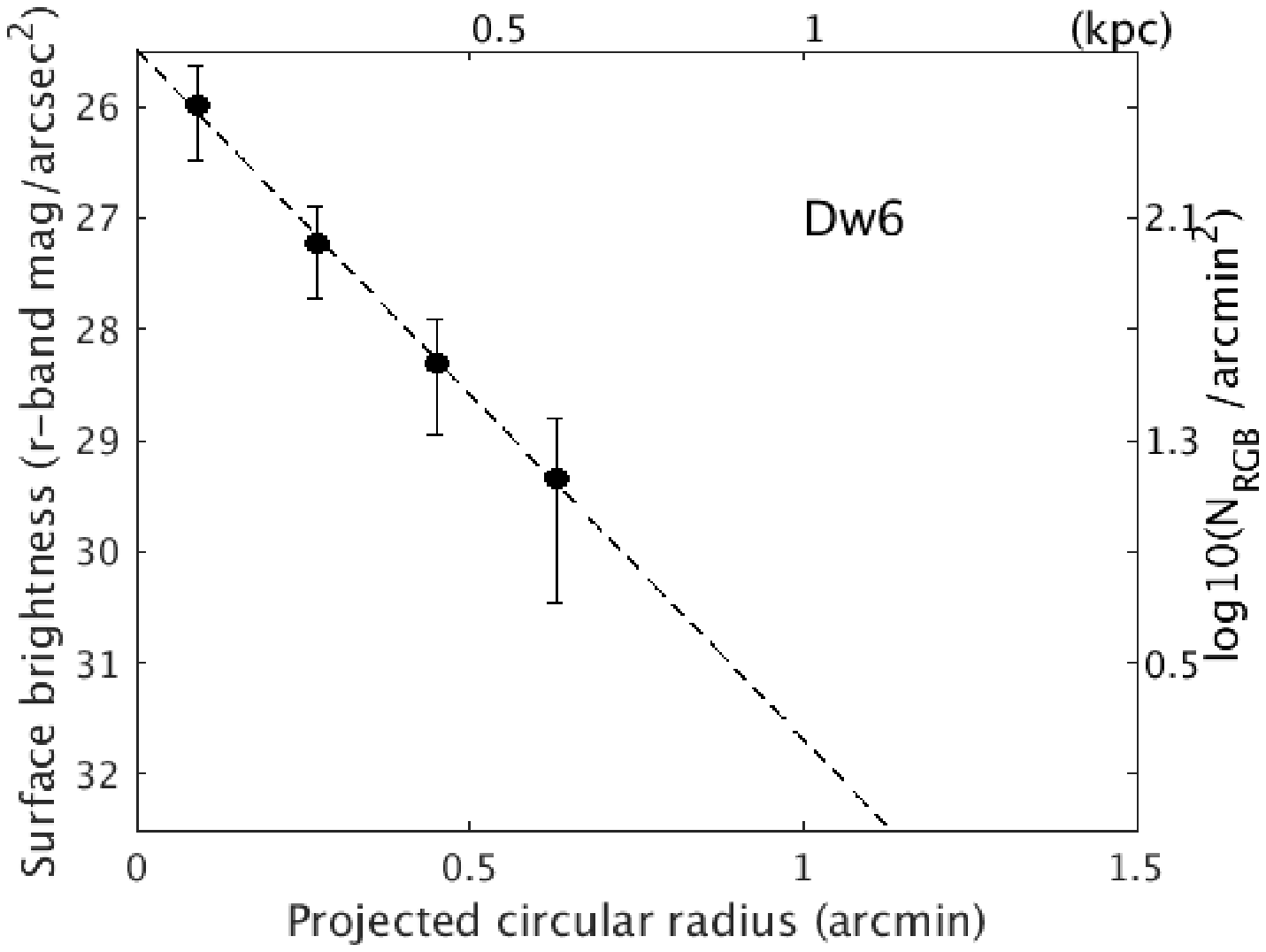}
\includegraphics[width=6.cm]{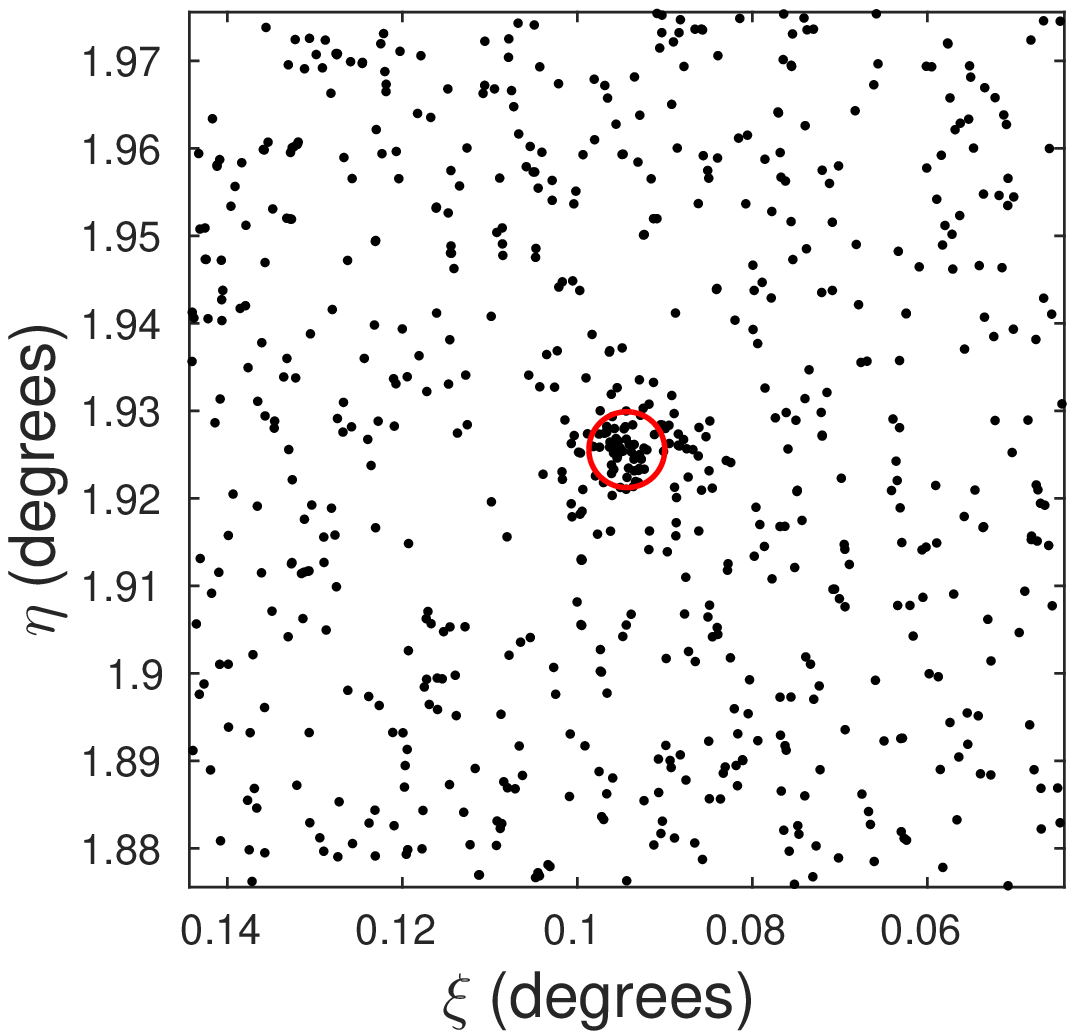}
\caption{Same as Fig~\ref{dw3_figs}, for CenA-MM-Dw6.
}
\label{dw5_figs}
\end{figure*}

\begin{figure*}
 \centering
\includegraphics[width=12cm]{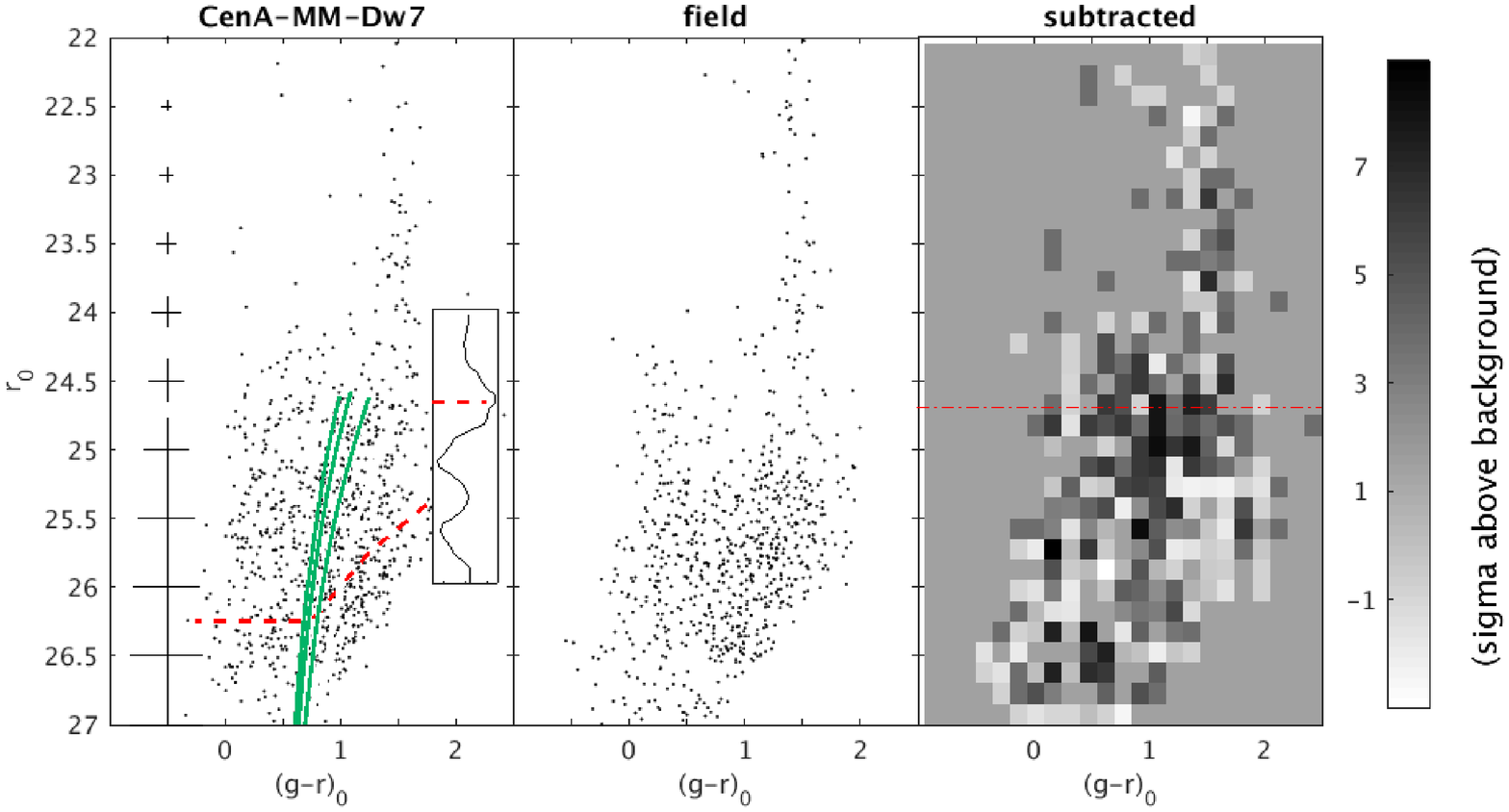}
\includegraphics[width=6.cm]{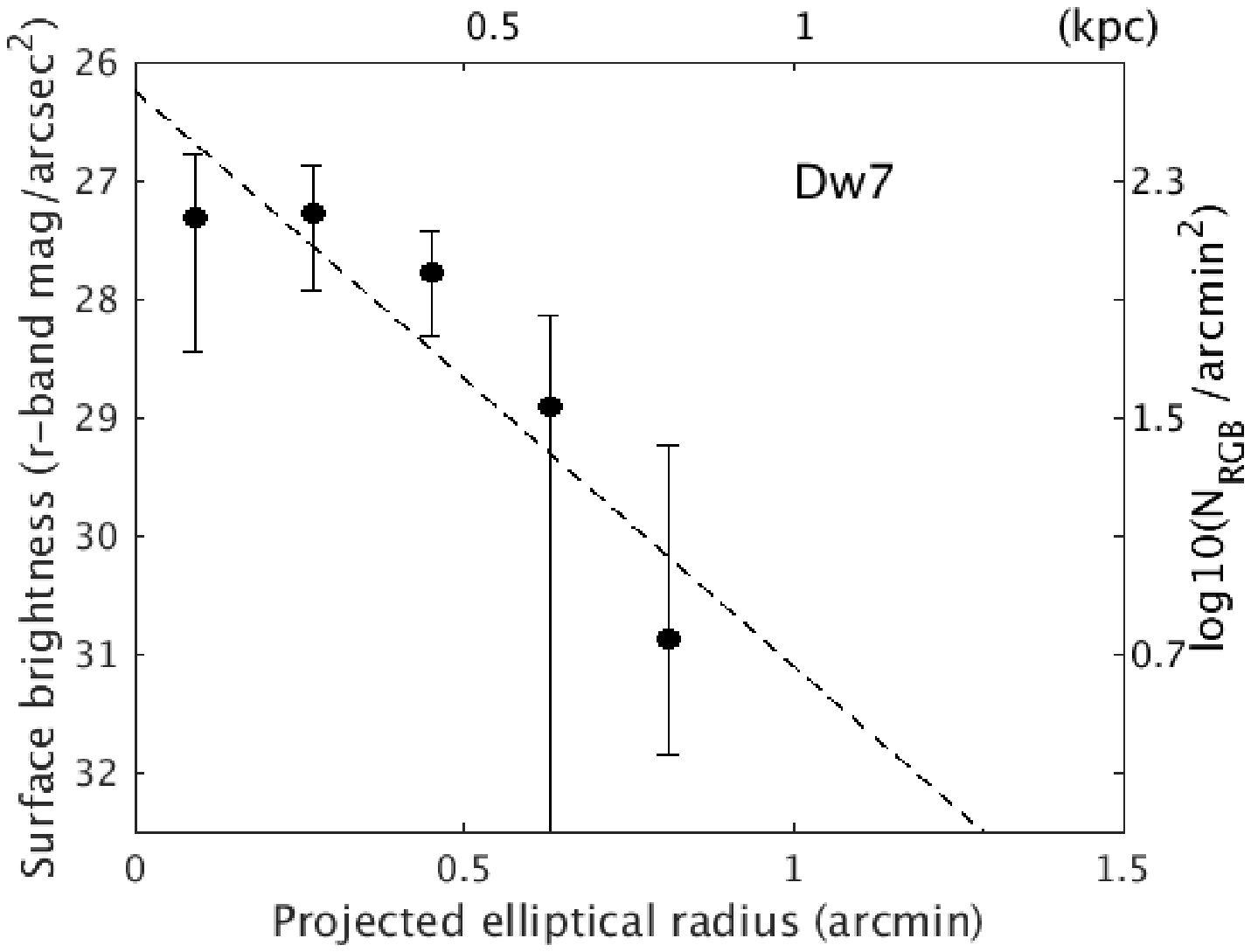}
\includegraphics[width=6.cm]{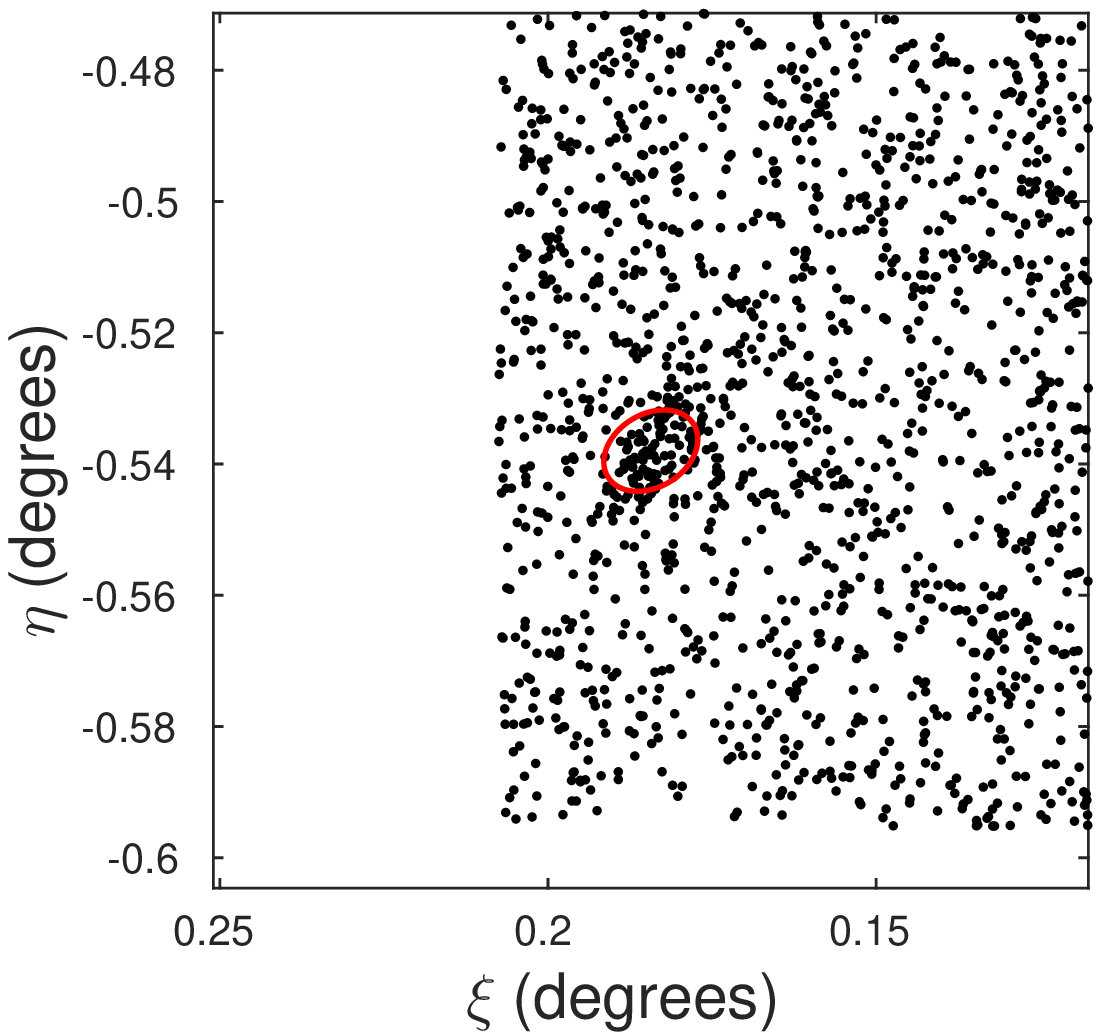}
\caption{Same as Fig~\ref{dw3_figs}, for CenA-MM-Dw7. The dwarf CMD
contains stars within a box of $0.9\times0.9$~arcmin$^2$ centered on 
CenA-MM-Dw7, and the RGB spatial distribution is shown for a
$4\times4$~arcmin$^2$ region. The contamination from Cen~A halo
stars is very strong for this dwarf and thus its CMD appears washed out.
}
\label{dw8_figs}
\end{figure*}

\begin{figure*}
 \centering
\includegraphics[width=12cm]{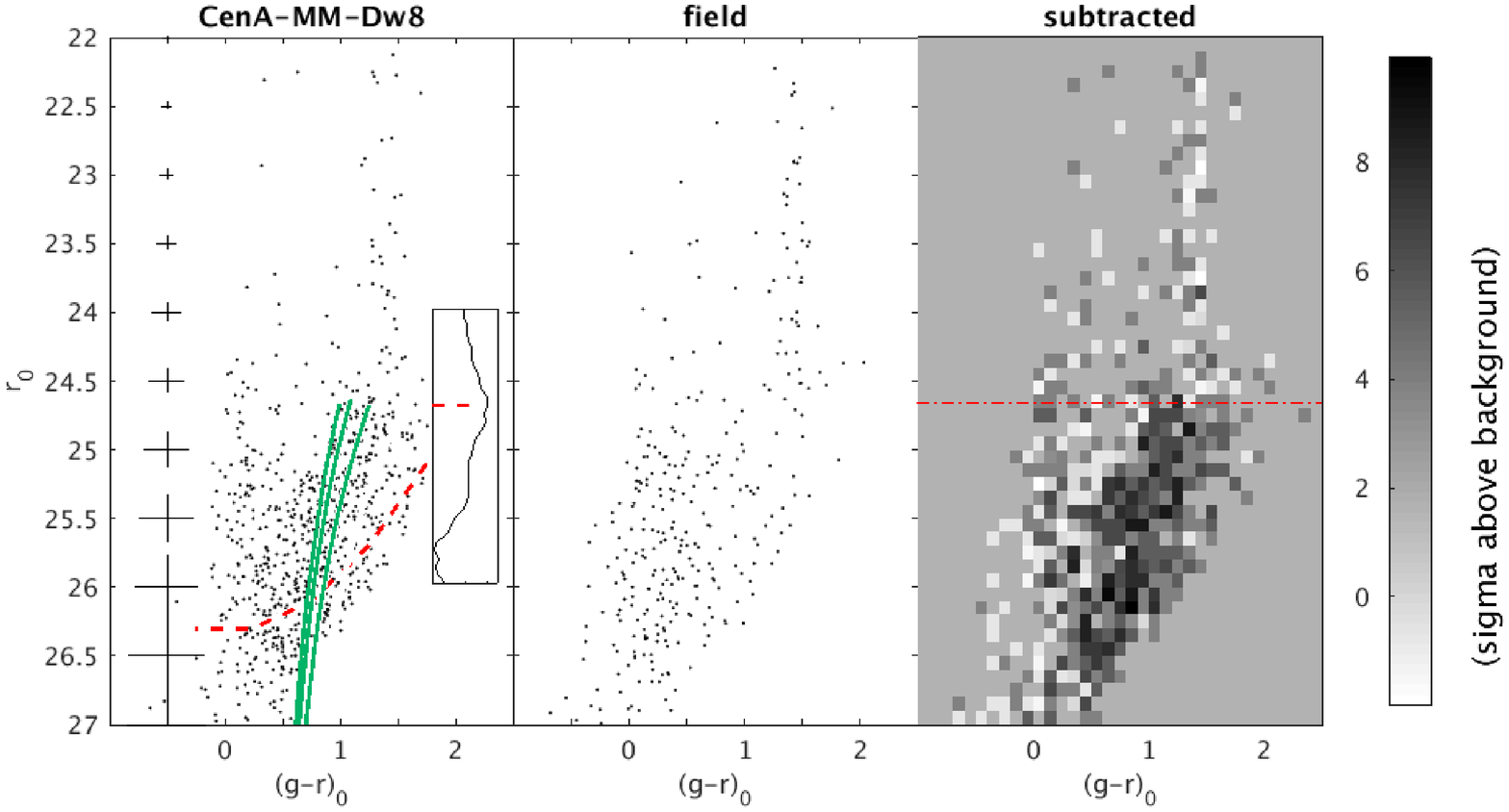}
\includegraphics[width=6.cm]{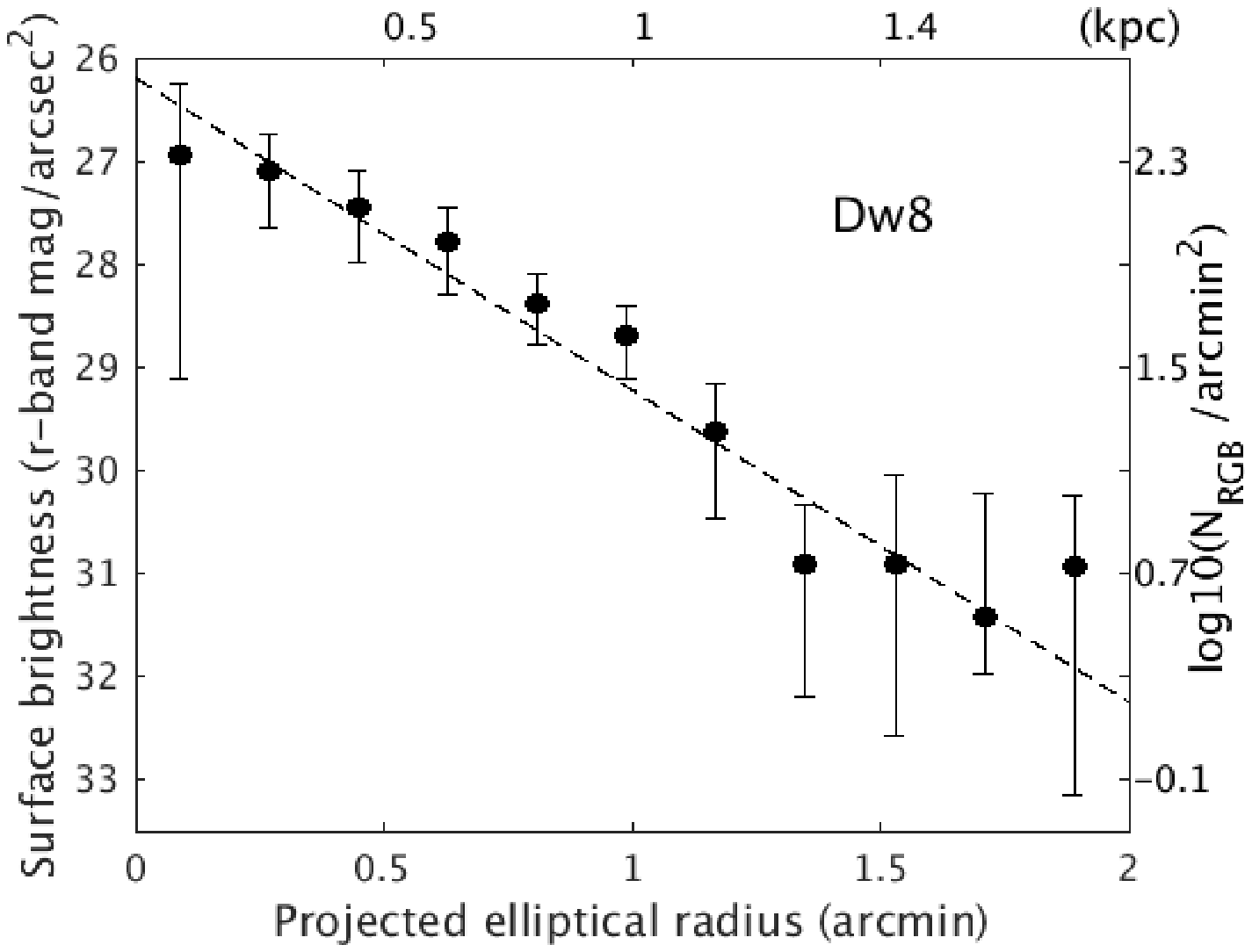}
\includegraphics[width=6.cm]{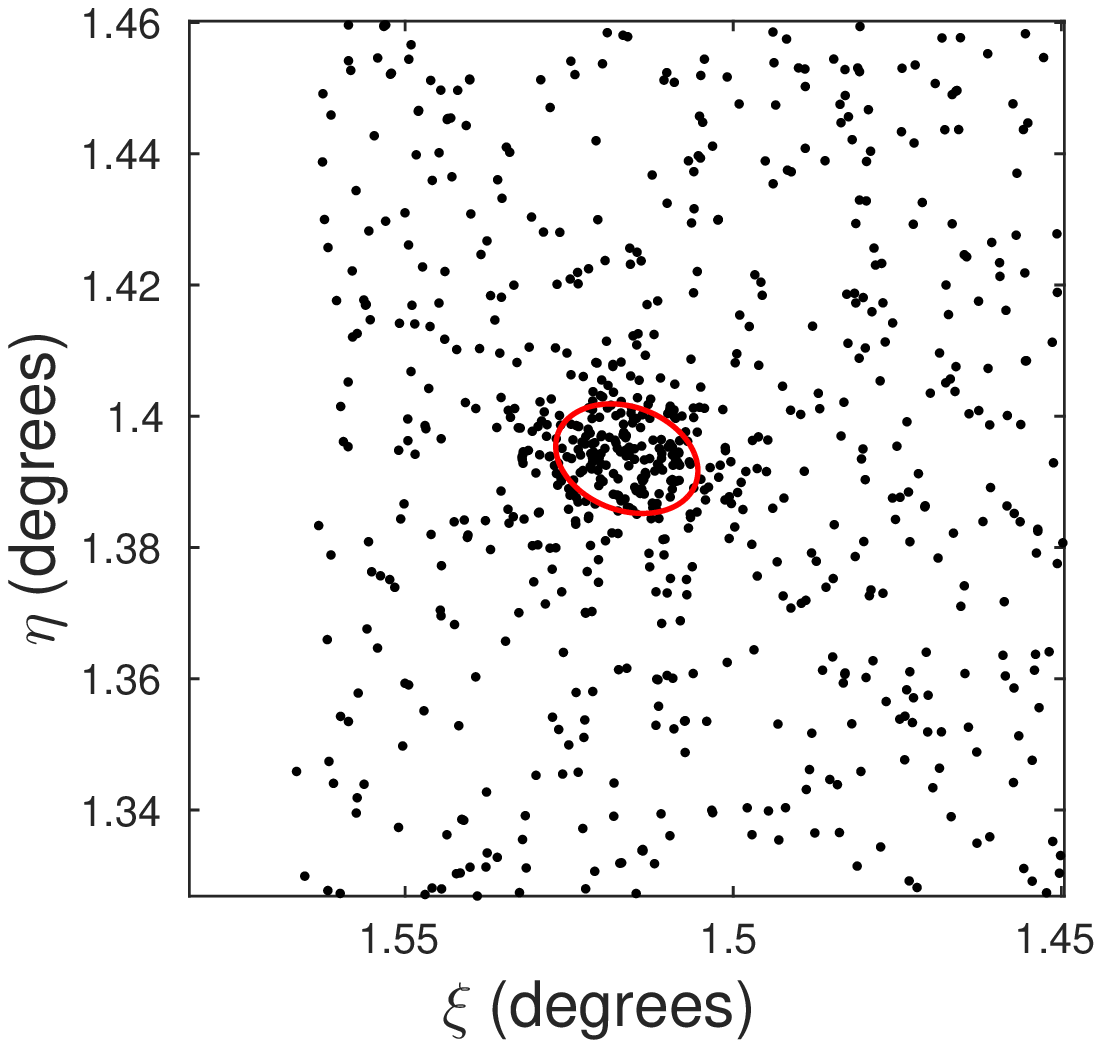}
\caption{Same as Fig~\ref{dw8_figs}, for CenA-MM-Dw8.
}
\label{dw17_figs}
\end{figure*}

\begin{figure*}
 \centering
\includegraphics[width=12cm]{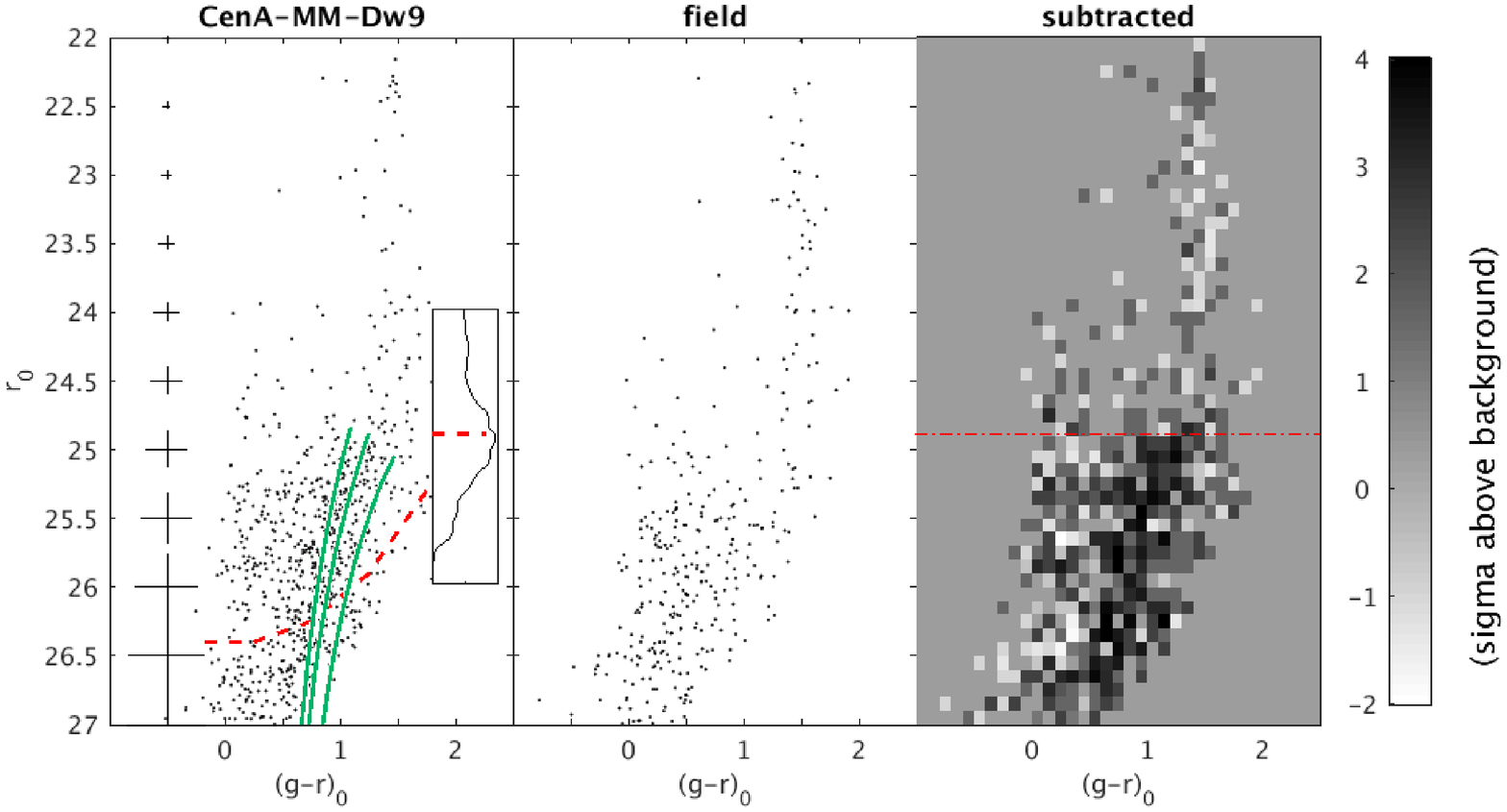}
\includegraphics[width=6.cm]{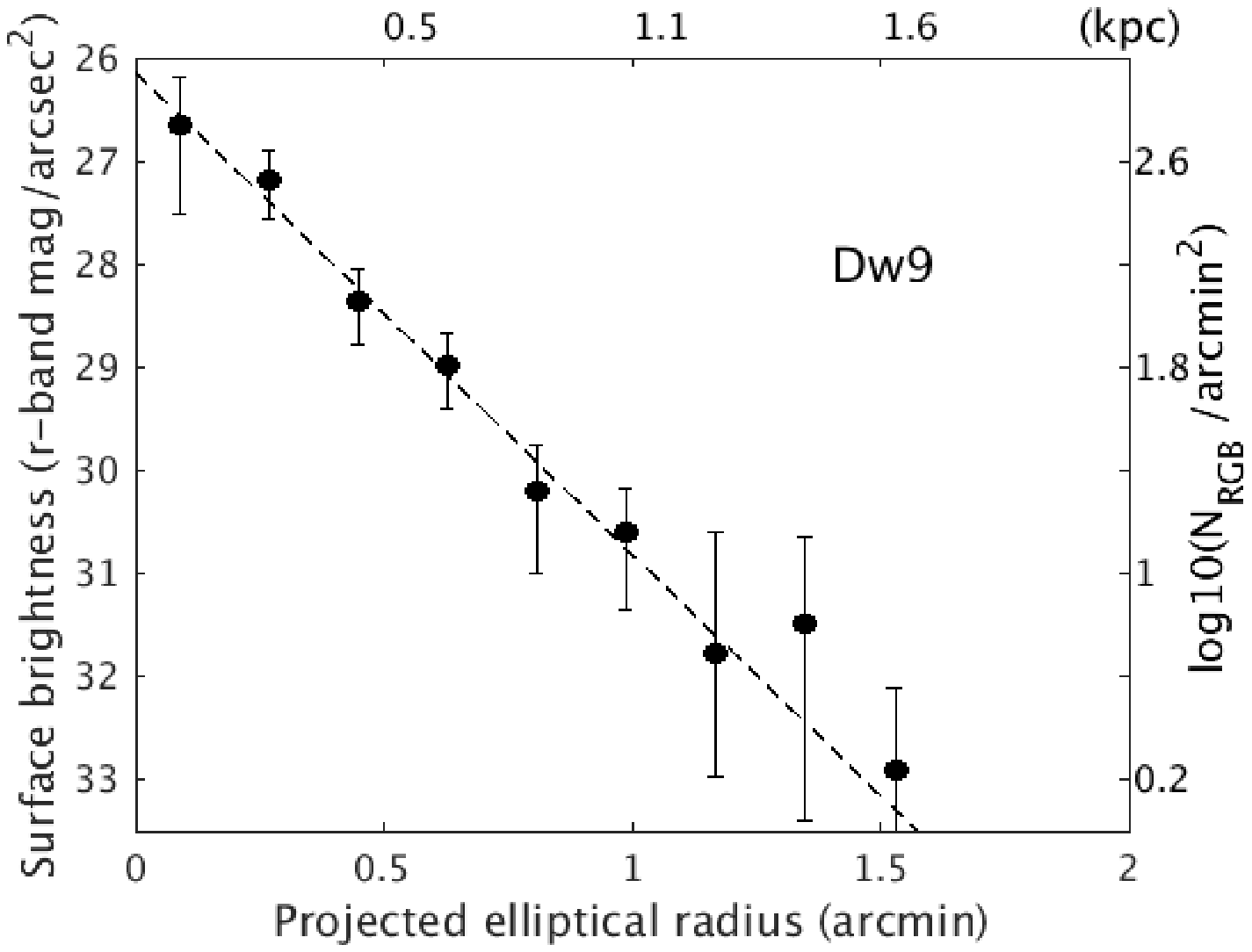}
\includegraphics[width=6.cm]{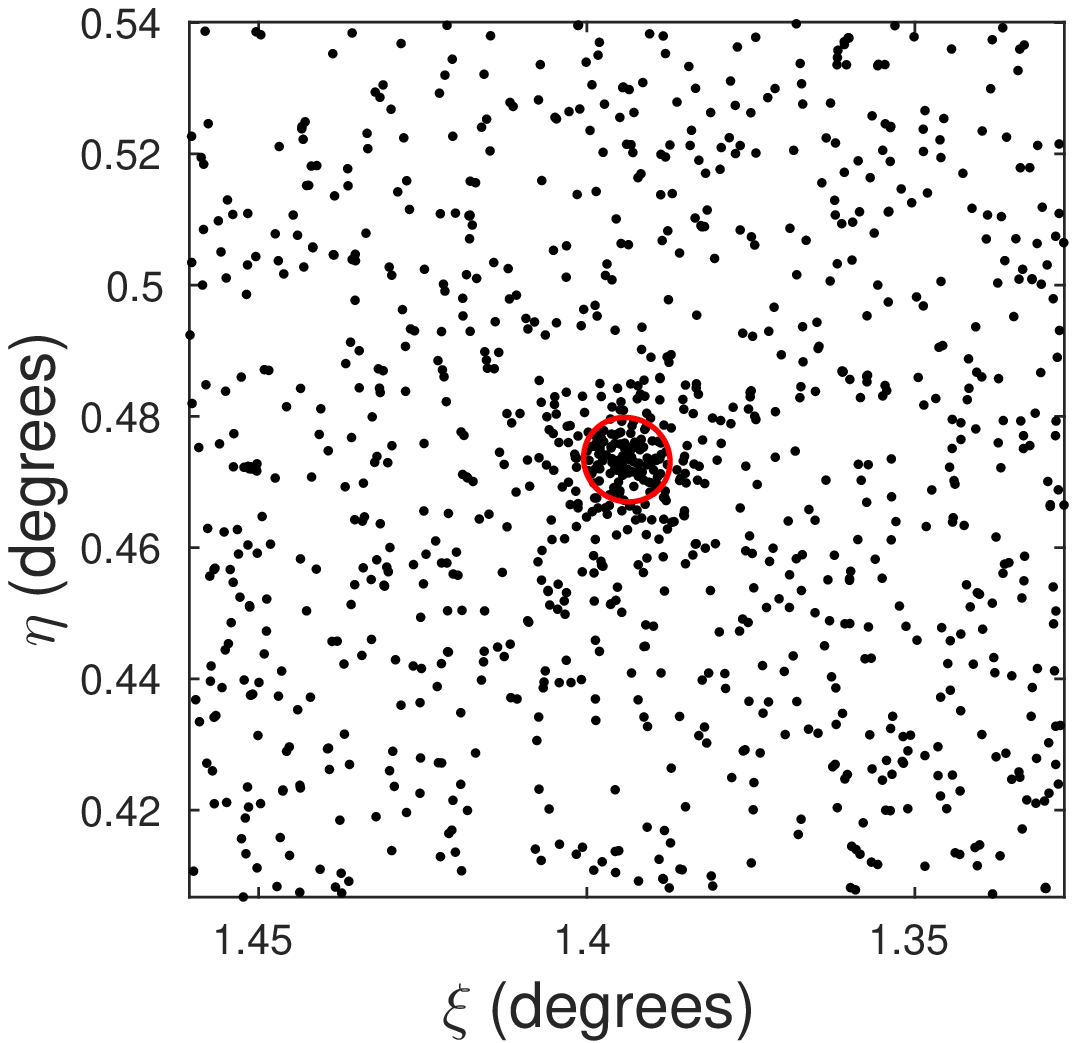}
\caption{Same as Fig~\ref{dw8_figs}, for CenA-MM-Dw9.
}
\label{dw18_figs}
\end{figure*}

\subsection{Properties of the most robust newly discovered
  satellites} \label{sec:cmds_sats}

In this subsection, we present the discovery CMDs, stellar 
spatial distributions and structural properties for our six robust 
new satellite discoveries, besides CenA-MM-Dw1, CenA-MM-Dw2 and CenA-MM-Dw3.
Their derived properties are listed in Table~\ref{tab1}.

The CMDs of the newly discovered dwarfs show predominantly 
old populations, and none of them presents significant 
signs of more recent star formation (i.e., blue stars or luminous AGB stars;
upper panels in Figs.~\ref{dw3_figs}--\ref{dw18_figs}). Their RGBs 
are consistent with relatively metal-poor isochrones (indicatively 
[Fe/H]$\sim-2.0/-1.5$), even though the low number of detected 
stars renders a more precise quantification difficult from this dataset.
Because of the sparsity of their RGBs, for each of these six dwarfs
we also report their luminosity functions convolved with the Sobel filter
(see Sect.~\ref{sec:distance}) alongside their CMDs. In most cases this
returns a clear maximum which corresponds to the TRGB, however for
CenA-MM-Dw5 (the least populated in the sample) the luminosity
function presents two comparable maxima. We take the TRGB value to be the average
of the maxima's magnitudes, which translates into a larger uncertainty 
in CenA-MM-Dw5's distance (see Table~\ref{tab1}).
 
As for CenA-MM-Dw3, we find exponential profiles to be a good
fit for all the other dwarfs (lower panels in Figs.~\ref{dw3_figs}--\ref{dw18_figs}).
From the exponential profiles we derive their absolute magnitudes, 
which cover three magnitudes, from $M_V=-7.2$ 
(CenA-MM-Dw5) to $-9.8$ (CenA-MM-Dw4). The best-fit profiles also
return values for their central surface brightness 
($\mu_{V,0}=25.4-26.9$~mag~arcsec$^{-2}$) 
and half-light radii ($r_h=0.22-0.58$~kpc).
None of these galaxies are visible in Digitized Sky Survey (DSS) 
images because of their low central surface
brightness and/or luminosity.

Within these six newly discovered dwarfs, three objects are
relatively small and compact ($r_h<0.31$~arcmin, or $r_h<0.35$~kpc),
namely CenA-MM-Dw4, CenA-MM-Dw5 and CenA-MM-Dw6.
The magnitude, half-light radius, and surface brightness of CenA-MM-Dw4 are 
comparable to those of Carina, AndV, and AndXVII in the LG
\citep[see][for LG dwarfs' properties]{mcconnachie12}.
The properties of CenA-MM-Dw5 closely compare to those of AndXIII and
AndXXVI; this is the faintest PISCeS discovery among our dwarf
candidates' sample, as well as the least luminous galaxy 
discovered beyond the LG, with $M_V=-7.2$ and
 $\mu_{V,0}=26.9$~mag~arcsec$^{-2}$.
CenA-MM-Dw6 is similar in its properties to AndX and AndXXVIII.

The remaining three dwarfs, CenA-MM-Dw7, CenA-MM-Dw8 and
CenA-MM-Dw9, are distinguished by fairly large half-light radii 
($r_h>0.37$~arcmin, or $r_h>0.36$~kpc), and they all have
elongated shapes. In particular, CenA-MM-Dw7, despite 
having a similar luminosity to that of
CenA-MM-Dw6, has a surface brightness one magnitude 
lower and a half-light radius almost twice as large, and it is
comparable to CVnI, AndIX and AndXXVII. Given its
proximity to the center of Cen~A (it is the closest dwarf in projection from
our sample) and its rather elongated shape (Fig.~\ref{dw8_figs}), 
CenA-MM-Dw7 might be in the process of being tidally affected by its
giant host, even though our density maps/surface brightness profile 
do not reveal any signs of structure around the main body of CenA-MM-Dw7
(the significant contamination from the halo of Cen~A
stars at the position of CenA-MM-Dw7 is however evident from its CMD).
The luminosity, surface brightness and half-light radius of CenA-MM-Dw8
closely resemble the recently discovered AndXXV and AndXXI; 
this dwarf is both brighter ($M_V=-9.7$) as well as more diffuse 
($r_h=0.58$~kpc for $\mu_{V,0}=26.6$~mag~arcsec$^{-2}$)
than CenA-MM-Dw7 and CenA-MM-Dw9.
Finally, CenA-MM-Dw9 and AndXIV have similar properties.

\tabletypesize{\scriptsize}
\begin{deluxetable*} {lccccccc}
\tablecolumns{8}
\tablecaption{Properties of the newly discovered dwarfs.}

 \tablehead{\colhead{Parameter}  & \colhead{Dw3} &\colhead{Dw4 } &\colhead{Dw5 } &\colhead{Dw6} &\colhead{Dw7 } &\colhead{Dw8 } &\colhead{Dw9 }}\\

\startdata
RA (h:m:s) & 13:30:21.5$\pm1$'' & 13:23:02.6$\pm1$'' & 13:19:52.4$\pm1$'' & 13:25:57.7$\pm1$''& 13:26:28.7$\pm1$''& 13:33:34.1$\pm1$''& 13:33:01.5$\pm1$''\\
Dec (d:m:s) & $-42$:11:33$\pm9$'' & $-41$:47:11$\pm8$'' & $-41$:59:38$\pm9$'' & $-41$:05:39$\pm9$''& $-43$:33:25$\pm9$''& $-41$:36:29$\pm8$''& $-42$:31:49$\pm9$''\\
$(m-M)_0$ (mag) & $28.32\pm0.19$ & $27.96\pm0.25$ & $27.67\pm0.38$ & $27.78\pm0.22$ & $27.65\pm0.19$ & $27.70\pm0.20$ & $27.90\pm0.20$ \\
D (Mpc) & $4.61\pm0.42$& $3.91\pm0.48$& $3.42\pm0.65$ & $3.61\pm0.38$& $3.38\pm0.32$ & $3.47\pm0.33$& $3.81\pm0.36$\\
$\epsilon$ & $0.29\pm0.19$& $<$0.30\tablenotemark{c} & $<$0.61\tablenotemark{c} & $<$0.56\tablenotemark{c} & $0.28\pm0.14$ & $0.26\pm0.22$ & $0.13\pm0.12$ \\
$r_{h}$ (arcmin) & $2.21\pm0.15$ & $0.31\pm0.08$ & $0.21\pm0.04$ & $0.29\pm0.01$ & $0.37\pm0.10$ & $0.60\pm0.06$& $0.39\pm0.02$ \\ 
$r_{h}$ (kpc) & $2.92\pm0.20$\tablenotemark{a} & $0.35\pm0.10$ & $0.22\pm0.04$ & $0.30\pm0.01$ & $0.36\pm0.09$ & $0.58\pm0.05$& $0.42\pm0.03$ \\ 
$\mu_{r,0}$ (mag~arcsec$^{-2}$) & $26.3\pm0.1$ & $25.0\pm0.5$ & $26.5\pm0.5$ & $25.5\pm0.1$ & $26.3\pm0.7$ &$26.2\pm0.3$& $26.1\pm0.2$ \\ 
$\mu_{V,0}$ (mag~arcsec$^{-2}$) & $26.7\pm0.1$ & $25.4\pm0.7$ & $26.9\pm0.7$ & $25.9\pm0.1$ &$26.7\pm0.9$ & $26.6\pm0.4$& $26.6\pm0.3$ \\ 
$M_V$ (mag) &$-13.0\pm0.4$\tablenotemark{b} &$-9.8\pm1.1$ &$-7.2\pm1.0$ &$-9.0\pm0.4$& $-8.6\pm1.3$& $-9.7\pm0.5$&$-9.1\pm0.4$ \\
$L_*$ ($10^5 L_\odot$) &$129.2\pm44.3$ &$6.8\pm8.4$ &$0.6\pm0.7$ &$3.4\pm1.1$ &$2.4\pm3.4$&$6.7\pm3.5$&$3.6\pm1.4$\\
$M_{HI}$\tablenotemark{d} ($10^6 M_\odot$) & $\lesssim5.7$ & $\lesssim4.6$ & $\lesssim3.6$ & $\lesssim3.4$ & $\lesssim4.2$ & $\lesssim2.3$ & $\lesssim4.4$\\
$M_{HI}/L_*$ ($M_\odot/L_\odot$) & $\lesssim0.4$ & $\lesssim6.8$ & $\lesssim59.7$ & $\lesssim10.0$ & $\lesssim17.4$ & $\lesssim3.4$ & $\lesssim12.3$\\
\enddata

\tablenotetext{a}{This is an indicative value only, since the studied galaxy is being heavily disrupted.}
\tablenotetext{b}{Excluding tidal tails.}
\tablenotetext{c}{Only upper limits on the ellipticity, $\epsilon$, were measurable.}
\tablenotetext{d}{5~$\sigma$ upper limits from HIPASS.}
\label{tab1}

\end{deluxetable*}

\begin{figure*}
 \centering
\includegraphics[width=12cm]{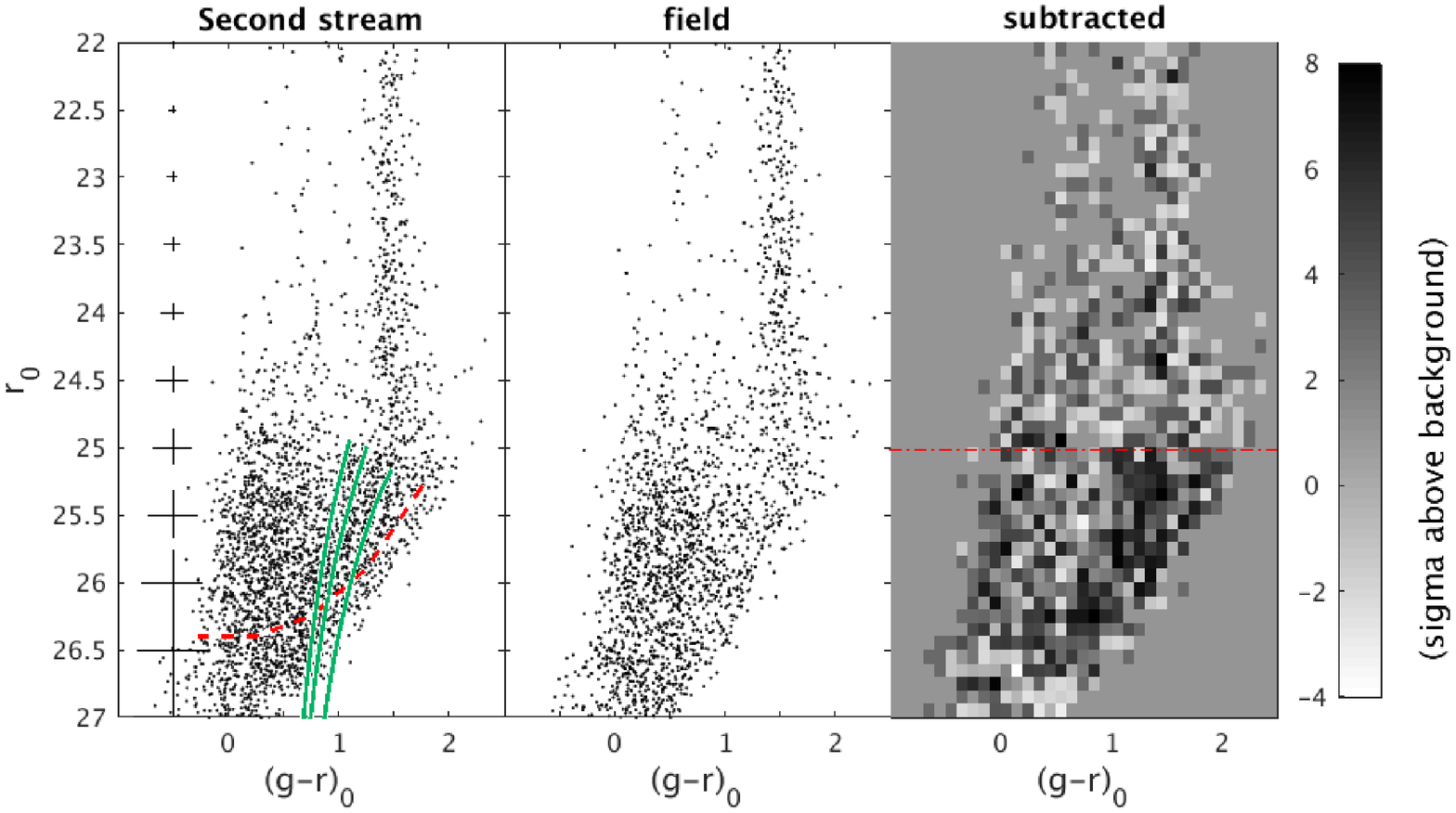}
\includegraphics[width=12cm]{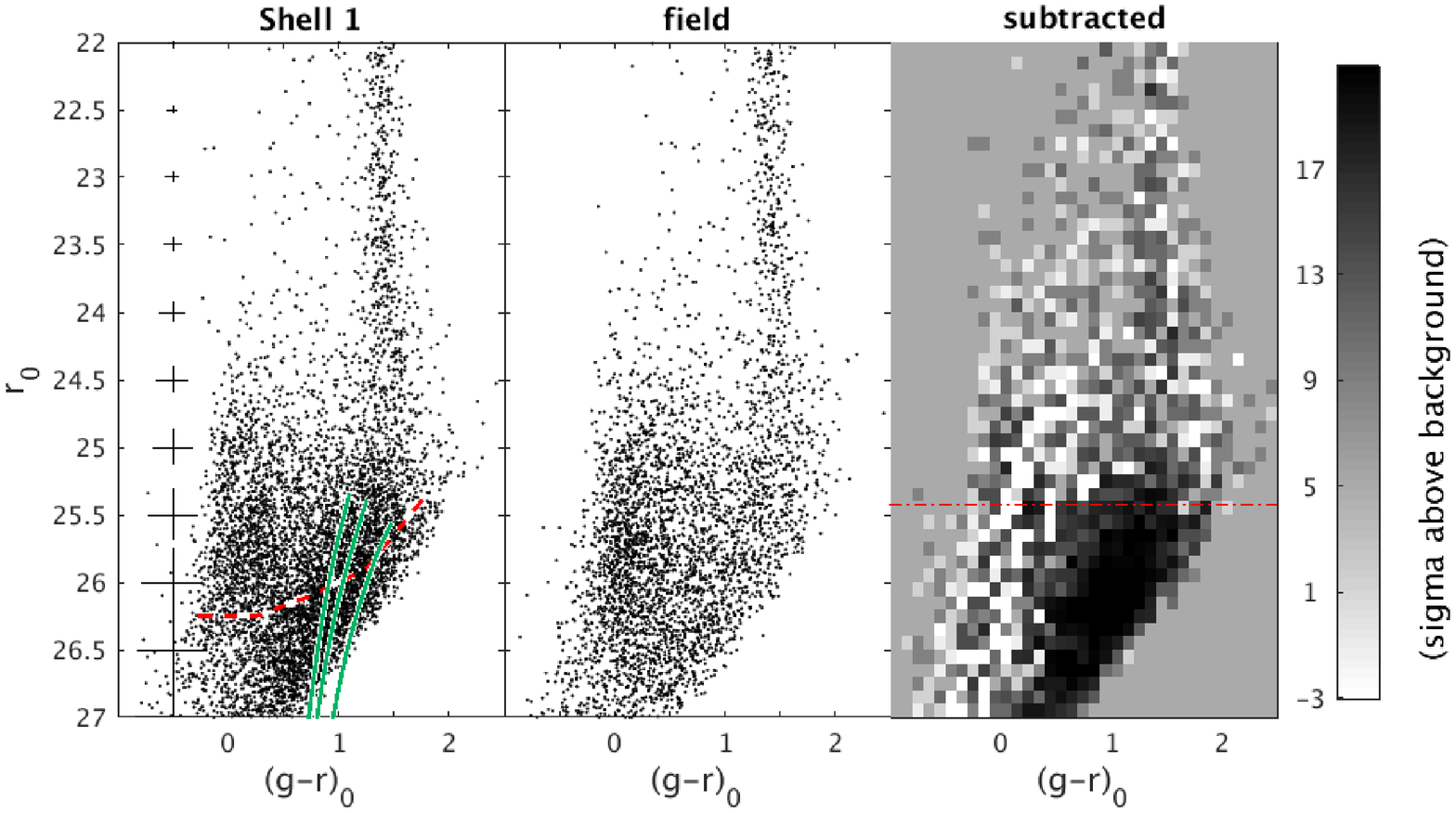}
\includegraphics[width=12cm]{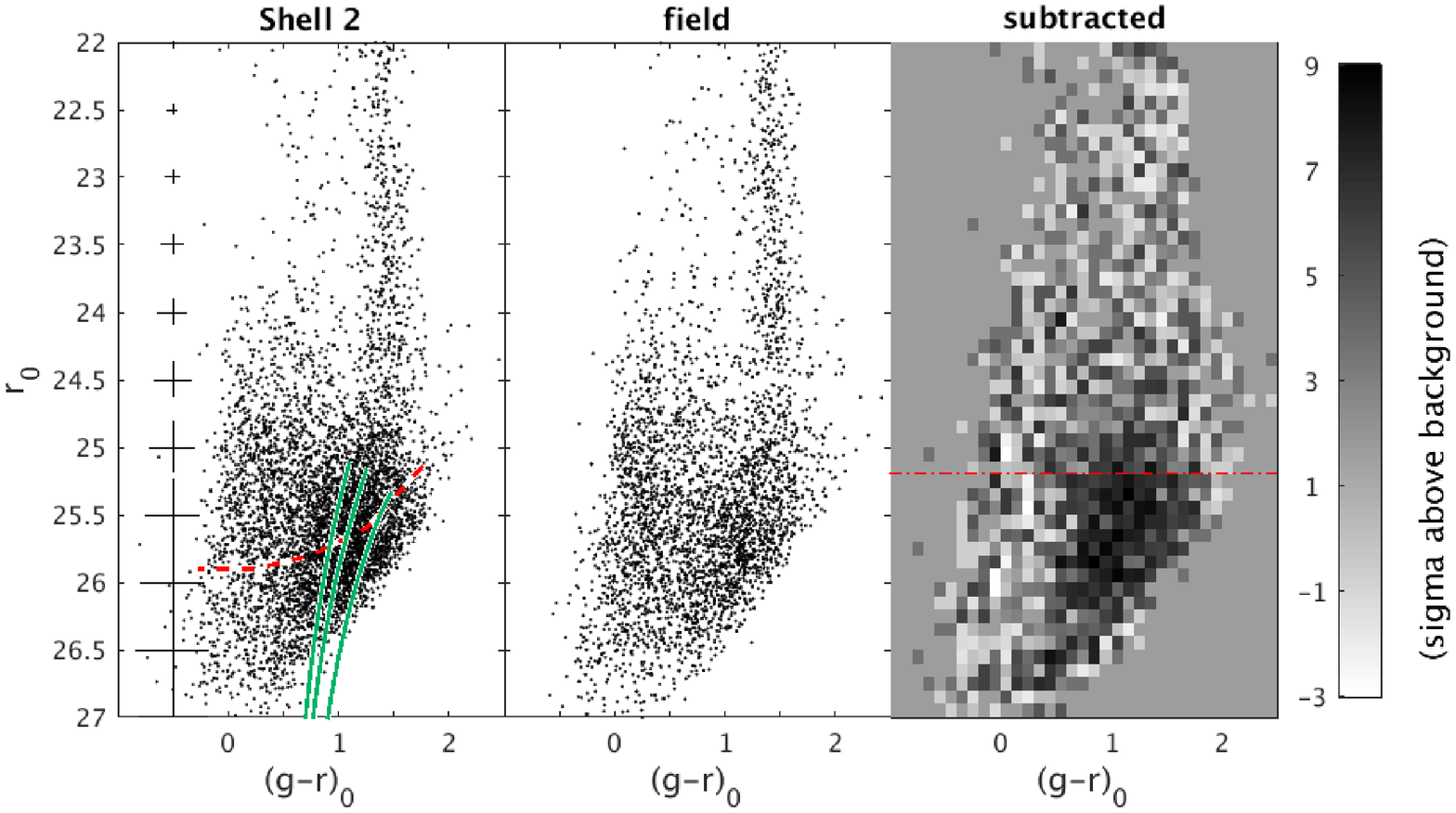}
\caption{Further CMDs of prominent Cen~A halo substructures: 
the apparently progenitor-less Second stream and the 
shells located to the North of Cen~A (Shell 1) 
and to its South-West (Shell 2; see Fig.~\ref{spat_box}). 
}
\label{cmds_spati3}
\end{figure*}

\subsection{Additional shells and streams} 

Besides the prominent tidal structure of CenA-MM-Dw3, the halo of Cen~A
hosts a variety of other previously unknown overdensities.
We focus on four of them, as selected from Figs.~\ref{spat_zoom}
and \ref{spat_box} (coordinates for these features are found in
Table~\ref{tab0}).

\emph{Second stream}. This stream is clearly identified to the East of 
the main body of Cen~A, and South of CenA-MM-Dw3. This stream, thinner with respect to 
the tails of CenA-MM-Dw3, does not host a clear compact remnant. We draw a CMD
from its densest portion (upper panel of  Fig.~\ref{cmds_spati3}),
which might contain what is left of its progenitor. At this point, 
the stream seems to bend back towards Cen~A, and 
significantly drops in stellar density. On the opposite end, the
stream points towards the northern
part of the Cen~A major axis overdensity \citep{crnojevic13} but fades
before reaching it, such that a possible connection between the two
substructures cannot be assessed. 
With the TRGB method described in Sect.~\ref{sec:distance},
we derive the distance to this second stream, finding 
$r_{0,\rm TRGB}=25.02\pm0.16$, or $(m-M)_0=28.03\pm0.19$.
From its CMD, the second stream seems to host an old RGB without the
presence of younger populations.
We argue that the progenitor of this second stream must have not been
very massive, and its remnant may be already stretched to the
point that we cannot recognize any signs of its original appearance.

\emph{Shell 1}. We call the feature directly to the North of Cen~A  Shell 1, which
becomes apparent to the West of the major axis and assumes a loop-like
shape for $\sim0.25$~deg. 
The substructure apparently does not extend beyond this one pointing,
but this may be due to incompleteness effects.
The adjacent pointing to its West has a lower completeness,
but even after our incompleteness correction (see Fig.~\ref{spat_box})
the Shell 1 structure does not seem to be detected further in this direction.
The CMD in Fig.~\ref{cmds_spati3} shows a very high stellar density,
such that crowding introduces significant incompleteness. The RGB
along Shell 1 appears broad, likely extending to large colors/metallicities 
given the rapidly dropping completeness. 
The TRGB is found at $r_{0,\rm TRGB}=25.42\pm0.17$, giving $(m-M)_0=28.43\pm0.20$,
a difference of $\sim2.5$~$\sigma$ from the distance of Cen~A, but this value
requires confirmation due to the high uncertainty.

\emph{Shell 2}. Located directly to the West of the center of Cen~A, Shell 2 resembles the
shape of Shell 1. We can trace this substructure extending North from
its highest density point (a possible remnant, see
Figs.~\ref{spat_glob} and \ref{spat_box}), the latter being
clearly separated by a density gap from the inner halo of Cen~A and 
running parallel to it. Beyond this pointing, the shape of Shell 2
becomes less defined and constitutes a lower density ``envelope'' to
the smooth component of Cen~A, but it does retain a sharp break with respect
to the region to its immediate West (see Fig.~\ref{spat_glob}).
The CMD for Shell 2 returns $r_{0,\rm TRGB}=25.19\pm0.17$, and
$(m-M)_0=28.20\pm0.20$.
As for Shell 1, the incompleteness due to crowding is high at this
galactocentric distance.

\emph{Cloud S}. We additionally detect a stellar density enhancement to the South of
Cen~A, i.e., a broad cloud-like structure extending from
$\xi\sim-0.8^{\circ}$ to $+0.4^{\circ}$ at $\eta\sim-1^{\circ}$, which we call Cloud S 
(labeled in Fig.~\ref{spat_zoom}). The stellar density with respect to the background halo 
is very low and we do not show a CMD for this region, but its extension
across a number of Megacam pointings suggests that it is
a real structure. A similar feature can be seen in the halo of M31, although the
latter has a higher stellar density (SW Cloud, e.g.,
\citealt{bate14}). From its shape, it does not seem to be connected
to any of the other substructures presented here.


\section{Discussion and Conclusions} \label{sec:concl}

We have presented the widest-field RGB map to date of the extended 
halo of Cen~A (Figs.~\ref{spat_glob} and \ref{spat_zoom}), the closest 
elliptical galaxy. Our deep Magellan/Megacam 
imaging allows us to resolve old red giant branch populations out to a 
projected galactocentric radius of $\sim150$~kpc, reaching very low
surface brightness values ($\mu_{V,0}\sim32$~mag~arcsec$^{-2}$) and 
uncovering previously unknown tidally induced features
and low-mass galaxies.
In this contribution we described the most prominent of these
features, including a disrupting satellite, pointing to the
active interaction/accretion history for Cen~A.
We additionally introduced 13 newly discovered candidate satellites 
of which 9 are confirmed to be at the 
distance of Cen~A (including the disrupting object), with absolute magnitudes 
in the range $M_V=-7.2$ to $-13.0$, central surface brightness values of 
$\mu_{V,0}=25.4-26.9$~mag~arcsec$^{-2}$, and half-light radii of $r_h=0.22-2.92$~kpc. 
The remaining 4 (unresolved) candidate satellites may be fainter than
our detection limits ($M_V<-7$), or alternatively background 
unresolved galaxies. 
Two HST imaging programs (GO-13856 and GO-14259, PI: Crnojevi\'c)
have already been approved to follow-up the newly discovered
satellites (15 orbits) and the most prominent substructures in the 
halo of Cen~A (20 orbits), respectively.

We can qualitatively compare the number and properties of the most
prominent Cen~A substructures to the ones found in M31 by PAndAS. 
A direct comparison is not straightforward due to the fact that the PAndAS
survey reaches $\sim3.5$~mag below the TRGB \citep[e.g.,][]{ibata14},
while PISCeS is limited to $\sim1.5$ -- 2~mag because of the larger distance
of our target galaxy ($D\sim3.8$~Mpc).
Prior to PAndAS, the INT/WFC survey of the inner regions of M31 
\citep{ferguson02} led the charge with a combination of wide-field
ground-based data and deep HST follow-up imaging 
\citep{ferguson05, richardson08, richardson09, bernard15}.  We seek to 
emulate this model with the current generation of wide-field imagers 
(e.g., Magellan/Megacam, Subaru/HyperSuprimeCam).

The PISCeS coverage of Cen~A to date reveals two clear tidal
streams, of which one still hosts an easily recognisable remnant
(CenA-MM-Dw3), at least two shell-like features in the inner 
regions of the galaxy (Shell 1 and 2), and one diffuse ``cloud'' 
(Cloud S). The additional inhomogeneity in the central regions of Cen~A 
is certainly due to accretion/interaction events, but caution should be used
when analyzing this evidence in more detail, as the observing 
conditions for the pointings of interest vary greatly. 
As a first, crude comparison to the PAndAS
discoveries, our Shell 1 and 2 resemble a scaled-up version of 
the NGC205 Loop or the NE Shelf \citep{ferguson05}.
While our disrupting dwarf CenA-MM-Dw3 does not have a 
counterpart in M31 (NGC147 is more luminous and its tails are
fainter and less elongated; e.g., \citealt{crnojevic14a}), the second
stream uncovered by PISCeS is similar to Stream C or D
to the South-East of M31 \citep[e.g.,][]{ibata07, lewis13}.
As already mentioned, our Cloud S can be compared to the feature
called the SW Cloud \citep{bate14}, even though the former is fainter
and less spatially defined. The faintest feature in PAndAS is
Stream A, at $\mu_{V,0}\sim32$~mag~arcsec$^{-2}$ \citep{ibata07},
while our faintest prominent feature (which we dub Dw3 NW) is located to 
the North-West of the stream of CenA-MM-Dw3, 
although it probably does not belong to the latter.
We estimate Dw3 NW to have 
$\mu_{V,0}\sim30.5$~mag~arcsec$^{-2}$ based on the conversion from
RGB number density to surface brightness of CenA-MM-Dw4
(see Fig.~\ref{dw3_figs}), i.e., the closest dwarf to Dw3 NW.
The overall number of stream-like features
in the halo of M31 seems to be higher than that found for Cen~A,
down to the PISCeS surface brightness limit, which is
surprising given that the mass of Cen~A is likely higher 
($\sim10^{12} M_{\odot}$ within the innermost 90~kpc, 
\citealt{woodley07}).
However, simulations predict a large variety in the number of accretion 
and merger events experienced by a MW-mass halo, and our
PISCeS Cen~A map roughly compares to ``Halo A'' in Fig.~6 
of \citet{cooper10}, in terms of number of features above a 
surface brightness of $\mu_{V,0}\sim30.5$~mag~arcsec$^{-2}$.

\begin{figure*}
 \centering
\includegraphics[width=8.5cm]{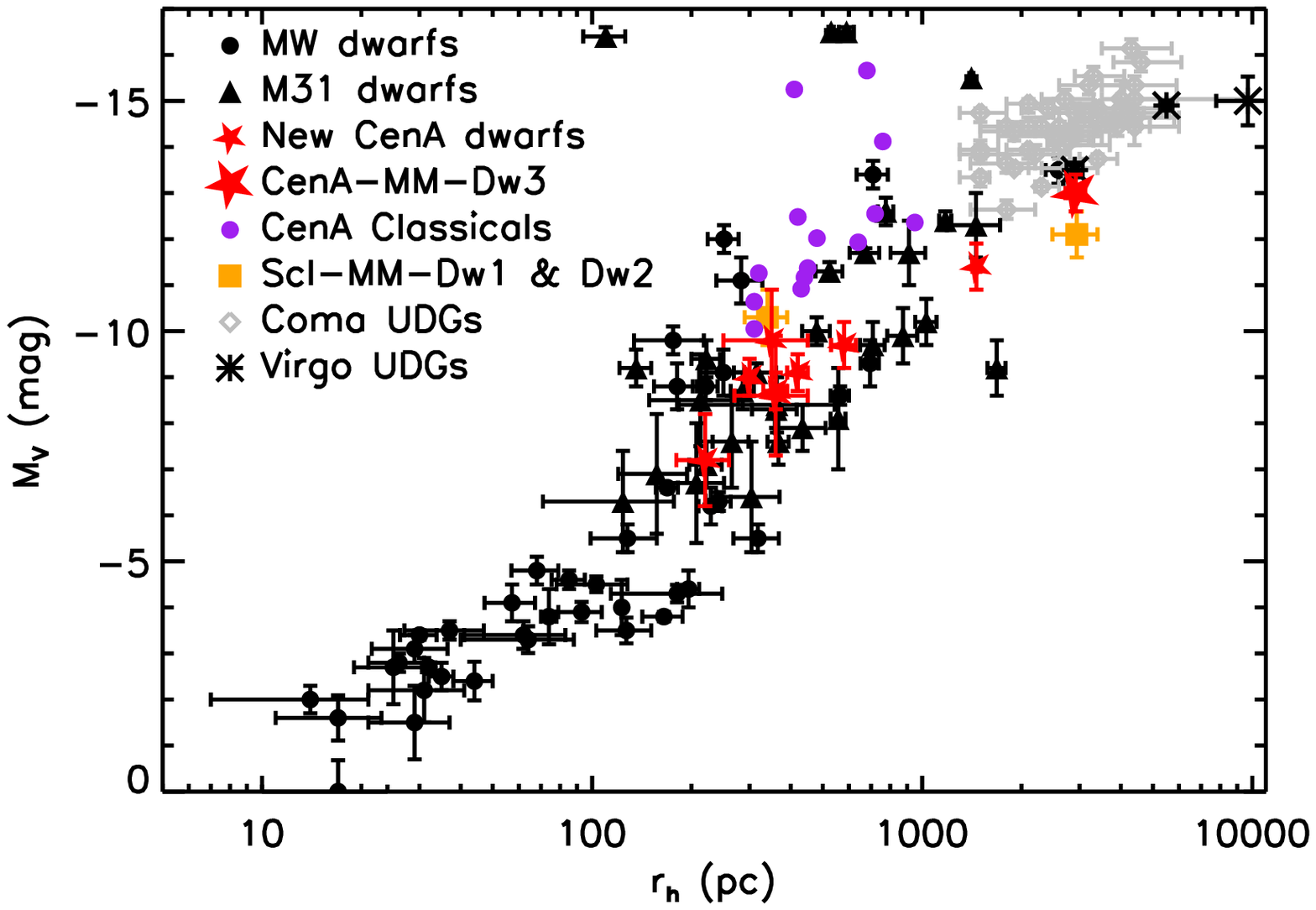}
\includegraphics[width=8.5cm]{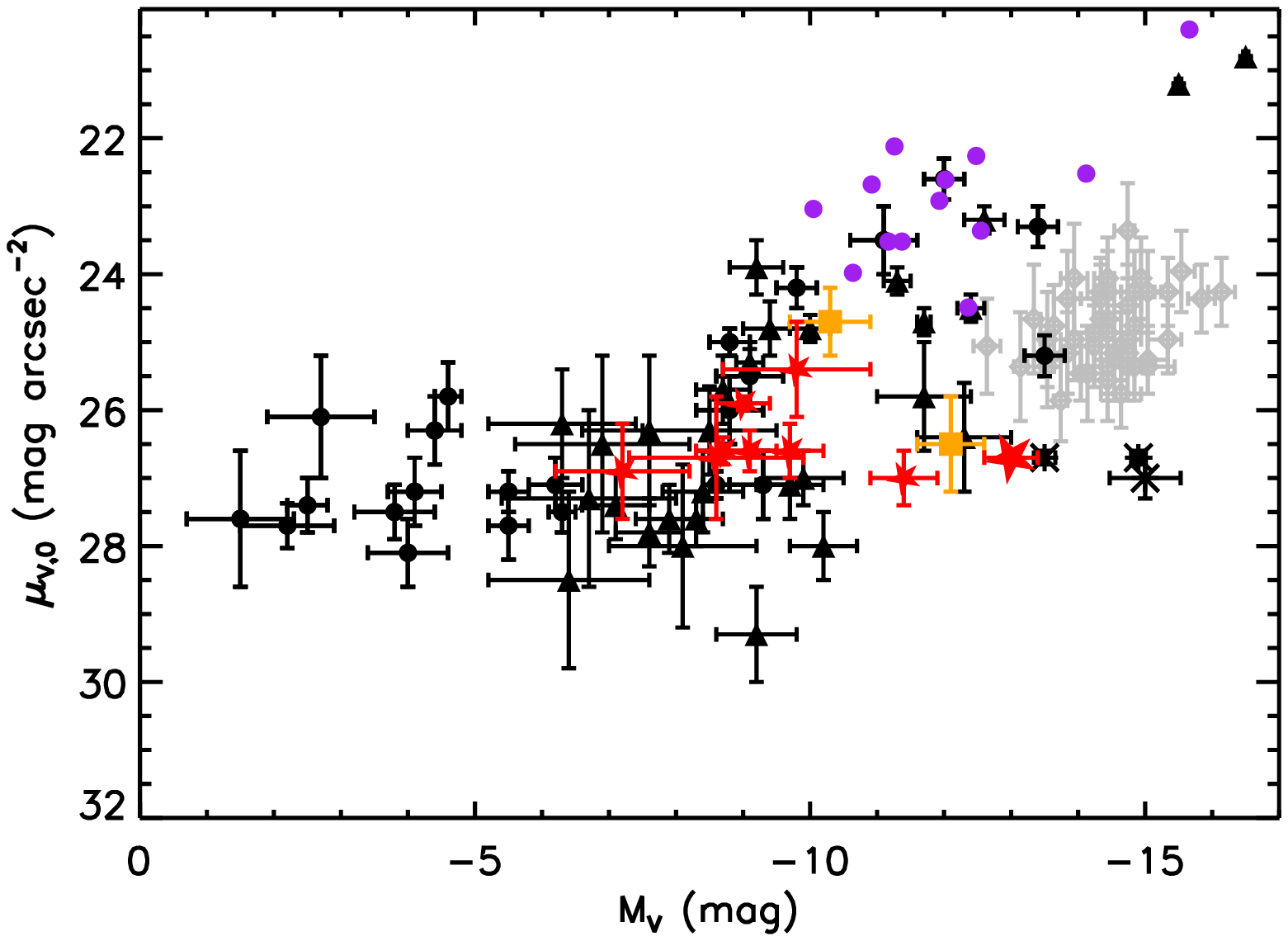}
\caption{\emph{Left panel}: Absolute $V$-band magnitude
as a function of half-light radius for: MW/M31 dwarf galaxies 
(black points/triangles, from \citealt{mcconnachie12, sand12,
crnojevic14a, drlica15, kim15a, kim15b, koposov15, laevens15a, laevens15b, martin15});
previously known (``classical'') Cen~A satellites (purple points, updated from 
\citealt{sharina08}); our newly discovered Cen~A dwarfs (red stars);
the remnant of the tidally disrupting CenA-MM-Dw3 (large red star);
the newly discovered NGC~253 satellites found in PISCeS
(yellow squares, \citealt{sand14} and \citealt{toloba15});
and the new faint, diffuse galaxies found in Virgo
and Coma by \citet{mihos15} and \citet{vandokkum15}, respectively
(black asterisks and grey diamonds).
\emph{Right panel}: Central $V$-band surface brightness as
a function of absolute magnitude (symbols as before). 
The newly discovered Cen~A companions have similar properties to those 
of LG dwarfs, except for the tidally disrupting CenA-MM-Dw3 which
has a large half-light radius and low surface brightness for its
absolute magnitude, similarly to Sagittarius in the Local Group,
Scl-MM-Dw2 in Sculptor, and to the recently discovered diffuse Virgo 
galaxies. The new Cen~A satellites have a consistently lower
surface brightness than the ``classical'' Cen~A satellites, and most
of them have lower luminosity as well.
}
\label{rhmv}
\end{figure*}

Moving on to satellites, our discovery of 9 robust Cen~A candidates
from PISCeS can be compared to the M31 dwarfs discovered by the 
PAndAS survey.
Once our survey is completed, by simply rescaling the number of discoveries to date
to the final area we expect to find $\sim6\pm2$ more satellites, 
which is fully consistent with the 14 dwarfs found in PAndAS 
over a similar projected area and within the same magnitude range 
\citep[$-13\lesssim M_V\lesssim-7$; see, e.g.,][]{richardson11}.
A more quantitative estimate of the faint end of the 
satellite luminosity function for Cen~A will be presented
once the full dataset is acquired.

Fig.~\ref{rhmv} shows the global properties of the robust newly discovered
satellites in the context of LG dwarfs (including CenA-MM-Dw1 and 
CenA-MM-Dw2 from \citealt{crnojevic14b}). The two samples cover similar regions
in the absolute magnitude vs central surface brightness and half-light radius 
vs absolute magnitude space. However, two of the PISCeS Cen~A satellites
have outlying properties in these distributions, i.e., they are very
diffuse, with low surface brightness and large half-light radii
for their absolute magnitude. Of these two objects, CenA-MM-Dw1 does
not show any sign of distortion, while CenA-MM-Dw3 is clearly
in an advanced tidal disruption phase. Despite this strong evidence,
the possible location of CenA-MM-Dw3 behind Cen~A and 
the uncertainty on its distance measurement 
imply that we cannot conclusively state that it is gravitationally bound to Cen~A;
our upcoming HST imaging will be crucial to 
definitively assess the distance of CenA-MM-Dw3.
With its large extent and very low surface brightness, CenA-MM-Dw3 
is one of the most extended and least dense objects known at its
luminosity, together with another recent discovery of the PISCeS
survey, i.e., the elongated Sculptor satellite Scl-MM-Dw2 
(\citealt{toloba15}, but see also \citealt{romanowsky16}).
These two satellites are comparable only to Sagittarius within the LG,
however their properties can be compared to the recently 
discovered faint and diffuse galaxies in the Virgo and 
Coma clusters by \citet{mihos15} and \citet{vandokkum15}, respectively.
In particular, the two PISCeS objects have lower absolute magnitudes, 
but similar half-light radii and central surface brightness values to 
these unusual Virgo galaxies (see Fig.~\ref{rhmv}). The latter have
been claimed to belong
to a new class of extremely faint galaxies, and we clearly demonstrate
that similar objects are also found in the Cen~A and Sculptor groups.
This result still holds even if we consider that the 
original (unperturbed) luminosity of CenA-MM-Dw3 might have
been $\sim2$~mag brighter than what we measured (see Sect.~\ref{sec:dw_lum}).
CenA-MM-Dw3 is a heavily disrupting galaxy, and Scl-MM-Dw2 is likely
ongoing a strong tidal interaction as well: this suggests that
the Virgo and Coma diffuse galaxies might as well be experiencing
disruption, but their tidal features, even if present, could not be detected
within the surface brightness limits of these surveys 
($\mu_{V,0}\sim29-27$~mag~arcsec$^{-2}$, respectively).

In the same plots, we also include the ``classical'' Cen~A satellites.
We choose to consider dwarfs for which resolved stellar 
studies are available from the literature, even though some of
them have also structural parameters derived from integrated light
\citep[see, e.g.,][]{jerjen00b}.
We include those dwarfs for which structural parameters have been derived 
by \citet{sharina08}, and for which TRGB
distance measurements are available from \citet{kara07}. There
are a dozen additional ``classical'' Cen~A satellites in
the magnitude range $-15\lesssim M_V\lesssim-10$ \citep[see][]{kara05,
kara07}, but no structural parameter measurements are available in the
literature for those. The original \citet{sharina08} distance moduli 
and extinction values are updated following \citet{kara07} and the
NASA Extragalactic Database\footnote{http://ned.ipac.caltech.edu/},
respectively. The new and previously known satellites are clearly distinct
in their properties. 
While part of this is due to the different techniques used
in deriving the parameters, the ``classical'' Cen~A satellites 
certainly have a significantly higher
central surface brightness/luminosity with respect to the new ones
(see the right panel in Fig.~\ref{rhmv}).
Interestingly, but perhaps unsurprisingly, 
all of the new objects, except for two, are fainter in absolute magnitude
than the ``classical'' ones, demonstrating the effectiveness of 
PISCeS in extending the faint end of the satellite luminosity function for this
elliptical galaxy. The newly discovered satellites bracket the previously known 
ones in terms of half-light radius,
i.e., a few of them have smaller values and others much larger values.
The two dwarfs with brighter magnitudes are very diffuse, which is likely
why they were discovered by PISCeS and not previous work.

Finally, none of the newly discovered dwarfs are detected in neutral gas,
as determined from the HI Parkes All Sky Survey (HIPASS; \citealt{barnes01}).
We derive 5~$\sigma$ upper HI mass limits for our dwarfs and report them in 
Table~\ref{tab1}. HIPASS is not sensitive enough to confirm or
exclude the presence of HI in most of our targets, but we find an interesting
upper limit of $M_{HI}/L_*\lesssim0.4M_{\odot}/L_{\odot,B}$ for 
CenA-MM-Dw3. Our luminosity estimate for this object is a 
lower limit given its disrupting state, and thus the HI mass to light
ratio is certainly even lower than the limit we derive, hinting at a 
genuine lack of gas for this dwarf.
Previous HI studies of Cen~A early-type satellites only find upper limits 
of $M_{HI}\lesssim7\times10^5 M_{\odot}$
with $M_{HI}/L_B\lesssim0.25M_{\odot}/L_{\odot,B}$ for this class of objects 
\citep{bouchard07}, suggesting that the environment of Cen~A
is very effective at removing gas reservoirs in these low-mass dwarfs
\citep[see also][]{crnojevic12}.

We explore the possibility that the new PISCeS dwarfs belong to one of the 
Cen~A planes of satellites recently described by \citet{tully15}.
In the cartesian Supergalactic coordinates plane 
\citep[][see the upper panel of their Fig.~1]{tully15}, these planes appear
as thin, almost parallel distributions of satellites. Most of our new dwarfs
seem to land on top of one of the planes:
CenA-MM-Dw1/2/3/4/6 belong to Plane 1, CenA-MM-Dw7/8/9 are closer to Plane 2, while
CenA-MM-Dw5 is found at an intermediate position between the two planes.
Until velocities are obtained for the new satellites,
the origin of these planes remains speculative.

In future contributions, we will further exploit this 
promising dataset in conjunction with our approved
HST follow-up imaging, which will be completed in mid-2017. 
In particular, we will investigate the smooth versus discrete halo profile 
and stellar content of this elliptical 
\citep[as done by the PAndAS and SPLASH surveys of M31, e.g.,][]{ibata14, gilbert12};
the satellite luminosity function of Cen~A as well as our observational detection
limits \citep[see, e.g.,][for M81 satellites]{chiboucas13};
the star formation histories and metallicities for the newly discovered dwarfs;
and the star forming regions in the inner part of Cen~A
\citep[e.g.,][]{kraft09, neff15b, neff15a}. The HST dataset
will also put firmer constraints on the distances to the newly discovered 
satellites and halo features.

We conclude by underlining the crucial role played by our
resolved PISCeS survey in characterizing the stellar populations
of nearby galaxy halos, including tidal features and faint
satellites, out to unprecented galactocentric distances.
Surveys such as this represent a first step in the 
observational census of nearby galaxy halos and their inhabitants,
allowing for the first time such investigations to be 
pushed beyond the limits of the LG and acting as 
prototype science cases for the next generation
of ground-based and space-borne telescopes.


\section*{Acknowledgements}

We are grateful to the referee for a careful reading of the
manuscript and for his/her useful suggestions that helped
improve this work.
We warmly thank Maureen Conroy, John Roll, Sean Moran and David Osip for 
their prolonged efforts and help related to Megacam.
DC wishes to kindly thank the hospitality of the Mullard Space Science
Laboratory, University College of London, where part
of this work has been carried out.  
DC, DJS, PG and ET acknowledge support from NSF grant AST-1412504;
PG and ET acknowledge additional support from NSF grant AST-1010039;
JDS acknowledges support from NSF grant AST-1412792.
KS acknowledges support from the Natural Sciences and Engineering
Research Council of Canada (NSERC).
This work was supported in part by National Science Foundation Grant No. 
PHYS-1066293 and the Aspen Center for Physics.
This paper uses data products produced by the OIR Telescope
Data Center, supported by the Smithsonian Astrophysical
Observatory.
This research has made use of the NASA/IPAC Extragalactic 
Database (NED) which is operated by the Jet Propulsion Laboratory, 
California Institute of Technology, under contract with the National 
Aeronautics and Space Administration. 


\bibliographystyle{apj}
\bibliography{biblio}

\clearpage


\end{document}